\documentclass[twocolumn,tighten]{aastex631}
\usepackage{tabto}
\usepackage{lineno}
\usepackage{capt-of}
\usepackage{upgreek}

\shorttitle{VaDAR}
\shortauthors{Schwartzman et al.}

\graphicspath{{./}{Figures/}}

\begin{document}

\title{VaDAR: Varstrometry for Dual AGN using Radio interferometry}

\author[0000-0002-6454-861X]{Emma Schwartzman}
\affiliation{U.S. Naval Research Laboratory, 4555 Overlook Ave SW, Washington, DC 20375, USA}
\affiliation{Department of Physics and Astronomy, George Mason University, 4400 University Drive, MSN 3F3, Fairfax, VA 22030, USA}

\author[0000-0001-6812-7938]{Tracy E. Clarke}
\affiliation{U.S. Naval Research Laboratory, 4555 Overlook Ave SW, Washington, DC 20375, USA}

\author[0000-0003-1991-370X]{Kristina Nyland}
\affiliation{U.S. Naval Research Laboratory, 4555 Overlook Ave SW, Washington, DC 20375, USA}

\author[0000-0002-4902-8077]{Nathan J. Secrest}
\affiliation{U.S. Naval Observatory, 3450 Massachusetts Avenue NW, Washington, DC 20392, USA}

\author[0000-0001-8640-8522]{Ryan W. Pfeifle}
\altaffiliation{NASA Postdoctoral Program Fellow}
\affiliation{X-ray Astrophysics Laboratory, NASA Goddard Space Flight Center, Code 662, Greenbelt, MD 20771, USA}

\author[0000-0003-2450-3246]{Henrique Schmitt}
\affiliation{U.S. Naval Research Laboratory, 4555 Overlook Ave SW, Washington, DC 20375, USA}

\author[0000-0003-2277-2354]{Shobita Satyapal}
\affiliation{Department of Physics and Astronomy, George Mason University, 4400 University Drive, MSN 3F3, Fairfax, VA 22030, USA}

\author[0000-0003-2283-2185]{Barry Rothberg}
\affiliation{U.S. Naval Observatory, 3450 Massachusetts Avenue NW, Washington, DC 20392, USA}
\affiliation{Department of Physics and Astronomy, George Mason University, 4400 University Drive, MSN 3F3, Fairfax, VA 22030, USA}

\begin{abstract} 
Binary and dual active galactic nuclei (AGN) are an important observational tool for studying the formation and dynamical evolution of galaxies and supermassive black holes (SMBHs). An entirely new method for identifying possible AGN pairs makes use of the exquisite positional accuracy of Gaia to detect astrometrically-variable quasars, in tandem with the high spatial resolution of the Karl G. Jansky Very Large Array (VLA). We present a new pilot study of radio observations of 18 quasars ($0.8 \leq z \leq 2.9$), selected from the SDSS DR16Q and matched with the Gaia DR3. All 18 targets are identified by their excess astrometric noise in Gaia. We targeted these 18 quasars with the VLA at 2-4 GHz (S-band) and 8-12 GHz (X-band), providing resolutions of 0.65$^{\prime\prime}$ and 0.2$^{\prime\prime}$, respectively, in order to constrain the origin of this variability. We combine these data with ancillary radio survey data and perform radio spectral modeling. The new observations are used to constrain the driver of the excess astrometric noise. We find that $\sim$44\% of the target sample is likely to be either candidate dual AGN or gravitationally lensed quasars. Ultimately, we use this new strategy to help identify and understand this sample of astrometrically-variable quasars, demonstrating the potential of this method for systematically identifying kpc-scale dual quasars.

\end{abstract}

\keywords{Radio active galactic nuclei (2134) --- Radio astronomy (1338) --- Double quasars (406)}
\section{Introduction} \label{sec:intro}

In the canonical model of galaxy evolution, the formation of more massive galaxies proceeds via a series of major and minor mergers of their smaller counterparts, in addition to gas and dark matter accretion \cite[e.g.,][]{schweizer1982, schweizer1996, toomre1972, barnes1992, hibbard1996, Somerville_2015}. Most massive galaxies contain a central supermassive black hole (SMBH; \cite{kormendy1995}), generally of mass $10^6 - 10^{10} M_{\odot}$. Thus, galaxy mergers should result in pairs of gravitationally-bound, synchronously feeding SMBHs that travel to the center of their host merger via dynamical friction, evolving from dual systems to binary systems that are expected to coalesce \cite[][]{kormendy1995,barnes1992,hopkins2008,volonteri2016}. Given the apparent scaling relations between the SMBHs and their host galaxy bulge properties, a potential co-evolution exists between an SMBH and its host galaxy \cite[][]{ferrassee2000,gebhardt2000,heckman2014}. 

The timescale of a galaxy merger is on the order of hundreds of millions to a few billion years \cite[][]{tremmel2018, Callegari_2009}. Dual SMBHs are precursor systems, defined as having two SMBHs with a separation beyond their mutual gravitational spheres of influence; i.e. separations of $\leq$ 100 kpc \cite[][]{ellison2011,liu2011,burkespolaor2018}. Binary systems are a more advanced stage of evolution, having separations within their mutual spheres of influence; i.e. separations $\leq$ 100 pc \cite[][]{ellison2011,liu2011,rodriguez2006}. 

Characterization of these systems at all stages of evolution is critical for understanding SMBH evolution and co-evolution with host galaxies. This is made difficult by the lack of robust detections of dual and binary SMBH systems, particularly in the binary SMBH regime. During the course of the merger, gravitational torques drive gas towards the nuclei \cite[e.g.][]{barnes1992,barnes1996,mihos1996,burkespolaor2018,khan2020}, which infalls and causes gas accretion onto the SMBHs. The SMBHs then ignite as active galactic nuclei \cite[AGN;][]{hopkins2008,hopkins2010,blecha2018}. 

AGN pair identification methods include multiple velocity peaks and/or coronal lines in optical/infrared spectroscopy \cite[][]{comerford2009,comerford2013,comerford2015,wang2009,wang2017,liu2010,liu2011,liu2013,lyu2016,smith2010,xu2009,barrows2013,u2013}, spatially-resolved imaging in the X-ray, optical, infrared, or radio regimes \cite[][]{liu2010,liu2013,barrows2013,liu2018,liu10b,komossa2003,koss2011,koss2012,bianchi2008,piconcelli2010}, and mid-infrared colors \cite[][]{pfeifle2019,pfeifle2019b,satyapal2017,ellison2011,ellison2019}. For dual AGNs, systematic searches have met with varying degrees of success. There are on the order of 50 robustly confirmed systems, many of which were originally identified serendipitously. The population is biased towards local (z $<$ 0.1) redshifts and physical separation $1 < r_p < 100$ kiloparsecs (where $r_p$ is the physical separation), as will be demonstrated in the forthcoming literature-complete catalog of multi-AGN systems (Pfeifle, in prep.).

Of particular interest are candidates and confirmed systems with redshifts at which both the number density of luminous quasars and the global star formation rate density peak, often referred to as ``cosmic noon," \citep[$1 \leq z \leq 3$, ][]{richards2006,madau2014}. However, the dynamical evolution timescales of dual and binary AGN systems are comparatively short with respect to other evolutionary stages \cite[][]{tremmel2018}, and thus there are a substantially lower numbers of these systems predicted on the scale of tens of kpc, as they quickly ($<$ 1 Gyr) advance to the next stage of evolution \cite[][]{tremmel2018, merritt2013, chen2020, yu2002}. This issue is further compounded by the sample pre-selection criteria and the limited resolutions and sensitivities of current telescopes, leaving an observational gap between $r_p < 6$ kpc at $z > 1$ (see Figure 1 in \citealt{chen_hst}). At small separations (i.e. sub-arcsecond), the resolutions of radio interferometry, space-based optical imaging, or ground-based imaging with large apertures, adaptive optics, or optical/infrared interferometry are required.

The focus of this paper is to describe a method for using radio interferometric observations to select dual AGN candidates. The National Science Foundation's Karl G. Jansky Very Large Array (VLA) can reach the high angular resolution necessary to detect dual and binary AGN systems with separations on sub-arcsecond scales. Given their rarity, a pre-selection strategy to select promising targets for radio follow-up is crucial.

The paper is organized as follows. In Section \ref{sec:vars}, we introduce the varstrometry method. In Section \ref{sec:targetselection}, we define our target selection process. In Section \ref{sec:observations}, we describe the observations, data reduction, and analysis. We present our sample properties in Section \ref{sec:results} and radio spectral modeling in Section \ref{sec:sedmodeling}. Our findings are discussed in Section \ref{sec:discussion}. Throughout this paper, all physical separations are the projected separation ($r_p$). A flat $\Lambda$CDM cosmology is adopted, with a $\Omega_{\Lambda}$ = 0.69, $\Omega_{m}$ = 0.31, and $H_{0}$ = 67.7 km s$^{-1}$ Mpc$^{-1}$ \cite[][]{planck2020}.

\section{Varstrometry} \label{sec:vars}

A novel astrometric technique, previously used to detect unresolved stellar binaries \citep[e.g.,][]{2016ApJS..224...19M} through photometric variability-induced photocenter pseudo-motion, was applied to search for unresolved dual AGNs by \cite{hwang_initial}. This technique, dubbed ``varstrometry'' in the context of dual AGN searches by \citet{hwang_initial}, leverages the unprecedented astrometric precision of Gaia \cite[][]{gaia2016}, which has mapped the positions, parallaxes, and proper motions of billions of stars in the Milky Way. It is sensitive enough to provide unprecedented sensitivity to the positions of hundreds of thousands of distant quasars \cite[][]{gaia2021}. While this sensitivity has confirmed the presence of wavelength-dependent positional offsets in jetted AGNs \cite[e.g.][]{Makarov_2017,Makarov_2019}, it has also revealed the presence of astrometrically variable AGN \cite[e.g.][]{shen_nature,chen_hst}. Because \textit{Gaia} is progressively source-confused for separations less than $\sim$ 2$\arcsec{}$ \cite[2$\arcsec{}$ corresponds to $>$ 12.3 kiloparsecs for z $>$ 0.5;][]{fabricius2021}, it is not capable of discerning a close secondary AGN or other extended phenomena associated with the immediate AGN environment, such as jet production. However, the sub-milliarcsecond astrometric precision of \textit{Gaia} allows for the motions of unresolved quasars to be detected. This makes astrometric variability a new discovery space for multi-AGNs and quasars. 

\begin{figure}[ht!]
    \centering
     \includegraphics[width=8cm]{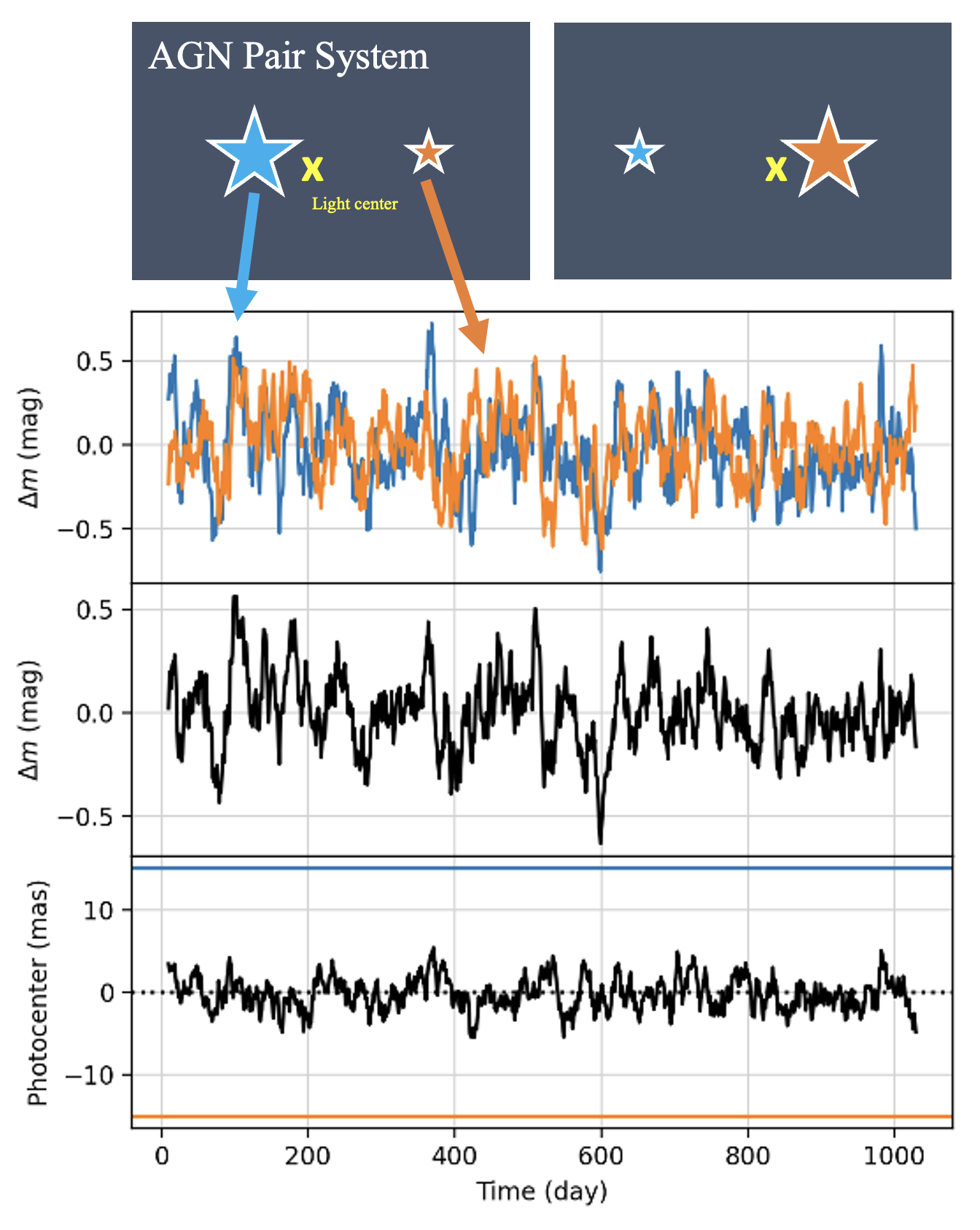}
        \vspace*{-1mm}
          \caption{\textbf{The Varstrometry Technique:} Top: A cartoon illustrating the stochastic variability of a dual AGN including the shifting of the light center between the two component AGN. Bottom: As an example, an AGN photometric lightcurve simulation, as seen by \textit{Gaia}, described by a first-order continuous autoregressive process \cite[e.g. damped random walk;][]{macleod2010randomwalk} with $\tau=10$ day. A typical variability amplitude for AGN of about 0.1 magnitudes was assumed \cite[e.g.][]{sesar2011variability}, as well as a pair angular separation of 30 milliarcseconds. The middle plot shows a joint variability lightcurve, which in itself indistinguishable from systems where the apparent photocenter of the AGN is different from that of its host galaxy. The bottom plot shows the astrometric ``jitter:'' the photocenter of the pair appears to wander back and forth. Top figures adapted from \cite{shen_nature}.}
   \label{fig:varstrometry}
\end{figure}

The orbital periods of binary and dual AGNs are respectively hundreds to millions of years \cite[e.g., Figure 2 in][]{dorland2020}, and thus their positions are essentially fixed on the sky. This precludes a direct motion measurement by an astrometric mission. However, AGN exhibit intrinsic, stochastic variability measurable on timescales as short as hours \cite[][]{sesar2011variability,MacLeod_2012_variability}. In a dual AGN with a separation less than the effective angular resolution of \textit{Gaia}, two components with varying brightnesses will appear to \textit{Gaia} to have a shifting centroid \cite[][]{hwang_initial}.

\textit{Gaia}'s resolution limits are such that, for an AGN pair system, individual light curves cannot be observed. Instead, a joint variability light curve encodes the stochastic variability of the system (see Figure \ref{fig:varstrometry}). In cases where the apparent photocenter of the AGN is different from that of its host galaxy (e.g., a disturbed/merger system or a system in which the AGN obscuration level changes rapidly), the joint variability lightcurve of that system will be indistinguishable from that of a dual AGN system. Additionally, the varstrometry technique is not only sensitive to AGN pair systems but can also select samples contaminated with star+quasar superposition systems \cite[][]{pfeifle2023}, lensed quasar systems \cite[e.g.][]{Mannucci_2022,Ciurlo_2023,Inada_2012,Inada_2014,O_Dowd_2015}, and any other morphology that might also drive the excess astrometric noise seen in AGN pair systems (e.g.\ single AGN in a host galaxy with bright stellar features).

In the case of AGN pair systems, as the mutual photocenter of the AGN pair wanders between the individual AGNs (again depending on their variability), \textit{Gaia} measures a positional ``jitter", which is representative of the astrometric variations. This measurement also provides a lower limit on the possible physical separation for the AGN pair \cite[$\sigma_{astro}$;][]{hwang_initial}. For a pair of quasars with similar mean flux densities and rms fluctuations, the expected variability amplitude $\sigma_{astro}$ is
\begin{center}
\begin{equation} \label{eq:1}
    \sigma_{astro} \approx \frac{D}{2} \frac{\sqrt{\langle \Delta f^{2} \rangle}}{\bar{f}},
\end{equation}
\end{center}

where \textit{D} is the separation of the component sources, and $\bar{f}$ and $\sqrt{\langle \Delta f^{2} \rangle}$ are the total mean and rms fluxes of the unresolved system. Assuming a typical fractional rms of $\sim$10\% \cite[e.g.][]{MacLeod_2012_variability}, one can calculate the expected astrometric ``jitter" to be $\sim10$ milliarcseconds for every 0.2$\arcsec{}$ of angular separation. Note that this measurement is a lower limit on the angular separation.

High spatial resolution follow-up has been successful for varstrometry-selected samples. \cite{chen_hst} followed-up a sample of 84 \textit{Gaia}-identified AGN pair candidates with the Hubble Space Telescope (HST) and Gemini GMOS optical spectroscopy. Their search revealed two previously unknown dual AGN candidates \cite[][]{chen_hst,shen_nature,chen2022}. They conclude that $\sim40\%$ of their HST resolved pairs are likely to be physical quasar pairs or gravitationally-lensed quasars \cite[][]{chen_hst}.

In this work, we present the first results from a pilot VLA program of varstrometry-selected sources \cite[][]{hwang_initial}. Each source has excessive astrometric variability as measured by \textit{Gaia}. Our primary goal is to place constraints on the morphology of the targets and fully characterize the sample using high-resolution radio interferometric follow-up observations. In particular, we aim to identify any potential AGN pair candidate systems, to robustly constrain the morphologies, luminosities, and radio spectra of the sample in order to understand the sample that the radio+varstrometry method selects for, constrain the driver of the excess astrometric variability, and to gauge the efficiency of the radio+varstrometry method as a pre-selection for more systematic searches for AGN pairs.

\section{Target Selection}
\label{sec:targetselection}

The pilot sample of quasars for this study was selected from the SDSS quasar catalog for DR16 \cite[DR16Q;][]{sdss}, which contains the most complete sample of bona fide spectroscopically-confirmed quasars from SDSS/BOSS. DR16Q was cross-matched to \textit{Gaia}'s Early Data Release 3 \cite[EDR3;][]{gaia2016} to within $1.5^{\prime\prime}$, ensuring spectroscopic fiber coverage of the \textit{Gaia} counterpart \cite[][]{fabricius2021}.

\begin{figure}[ht!]
    \centering
     \includegraphics[width=8cm]{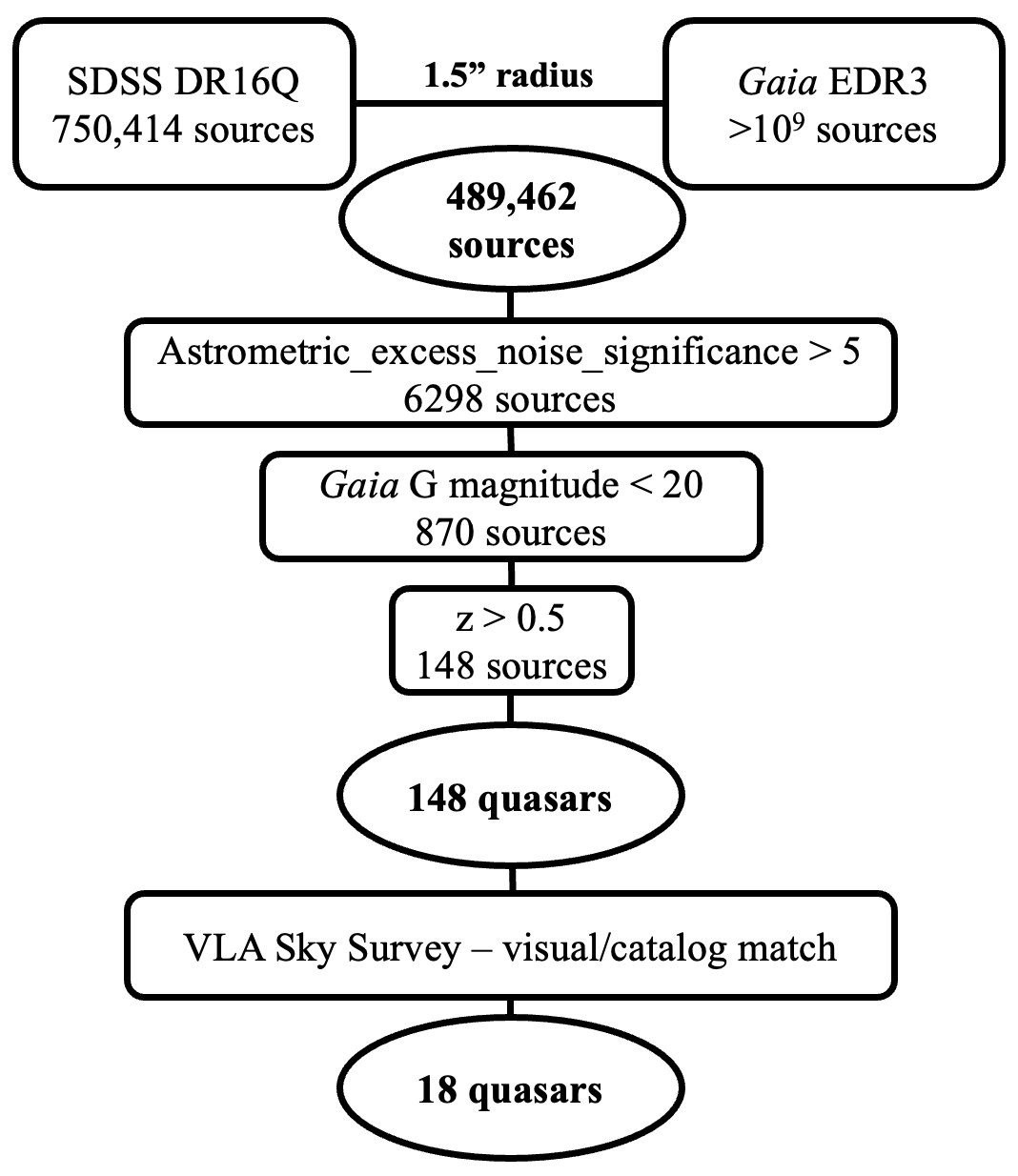}
        \vspace*{-3mm}
          \caption{\textbf{Target Selection:} A flowchart summarizing the process of sample selection.}
   \label{fig:targetflowchart}
\end{figure}

There are several \textit{Gaia} parameters that can act as indicators of positional noise in a target, as well as more that can be manipulated to search for AGN pair systems. For this sample, the important \textit{Gaia} parameter (see Table \ref{tab:paramdetails}) is \texttt{astrometric\_excess\_noise} (AEN; the previously mentioned astrometric ``jitter''), and it is defined as the amount of statistical dispersion required such that \textit{Gaia}'s astrometric solution for the source leaves no unexplained variance \cite[][]{lindegren2021}. \textit{Gaia} also calculates the statistical significance of this value: \texttt{astrometric\_excess\_noise\_significance} (AENS). Limiting to sources with AENS $> 5$ ensured a highly statistically significant measurement of excess astrometric noise in the \textit{Gaia} astrometric solution for each target.

\begin{deluxetable*}{cccc}[ht]
\tablenum{1}
\tablecaption{Parameter Details\label{tab:paramdetails}}
\tablewidth{0pt}
\tablehead{
\colhead{Parameter} & \colhead{Unit} & \colhead{Requirement} & \colhead{Meaning}\\
\colhead{(1)} & \colhead{(2)} & \colhead{(3)} &
\colhead{(4)}
}
\startdata
\texttt{astrometric\_excess\_noise\_sig} & & $\geq$ 5 & identify targets with significant AEN\\
\textit{Gaia} G magnitude & mag & $<$ 20 & mitigate spurious high-AEN targets near \textit{Gaia}'s sensitivity limit\\
Redshift & & $>$ 0.5 & eliminate targets with extended host galaxies\\
VLASS Detection & & & identify targets with radio signatures\\
\enddata
\caption{Note - Column 1: Parameter used in target selection. Column 2: Parameter unit. Column 3: Parameter constraint used to define target sample. Column 4: Requirement for constraint used to define target sample. }
\end{deluxetable*}
As \textit{Gaia}'s astrometry is inherently optimized for compact/unresolved sources \cite[][]{makarov2012quasometry, makarov2022}, quasars with large, extended host galaxies will exhibit spurious astrometric excess noise, as \textit{Gaia}'s scanning window ($\sim$18 pixels along-scan by $\sim$12 pixels across scan for the main astrometric field chips; \cite{gaia2016}) is filled by the host galaxy \cite[][]{makarov2022,souchay2022}. This can be mitigated by making a cut on redshift, which we do here, of z $>$ 0.5 (where 1$\arcsec{}$ is $\sim$ 6.1 kiloparsecs). We also make a cut on the \textit{Gaia} \texttt{G} magnitude of G $<$ 20, as there is an over-abundance of sources near \textit{Gaia}'s sensitivity limit with high AEN, likely due to the current limitations of \textit{Gaia}'s astrometric pipeline \cite[][]{hwang_initial}. After applying these limits on the sample, 148 systems remain.

It is generally accepted that between 1-10\% of AGN are typically radio bright \cite[][]{osterbrock1993,osterbrockbook}, so it was necessary to cross-match the 148 quasars remaining to various radio surveys before committing to pointed radio observations. The larger sample was matched to the VLA Sky Survey \cite[VLASS;][]{vlass} in order to confirm the existence of a radio signature. The final sample was composed of 18 targets (see Table \ref{tab:targetdetails}) exhibiting a VLASS detection.

\begin{deluxetable*}{cccccccc}[ht]
\tablenum{2}
\tablecaption{Target Details\label{tab:targetdetails}}
\tablewidth{0pt}
\tablehead{
\colhead{Source} & \colhead{Redshift} & \colhead{Scale} & \colhead{G} &
\colhead{$\sigma_{astro}$} & \colhead{AENS} & \colhead{$D_{pred}$} & \colhead{Self Cal}\\
\colhead{[SDSS]} & \colhead{} & \colhead{[kpc/arcsec]} & \colhead{[mag]} &
\colhead{[mas]} & \colhead{} & \colhead{[arcsec(kpcs)]} & \colhead{}\\
\colhead{(1)} & \colhead{(2)} & \colhead{(3)} &
\colhead{(4)} & \colhead{(5)} & \colhead{(6)} & \colhead{(7)} & \colhead{(8)}
}
\startdata
J011114.41+171328.5 & 2.198 & 8.5 & 19.29 & 4.46 & 42.3 & 0.08(0.73) & P\\
J024131.89-053139.6 & 0.837 & 7.9 & 17.51 & 1.51 & 37.6 & 0.03(0.23) & P\\
J074922.97+225511.8 & 2.166 & 8.5 & 18.52 & 1.48 & 12.8 & 0.02(0.25) & AP\\
J080009.98+165509.4 & 0.708 & 7.4 & 18.30 & 4.51 & 175 & 0.09(0.64) & P\\
J081830.46+060138.0 & 2.381 & 8.4 & 17.80 & 0.66 & 5.22 & 0.013(0.11) & N\\
J095031.63+432908.4 & 1.771 & 8.7 & 17.90 & 3.27 & 139 & 0.06(0.55) & P\\
J095122.57+263513.9 & 1.247 & 8.6 & 17.48 & 1.65 & 76.3 & 0.03(0.27) & N\\
J104406.33+295900.9 & 2.983 & 7.9 & 18.88 & 1.61 & 10.1 & 0.03(0.25) & AP\\
J105811.94+345808.7 & 1.061 & 8.3 & 18.21 & 0.83 & 5.20 & 0.016(0.14) & N\\
J112818.49+240217.4 & 1.607 & 8.7 & 18.53 & 1.09 & 5.51 & 0.02(0.19) & N\\
J121544.36+452912.7 & 1.132 & 8.4 & 19.09 & 1.22 & 5.41 & 0.02(0.20) & P\\
J141546.24+112943.4 & 2.519 & 8.3 & 17.56 & 8.03 & 1050 & 0.16(1.29) & N\\
J143333.02+484227.7 & 1.357 & 8.6 & 19.08 & 1.94 & 13.2 & 0.03(0.33) & P\\
J155859.13+282429.3 & 2.300 & 8.4 & 19.04 & 2.92 & 35.1 & 0.05(0.48) & N\\
J162501.98+430931.6 & 1.653 & 8.7 & 19.23 & 8.20 & 157 & 0.16(1.39) & N\\
J172308.14+524455.5 & 2.568 & 8.2 & 17.72 & 0.64 & 6.50 & 0.013(0.11) & P\\
J173330.80+552030.9 & 1.201 & 8.5 & 18.58 & 1.73 & 17.3 & 0.03(0.29) & P\\
J210947.09+065634.7 & 2.939 & 7.9 & 19.49 & 10.2 & 196 & 0.20(1.60) & P\\
\enddata
\caption{Note - Column 1: Coordinate names in the form of ``hhmmss.ss$\pm$ddmmss.s" based on SDSS DR16Q. Column 2: Spectroscopic redshift from SDSSDR16Q \cite[][]{sdss}, described in detail in \cite{bolton2012}. Column 3: Cosmological scale in kiloparsecs per arcsecond. Column 4: \textit{Gaia} \textit{G}-band mean magnitude. Column 5: \textit{Gaia} \texttt{astrometric\_excess\_noise}. Column 6: \textit{Gaia} \texttt{astrometric\_excess\_noise\_significance}. Column 7: Predicted angular separation calculated assuming a typical fractional rms of $10\%$, which gives a predicted angular separation of 0.2$\arcsec{}$ for every 10 milliarcseconds of $\sigma_{astro}$. Column 8: Extent of self-calibration; N indicates no successful self-calibration, P indicates phase-only, and AP indicates amplitude and phase.}
\end{deluxetable*}

\section{Observations and Data Analysis}
\label{sec:observations}
\subsection{VLA Observations}
\label{sec:VLAobs}

High sensitivity radio observations of all 18 targets were made with the VLA at S-band (2-4 GHz, central frequency 3 GHz), and X-band (8-12 GHz, central frequency 10 GHz), with the telescope in A configuration. This provides a resolution of 0.65$\arcsec{}$ at S-band and 0.2$\arcsec{}$ at X-band (Project Code 22A-384, PI Schwartzman). The observations reach a sensitivity of 18.1 $\mu$Jy/beam at S-band, with a maximum recoverable scale of 18 arcseconds, and a sensitivity of 8.9 $\mu$Jy/beam at X-band, with a maximum recoverable scale of 5 arcseconds. Table \ref{tab:obsdetails} lists the observational details for each target, including the primary flux density and complex gain calibrators, and synthesized beams. X-band targets were observed in repeated phase calibrator-target cycles, with a total on-source integration time of about 350 seconds per target, while S-band targets were observed in repeated phase calibrator-target-phase calibrator cycles, with a total on-source integration time of about 450 seconds per target. 

\begin{deluxetable*}{ccccccccc}[ht]
\tablenum{3}
\tablecaption{Observation Details\label{tab:obsdetails}}
\tablewidth{0pt}
\tablehead{
\colhead{Source} & \colhead{Obs. Date} & \colhead{Flux cal} &
\colhead{Gain cal} & \colhead{$B_{maj}, B_{min}, PA$} & \colhead{$T_{int}$(S)} & \colhead{$T_{int}$(X)} & \colhead{$\sigma_{3 GHz}$} & \colhead{$\sigma_{10 GHz}$}\\
\colhead{[SDSS]} & \colhead{[UT]} & \colhead{} &
\colhead{} & \colhead{[asec, asec, $^\circ$]} & \colhead{[s]} & \colhead{[s]} & \colhead{[$\upmu$Jy/bm]} & \colhead{[$\upmu$Jy/bm]}\\
\colhead{(1)} & \colhead{(2)} & \colhead{(3)} &
\colhead{(4)} & \colhead{(5)} & \colhead{(6)} & \colhead{(7)} & \colhead{(8)} & \colhead{(9)}
}
\startdata
J011114.41+171328.5 & 29-03-2022 & 3C138 & J0122+2502 & 0.27, 0.21, 46 & 318 & 436 & 23 & 18 \\
J024131.89 - 053139.6 & 29-03-2022 & 3C138 & J0239+0416 & 0.19, 0.17, -72 & 356 & 472 & 19 & 10 \\
J074922.97+225511.8 & 21-03-2022 & 3C286 & J0738+1742 & 0.24, 0.17, -77 & 328 & 474 & 21 & 9.4 \\
J080009.98+165509.4 & 21-03-2022 & 3C286 & J0738+1742 & 0.23, 0.17, 49 & 282 & 476 & 22 & 19 \\
J081830.46+060138.0 & 21-03-2022 & 3C286 & J0825+0309 & 0.21, 0.18, -46 & 358 & 416 & 20 & 11 \\
J095031.63+432908.4 & 06-03-2022 & 3C147 & J1018+3542 & 0.21, 0.17, 53 & 282 & 474 & 22 & 12 \\
J095122.57+263513.9 & 06-03-2022 & 3C147 & J0956+2515 & 0.21, 0.17, 85 & 323 & 474 & 20 & 9.3 \\
J104406.33+295900.9 & 29-03-2022 & 3C147 & J1018+3542 & 0.18, 0.15, 20 & 356 & 474 & 18 & 11 \\
J105811.94+345808.7 & 29-03-2022 & 3C147 & J1018+3542 & 0.21, 0.16, 70 & 358 & 476 & 19 & 10 \\
J112818.49+240217.4 & 06-04-2022 & 3C286 & J1125+2510 & 0.25, 0.17, 86 & 402 & 474 & 42 & 12 \\
J121544.36+452912.7 & 06-04-2022 & 3C286 & J1219+4829 & 0.23, 0.16, 88 & 338 & 476 & 20 & 9.3 \\
J141546.24+112943.4 & 06-04-2022 & 3C286 & J1415+1320 & 0.19, 0.16, 23 & 358 & 476 & 20 & 10 \\
J143333.02+484227.7 & 06-04-2022 & 3C286 & J1419+5423 & 0.21, 0.17, -75 & 328 & 476 & 19 & 8.9 \\
J155859.13+282429.3 & 24-03-2022 & 3C286 & J1613+3412 & 0.31, 0.18, -62 & 386 & 472 & 36 & 10 \\
J162501.98+430931.6 & 24-03-2022 & 3C286 & J1613+3412 & 0.33, 0.18, 77 & 328 & 446 & 72 & 19 \\
J172308.14+524455.5 & 01-04-2022 & 3C286 & J1740+5211 & 0.28, 0.16, -77 & 306 & 476 & 36 & 26 \\
J173330.80+552030.9 & 01-04-2022 & 3C286 & J1740+5211 & 0.19, 0.17, 57 & 298 & 476 & 28 & 15 \\
J210947.09+065634.7 & 07-03-2022 &  3C48 & J2123+0535 & 0.26, 0.16, -81 & 386 & 476 & 18 & 11 \\
\enddata
\caption{Note - Column 1: Coordinate names in the form of ``hhmmss.ss$\pm$ddmmss.s" based on SDSS DR16Q. Column 2: VLA observation date. Column 3: VLA flux density calibrator. Column 4: VLA complex gain calibrator. Column 5: Restoring beam at 10 GHz: major axis, minor axis, position angle. Column 6: 3 GHz (S-band) VLA observation integration time. Column 7: 10 GHz (X-band) VLA observation integration time. Column 8: 1-$\sigma$ sensitivity at 3 GHz, measured near the phase center. Column 9: 1-$\sigma$ sensitivity at 10 GHz, measured near the phase center.}
\end{deluxetable*}

\subsection{Data Reduction} \label{sec:datareduction}

At S-band, the data were recorded with 16 spectral windows, each having 64 channels of 2 MHz width, and covering the 2-4 GHz band. For the X-band, the data were recorded with 32 spectral windows, each having 64 channels of 2 MHz width, and covering the 8-12 GHz band. The data were reduced and calibrated with the National Radio Astronomy Observatory's (NRAO) Common Astronomy Software Applications \cite[CASA;][]{casanew2022} VLA Pipeline 1.4.2, using CASA version 5.3.0. 

The initial pipeline steps followed standard procedures; the data were Hanning smoothed, antenna position corrections were applied, in addition to ionospheric TEC corrections and requantizer gains. Calibration was performed similarly, using antenna delay, bandpass, and complex gain solutions. The flux densities for the primary calibrations were taken from the \cite{perley2017} extension to the \cite{baars1977} scale. The gain solutions were then transferred to the target sources. 

Once pipeline calibration was completed, radio frequency interference (RFI) was removed where necessary using a combination of the CASA tasks \texttt{RFlag} and \texttt{TFCrop}. To refine the gain calibration, several rounds of self-calibration were performed, beginning with phase-only, and proceeding to amplitude and phase (see Table \ref{tab:targetdetails}). In certain targets, self-calibration was limited (or not possible) due to low signal-to-noise. All imaging was completed with CASA, and was performed with a multi-term multi-frequency synthesis (MTMFS) deconvolver \cite[][]{rau2011} and Briggs weighting, with a robust parameter of 0.5. Two Taylor terms were used to model the frequency dependence of the sky, and to account for the VLA's wideband receivers. When necessary, W-projection with 64 w-planes \cite[][]{cornwell2005,cornwell2008} was employed to account for the wide-field errors. Clean masks were employed at all stages.

\subsection{Flux Density Measurements}
\label{sec:fluxmeasurements}

Flux densities were measured with the Python Blob Detection and Source Finder package \cite[PyBDSF;][]{pybdsf}. Every source in an image was fitted with a Gaussian, and the flux density properties were recorded. All unresolved, point-like sources were fitted with a single Gaussian. For extended sources, a multi-Gaussian source was fitted. In some cases, for a cluster of closely associated peaks within an island (sources with multiple components), a single-Gaussian source was fitted to each peak.

Flux density scale errors were also taken into account. Following the VLA Observing Guide, a flux density scale calibration accuracy of 5\% was assumed for both bands. Flux density calibrators 3C48 and 3C138 are currently undergoing flares. While the 3C138 flare does not appear to impact measurements at the observed bands, the single target observed with 3C48 as a flux density calibrator was assumed to have a flux density scale calibration accuracy of 10\%.

\section{Sample Properties} 
\label{sec:results}

We present new VLA observations of 18 quasars with significantly high astrometric excess noise. Targets were initially categorized by radio morphology, though we stress that morphology is not necessarily equivalent to the driver of the astrometric variability. An unresolved source features a single source that remains unresolved at sub-arcsecond scales. A resolved source features some structure at sub-arcsecond scales, and those targets have been further categorized as multi-component or extended targets. A more detailed discussion of each category follows, as well as a more in-depth look at each individual target.

\subsection{Unresolved Targets}
\label{sec:unresolved}

Nine of the 18 targets display an unresolved point-like morphology at scales probed with the new VLA observations. Four of these are further identified as star+quasar superposition via an analysis of the SDSS spectra (see Section \ref{sec:star-quasars}). Table \ref{tab:unresolved} presents the properties of these nine targets, including flux densities, luminosities, and source angular and projected physical size, which is equivalent to an upper limit on the separation of any possible components. The new VLA images are presented in Appendix A, Figure \ref{fig:radiounresolved1}.

While unresolved at sub-arcsecond scales, these sources may each be hiding more detailed structure at smaller scales. Upper and lower limits on the separation of any possible components were calculated from the new VLA observations and the varstrometry method respectively. As these sources are unresolved at sub-arcsecond scales, we may calculate an upper limit on their separation as the full extent of the unresolved source. As described in Section \ref{sec:vars}, AEN can be used, in combination with other quasar properties, to calculate a lower limit on the separation of any possible components.

While we may place constraints on these targets, without smaller-scale very long baseline interferometry (VLBI) observations, the astrometric driver of each system will likely remain uncharacterized. Some unresolved targets will likely be single AGN with a slightly increased intrinsic variability driving the AEN. Others, with follow-up observations, may be identified as quasar pair systems with separations that are not resolvable at these current scales. It is also possible for these sources to exhibit a variety of sub-milliarcsecond scale structure (i.e.\ jets or other extended emission associated with a single AGN) that is similarly unresolved. All of these possibilities could give rise to astrometric variability. Higher angular resolution follow-up will be required in order to confirm the morphology of these sources.

\begin{deluxetable*}{cccccc}[ht!]
\tablenum{4}
\tablecaption{Sample Properties for Unresolved Sources\label{tab:unresolved}}
\tablewidth{0pt}
\tablehead{
\colhead{Source} &
\colhead{$S_{3 GHz}$} & \colhead{log($L_{3 GHz}$)} & \colhead{$S_{10 GHz}$} &
\colhead{log($L_{10 GHz}$)} & \colhead{$\Delta \theta r_{p}$}\\
\colhead{[SDSS]} &
\colhead{[mJy]} & \colhead{[W Hz$^{-1}$]} & \colhead{[mJy]} & \colhead{[W Hz$^{-1}$]} & \colhead{[arcsec(kpcs)]}\\
\colhead{(1)} & \colhead{(2)} & \colhead{(3)} &
\colhead{(4)} & \colhead{(5)} & \colhead{(6)}
}
\startdata
J024131.89-053139.6$\dagger$ & 0.57$\pm$0.025 & 24.26 & 0.15$\pm$0.02 & 23.68 & 1.1(8.9) \\
J080009.98+165509.4 & 1.81$\pm$0.03 & 24.62 & 2.23$\pm$0.04 & 24.95 & 0.28(2.1) \\
J081830.46+060138.0 & 0.28$\pm$0.02 & 25.08 & 0.061$\pm$0.01 & 24.42 & 0.27(2.3) \\
J104406.33+295900.9 & 156.5$\pm$0.08 & 28.06 & 125.8$\pm$0.09 & 27.97 & 0.31(2.5) \\
J105811.94+345808.7 & 0.32$\pm$0.03 & 24.35 & 0.069$\pm$0.03 & 23.68 & 0.24(2.0) \\
J155859.13+282429.3$\dagger$ & 1.17$\pm$0.06 & 25.69 & 0.56$\pm$0.06 & 25.38 & 0.29(2.4) \\
J172308.14+524455.5$\dagger$ & 1.06$\pm$0.018 & 25.39 & 2.91$\pm$0.01 & 25.83 & 0.27(2.2) \\
J173330.80+552030.9 & 7.03$\pm$0.03 & 25.78 & 4.68$\pm$0.01 & 25.61 & 0.28(2.4) \\
J210947.09+065634.7$\dagger$ & 5.58$\pm$0.02 & 26.62 & 1.57$\pm$0.01 & 26.07 & 0.25(1.9) \\
\enddata
\caption{Note - Column 1: Coordinate names in the form of ``hhmmss.ss$\pm$ddmmss.s" based on SDSS DR16Q; $\dagger$ indicates target identified as star+quasar superposition via SDSS spectrum. Column 2: VLA 3 GHz (S-band) peak flux density $\pm$ error. Column 3: VLA 3 GHz (S-band) peak luminosity. Column 4: VLA 10 GHz (X-band) peak flux density $\pm$ error. Column 5: VLA 10 GHz (X-band) peak luminosity. Column 6: Upper limit on AGN pair separation where applicable, reported in arcseconds and kiloparsecs, measured from angular width of unresolved source at 10 GHz.}
\end{deluxetable*}

\subsection{Resolved Targets}
\label{sec:resolved}

Nine of the 18 original targets display a resolved morphology. This is defined as any source with two or more clear radio peaks at either 3 GHz, 10 GHz, or both, or otherwise extended or complex emission. In order to help sort out targets which display a morphology consistent with that of an AGN pair, this category can be further subdivided into multi-component and extended/complex targets, both of which are discussed below.

\subsubsection{Multi-component Targets}
\label{sec:multicomponent}

Six of the nine resolved targets have been identified as multi-component targets, defined as any source with two or more clear radio peaks at either 3 GHz, 10 GHz, or both. Table \ref{tab:multi} presents the properties of these targets, including flux densities, luminosities, and angular and projected physical separations of the components at 10 GHz (with the exception of J1128+2402 (see Appendix A, Figure \ref{fig:radiomc1}e) for which the separation was measured at 3 GHz due to the lack of a secondary radio peak in the 10 GHz observations). The new VLA images are presented in Appendix A, Figure \ref{fig:radiomc1}.

Fluxes of the multi-component targets were measured as integrated flux densities. Given the lower resolution of the 3 GHz observations, some targets with resolved components at 10 GHz remain unresolved at 3 GHz. In those cases, the total flux density of the 3 GHz signature has been reported, while the flux densities of each component at 10 GHz have been separately reported. 

Our multi-component targets range in angular separation from $0.33-1.11$ arcseconds, and in physical separation from $2.86-9.50$ kiloparsecs, at redshifts of 1.247 to 2.166. While further analysis is still necessary, this range of separations at these redshifts is a parameter space that represents an observational gap in the known population of AGN pair systems \cite[][]{chen_hst,pfeifle2023inprep}, and thus is of great interest. 

Given the breadth of the definition used for multi-component targets, it is expected that the majority of the sample will not be genuine dual quasar candidates. For example, J0951+2635 has been identified via spectroscopy as a gravitaionally-lensed quasar \cite[][]{Schechter_1998}. However, with the currently available data, we calculate that 44\% of the entire sample are targets identified as either lensed quasar systems or bonafide AGN pair systems. While the new VLA observations have directly observed the systems identified here as likely candidate dual quasar systems, follow-up observations (i.e.\ high angular resolution optical imaging and spectroscopy with \textit{HST}, as targets lack any nearby reference stars necessary for adaptive optics) will be necessary to confirm the nature of these systems.

\begin{deluxetable*}{ccccccc}[ht]
\tablenum{5}
\tablecaption{Sample Properties for Multi-component Sources\label{tab:multi}}
\tablewidth{0pt}
\tablehead{
\colhead{\textbf{Source}} &
\colhead{$S_{3 GHz}$} & \colhead{log($L_{3 GHz}$)} & \colhead{$S_{10 GHz}$} &
\colhead{log($L_{10 GHz}$)} & \colhead{$\Delta\theta(r_{p})$} & \colhead{Class}\\
\colhead{Comp. coord.} &
\colhead{} & \colhead{} & \colhead{} & \colhead{} & \colhead{} & \colhead{}\\
\colhead{[SDSS]} &
\colhead{[mJy]} & \colhead{[W Hz$^{-1}$]} & \colhead{[mJy]} & \colhead{[W Hz$^{-1}$]} & \colhead{[arcsec(kpcs)]} & \colhead{}\\
\colhead{(1)} & \colhead{(2)} & \colhead{(3)} &
\colhead{(4)} & \colhead{(5)} & \colhead{(6)} & \colhead{(7)} 
}
\startdata
\textbf{J011114.41+171328.5*} & & & & & 0.9(7.9) & Multi \\
01:11:14.42 +17:13:28.58 & 77.0$\pm$0.23 & 27.46 & 66.8$\pm$0.16 & 27.41 & -  & NE\\
01:11:14.37 +17:13:27.96 & 1.76$\pm$0.09 & 25.82 & 3.36$\pm$0.14 & 26.11 & - & SW \\
\textbf{074922.97+225511.8*} & & & & & 0.5(3.9) & Multi \\
07:49:22.99 +22:55:12.02 & 104.9$\pm$0.19 & 27.55 & 17.5$\pm$0.16 & 26.77 & -  & NE \\
07:49:22.97 +22:55:11.76 & - & - & 48.7$\pm$0.09 & 27.22 & - & SW \\
\textbf{J095031.63+432908.4*} & & & & & 0.3(2.9) & Multi \\
09:50:31.63 +43:29:08.59 & 2.61$\pm$0.06 & 25.73 & 1.42$\pm$0.04 & 25.46 & - & N \\
09:50:31.63 +43:29:08.26 & - & - & 0.44$\pm$0.02 & 24.95 & - & S \\
\textbf{J095122.57+263513.9*} & & & & & 1.1(9.5) & Lens \\
09:51:22.57 +26:35:14.03 & 0.57$\pm$0.04 & 24.69 & 0.81$\pm$0.02 & 24.85 & - & NW \\
09:51:22.64 +26:35:13.41 & 0.19$\pm$0.04 & 24.22 & 0.23$\pm$0.02 & 24.29 & - & SE \\
\textbf{112818.49+240217.40*} & & & & & 0.7(6.3) & Multi \\
11:28:18.52 +24:02:17.73 & 0.12$\pm$0.03 & 24.32 & 0.14$\pm$0.03 & 24.39 & - & NE\\
11:28:18.48 +24:02:17.33 & 0.075$\pm$0.02 & 24.12 & - & - & - & SW \\
\textbf{J162501.98+430931.6*} & & & & & 0.5(4.5) & Multi \\
16:25:01.99 +43:09:31.93 & 1.31$\pm$0.08 & 25.36 & 0.12$\pm$0.03 & 24.32 & - & N\\
16:25:01.99 +43:09:31.39 & - & - & 0.41$\pm$0.03 & 24.86 & - & S \\
\enddata
\caption{Note - Column 1: Coordinate names in the form of ``hhmmss.ss$\pm$ddmmss.s" based on SDSS DR16Q are bolded and starred, and individual component location as measured from 10 GHz VLA observations are listed beneath. Column 2: VLA 3 GHz (S-band) peak flux density $\pm$ error. Column 3: VLA 3 GHz (S-band) peak luminosity. Column 4: VLA 10 GHz (X-band) peak flux density $\pm$ error. Column 5: VLA 10 GHz (X-band) peak luminosity. Column 6: AGN pair separation where applicable, reported in arcseconds and kiloparsecs, measured at 10 GHz, with the exception of J1128+2402, for which the separation was measured at 3 GHz due to lack of a secondary peak in the 10 GHz observations. Column 7: Classification and component location.}
\end{deluxetable*}

\subsubsection{Extended and Complex Targets}
\label{sec:extendedandcomplex}

Three of the 18 targets display jets or other extended emission more commonly associated with a single AGN, though multi-AGN systems are capable of exhibiting jet activity \cite[e.g. 3C75;][]{owen1985}. This morphology generally provides an excellent indicator of the driver of the astrometric variability \cite[][]{hwang_initial}. Table \ref{tab:extended} presents the sample properties for the extended and jetted targets, including flux densities, luminosities, and the measured size of any extended emission on the sky. The new VLA images are presented in Appendix A, Figure \ref{fig:radioextended1}.

Final flux densities were measured as discussed in Section \ref{sec:fluxmeasurements}. J1215+4529 features a pair of canonical radio jets (see Appendix A, Figure \ref{fig:radioextended1}a). Table \ref{tab:extended} presents the integrated flux densities of the core and both lobes, in addition to angular and projected physical sizes of the jet extensions and of the core. J1415+1129 and J1433+4842 both exhibit other forms of extended emission associated with the quasar. Given the complex nature of the targets, we have assigned four components, each coincident with a peak in the 10 GHz radio signature (see Appendix A, Figures \ref{fig:radioextended1}b and \ref{fig:radioextended1}c), and measured the integrated flux density for each. At 3 GHz, the integrated flux density of the entire source was measured. Finally, the full extent on the sky of each target was measured (see Table \ref{tab:extended}).

\begin{deluxetable*}{ccccccccc}[ht]
\tablenum{6}
\tablecaption{Sample Properties for Extended Sources\label{tab:extended}}
\tablewidth{0pt}
\tablehead{
\colhead{Source} & \colhead{Class} &
\colhead{$S_{3 GHz}$} & \colhead{log($L_{3 GHz}$)} & \colhead{$S_{10 GHz}$} &
\colhead{log($L_{10 GHz}$)} & \colhead{$\Delta\theta(r_{p})$}\\
\colhead{[SDSS]} & \colhead{} &
\colhead{[mJy]} & \colhead{W Hz$^{-1}$} & \colhead{[mJy]} & \colhead{W Hz$^{-1}$} &\colhead{[arcsec(kpcs)]}\\
\colhead{(1)} & \colhead{(2)} & \colhead{(3)} &
\colhead{(4)} & \colhead{(5)} & \colhead{(6)} & \colhead{(7)}
}
\startdata
\textbf{J121544.36+452912.7*} & Jet &  & &  & & \\
12:15:44.36 +45:29:12.78 & Core & 23.4$\pm$0.05 & 26.21 & 16.4$\pm$0.03 & 26.06 & 0.25(2.1) \\
12:15:43.94 +45:29:20.20 & N Lobe & 1.38$\pm$0.13 & 24.98 & 0.57$\pm$0.06 & 24.59 & 8.8(74) \\
12:15:44.38 +45:29:05.12 & S Lobe & 4.61$\pm$0.16 & 25.51 & 0.708$\pm$0.09 & 24.69 & 11(97) \\
\textbf{J141546.24+112943.4*} & Lens & & & & & 1.7(14) \\
14:15:46.24 +11:29:43.40 & Unresolved & 2.61$\pm$0.19 & 26.13 & - & - & -\\
14:15:46.27 +11:29:43.81 & A & - & - & 0.31$\pm$0.03  & 25.21 & -\\
14:15:46.28 +11:29:43.23 & B & - & - & 0.16$\pm$0.04  & 24.92 &- \\
14:15:46.23 +11:29:43.13 & C & - & - & 0.21$\pm$0.04  & 25.04 & -\\
14:15:46.19 +11:29:44.12 & D & - & - & 0.09$\pm$0.02  & 24.67 & -\\
\textbf{J143333.02+484227.7*} & Jet & & & & & 1.4(12) \\
14:33:33.02 +48:42:27.70 & Unresolved & 34.9$\pm$0.22 & 26.56 & - & - & -\\
14:33:33.03 +48:42:27.76 & A & - & - & 1.1$\pm$0.07  & 25.06 & -\\
14:33:32.99 +48:42:27.58 & B & - & - & 7.24$\pm$0.15  & 25.88 & -\\
14:33:30.90 +48:42:27.56 & C & - & - & 1.64$\pm$0.09  & 25.24 & -\\
14:33:32.89 +48:42:27.85 & D & - & - & 0.12$\pm$0.03  & 24.10 & -\\
\enddata
\caption{Note - Column 1: Coordinate names in the form of ``hhmmss.ss$\pm$ddmmss.s" based on SDSS DR16Q are bolded and starred, and individual component locations as measured from new 10 GHz VLA observations are listed beneath. Column 2: Classification and component label. Column 3: VLA 3 GHz (S-band) peak flux density $\pm$ error. Column 4: VLA 3 GHz (S-band) peak luminosity. Column 5: VLA 10 GHz (X-band) peak flux density $\pm$ error. Column 6: VLA 10 GHz (X-band) peak luminosity. Column 7: AGN pair separation where applicable, reported in arcseconds and kiloparsecs, measured from 10 GHz observations.}
\end{deluxetable*}

\section{Radio Spectral Modeling}
\label{sec:sedmodeling}

We performed radio spectral modeling using data from a variety of public radio surveys (see Section \ref{sec:archivalsurveys}). Radio spectral shapes, in combination with morphological information, reveal important information about the quasar environment, the physical conditions of the jets/lobes (e.g. absorbed vs. optically-thin), and the evolutionary stage of the radio source \cite[][]{nyland2020,bicknell1998,odea1998,odea2021,orienti2014}. 

The observed radio spectrum is a combination of absorption and emission processes.  To quantify the distribution and the location of the spectral peak, least squares fitting was performed for the new VLA observations presented above, in addition to all available archival and commensal observations. Two basic synchrotron emission models were employed which will allow us to constrain important physical properties of each radio source. 

\begin{itemize}
    \item \textbf{Standard Power Law:} A standard power law model that is defined by $S_{\nu} = S_{o} \nu^{\alpha}$, where $S_{\nu}$ is the flux density at frequency $\nu$, $\alpha$ is the spectral index, and $S_{o}$ is the flux density at 1 GHz. The value and sign of the spectral index can be used to help distinguish between synchrotron and thermal emission \cite[e.g.,][]{patil2021,patil2022}.  
    \item \textbf{Curved Power Law:} A curved power law generated from the flux density at frequency $\nu$, $S_{\nu}$, the spectral index $\alpha$, and the flux density at 1 GHz $S_{o}$, defined as $S_{\nu} = S_{o} \nu^{\alpha} e^{q^{(ln \nu)^2}}$, where $q$ represents the width of the peak, characterizes the degree of curvature, and is defined as $\nu_{peak} = e^{-\alpha / 2q}$, where $\nu_{peak}$ is the peak frequency. Significant spectral curvature is typically defined as $|q| \geq 0.2$ \cite[][]{duffy2012}, and is indicative of absorption of low frequency emission. Absorption may be due to a high synchrotron optical depth within the source \cite[i.e., synchrotron self-absorption (SSA);][]{odea2021} or free-free absorption from surrounding ionized gas \cite[][]{bicknell1998}. Understanding which process is in play can further constrain the radio source and environment properties. 
\end{itemize}

Following the approach described in \cite{patil2022}, each source was fitted with a standard power law and a curved power law, making use of the \texttt{Radio Spectral Fitting}\footnote{https://github.com/paloween/Radio\_Spectral\_Fitting} tools \cite[][]{patil2021}. The new VLA data at 10 GHz were re-imaged and uv-tapered to produce similar resolutions in both the X- and S-bands. The uv-tapered image properties are described in Table \ref{tab:taper}. Least squares fitting was performed with the values weighted by their measurement uncertainties. The results of the fitting are summarized in Table \ref{tab:sed}, and the radio spectra are presented in Appendix D, Figures \ref{fig:radiospecunresolved}, \ref{fig:radiospecmulticomp}, and \ref{fig:radiospecextended}. Radio spectral classifications for the overall sample are discussed in Section \ref{sec:specclass}. Important results from the radio spectral modeling are included in Section \ref{sec:individualtargets}, as are spatially resolved spectral indices, where possible.

\subsection{Archival Radio Surveys}
\label{sec:archivalsurveys}

Radio spectral modeling with broad frequency coverage required adding survey data to our new observations. Radio flux densities and their uncertainties were taken from the following archival surveys. The flux density values are taken directly from the published catalogs with corrections applied for calibration uncertainties.  

\subsubsection{LOTSS}
\label{sec:lotss}

The Low Frequency Array Two-meter Sky Survey \cite[LOTSS;][]{shimwell2019, shimwell2022} is an ongoing all-sky survey at 144 MHz covering the northern sky above 34 degrees. The second data release covers about 27\% of the sky with a point-source sensitivity of 83 $\mu$Jy per beam. A calibration uncertainty of 10\% was applied. 

\subsubsection{TGSS ADR}
\label{sec:tgssadr}

The Tata Institute of Fundamental Research (TIFR) Giant Metrewave Radio Telescope (GMRT) Sky Survey Alternative Data Release \cite[TGSS ADR;][]{intema2017} is a 150 MHz survey covering about 90\% of the sky, or about 36900 deg$^2$. TGSS reaches a resolution of 25$^{\prime\prime}$, and has a 1$\sigma$ sensitivity of between 3 and 5 mJy per beam. A calibration uncertainty of 10\% was applied.

\subsubsection{VLITE}
\label{sec:vlite}

The VLA Low-band Ionosphere and Transient Experiment \cite[VLITE;][]{clarke2016,polisensky2016} is a commensal system on the VLA operating at all configurations that records of 6000 hours of commensal data per year at 350 MHz. With the VLA in A configuration (as it was for all targets in this project), VLITE reaches a resolution of about 5.6$^{\prime\prime}$. A calibration uncertainty of 15\% was applied.

\subsubsection{RACS}
\label{sec:racs}

The Rapid Australian Square Kilometer Array Pathfinder (ASKAP) Continuum Survey \cite[RACS;][]{mcconnell2020} is the first all-sky survey conducted with ASKAP. The first epoch observations at 887.5 MHz use 36 12-m antennas to cover the entire sky south of $+41$ degrees, reaching a sensitivity of between 25 and 40 $\mu$Jy per beam and a beam of about 15$^{\prime\prime}$. A calibration uncertainty of 5\% was applied.

\subsubsection{NVSS}
\label{sec:nvss}

The NRAO VLA Sky Survey \cite[NVSS;][]{condon1998} observed the entire northern sky about $\delta >$ -40 degrees at 1.4 GHz. The survey was conduction with the VLA in D configuration, resulting in an angular resolution of about 45$^{\prime\prime}$. A calibration uncertainty of 3\% was applied. 

\subsubsection{FIRST}
\label{sec:first}

The Faint Images of the Radio Sky at Twenty-one centimeters \cite[FIRST;][]{becker1995} survey is a 1.4 GHz survey covering 10,575 deg$^2$ of the sky. The observations reach a resolution of about 5$^{\prime\prime}$ and a sensitivity of $\sim$0.13 mJy per beam. A calibration uncertainty of 5\% was applied.

\subsubsection{VLASS}
\label{sec:vlass}

The VLA Sky Survey \cite[VLASS;][]{vlass} is an ongoing 3 GHz continuum survey covering 33,885 deg$^2$. Of the 3 planned epochs, the first two have been completed, and the third is ongoing. A source catalog\footnote{Quick-look image cutouts are also available at the Canadian Institute for Radio Astronomy Data Analysis: http://cutouts.cirada.ca/} is now available \cite[][]{gordon2021}. The typical 1$\sigma$ sensitivity of a single epoch image is $\sim$ 120 $\mu$Jy per beam with an angular resolution of 2.5$^{\prime\prime}$. A calibration uncertainty of 3\% was applied.

\subsubsection{VIPS}
\label{sec:vips}
The VLBA Imaging and Polarimetry Survey \cite[VIPS;][]{helmboldt2007} observed 1128 flat-spectrum sources brighter than 85 mJy in the Northern Cap region of SDSS at 5 GHz, covering about 14\% of the sky. The typical 1$\sigma$ sensitivity of a VIPS image is $\sim$ 0.2 mJy per beam with an angular resolution of 1.4 milliarcseconds. A calibration uncertainty of 5\% was applied.

\subsection{Radio Spectral Shape Classification}
\label{sec:specclass}

All radio spectra were inspected in order to best categorize their spectral classifications. The shapes of both the standard power law and the curved power law were considered, as were the reduced $\chi^2$ values for each fit. We note that, in the case of SDSS J104406.33+295900.9 (see Appendix D, Figure \ref{fig:radiospecunresolved}), neither the standard power law nor the curved power law accurately model the target's spectrum, possibly due to flux variability on years-long timescales\footnote{Other explanations for the spectral shape include source blending of multiple components at lower frequencies, though the source appears compact in all surveys investigated. Additionally, no nearby galaxies are apparent in either SDSS or PanSTARRS imaging.}. Additionally, we note that for the systems identified as star+quasar superposition (see Section \ref{sec:star-quasars}), there may be a stellar contribution to the radio spectrum \cite[see Figure 1 in][]{gudel2002}. In particular, we note the inverted spectrum in SDSS J172308.14+524455.6, indicating potential stellar contributions at higher frequencies \cite[][]{dulk1985,gudel2002}. Quantifying the contribution from the thermal regime is not possible without further high frequency observations. For each target, we adopt the approach described in \cite{patil2022}, and divide classifications into the following categories:

\begin{itemize}
    \item \textbf{Standard Power Law:} This is the standard power law defined in Section \ref{sec:sedmodeling}. The power law spans the full range of frequencies presented in our radio spectra, and the reduced $\chi^2$ is lower for the power law model, in comparison to the curved power law model. In this case, a spectral index of $<$ -0.7 is consistent with the optically thin synchrotron emission expected of most AGN.
    \item \textbf{Curved Power Law:} This is the curved power law defined in Section \ref{sec:sedmodeling}. The curved power law spans the full range of the frequencies presented in our radio spectra, but does not display a peak or turnover. The reduced $\chi^2$ is lower for the curved power law model, in comparison to the standard power law model, indicating significant deviation from the standard power law.
    \item \textbf{Peaked Spectrum:} A visual inspection of the radio spectrum reveals a peak or turnover within the range of frequencies.
    \item \textbf{Flat Spectrum:} The spectral index as measured from the best-fit standard power law or curved power law is $|\alpha| <$ 0.5.
    \item \textbf{Upturned Spectrum:} A visual inspection of the radio spectrum reveals a upturned concave spectrum.
    \item \textbf{Inverted Spectrum:} The spectral index as measured from the best-fit standard power law or curved power law is $\alpha >$ 0.5.
\end{itemize}

\begin{figure*}[!htb]
    \centering
     \includegraphics[width=13cm]{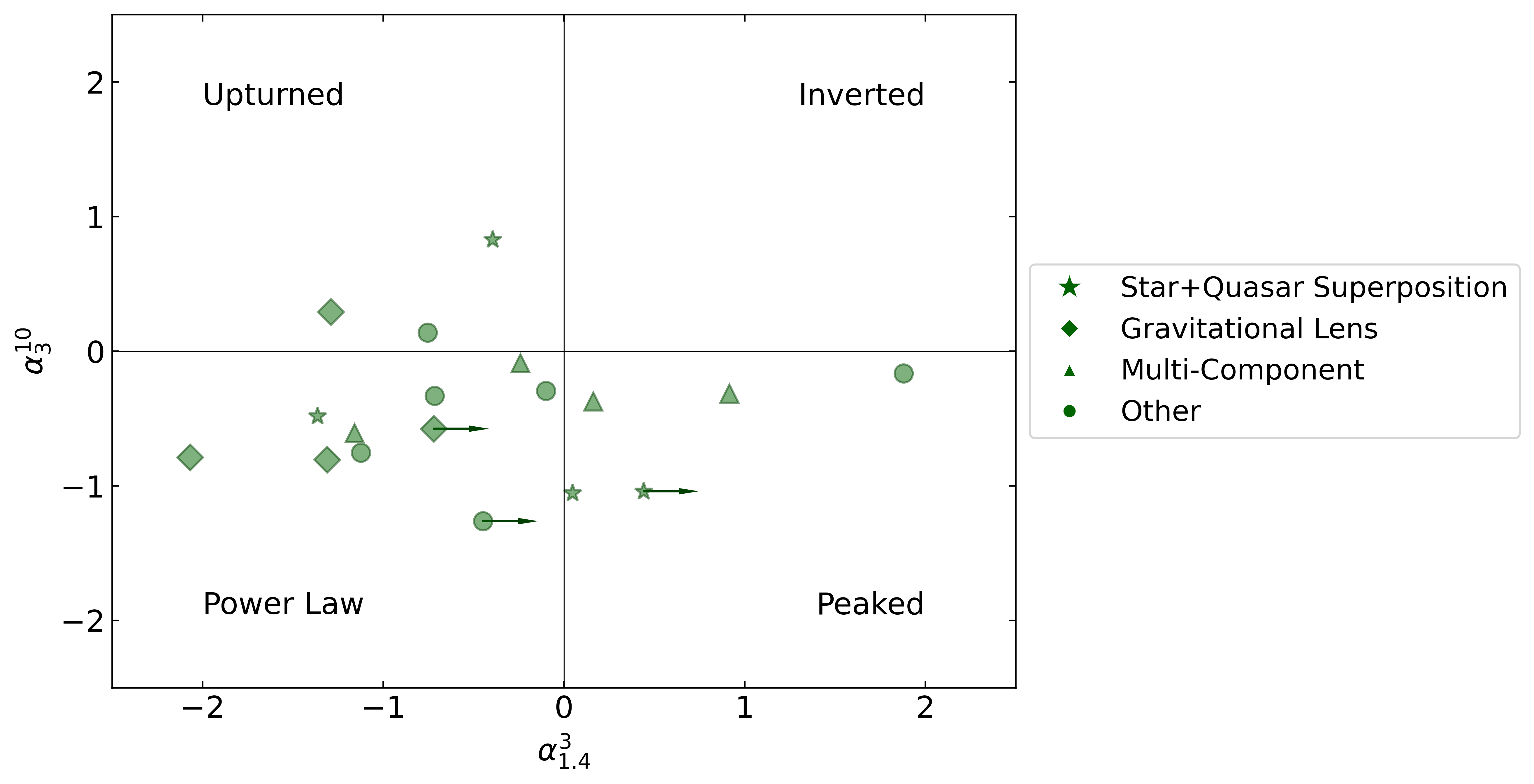}
        \vspace*{1mm}
          \caption{Radio color-color plot showing two-band spectral indices $\alpha^{3}_{1.4}$ and $\alpha^{10}_{3}$, calculated using flux densities from the new 3 GHz and 10 GHz VLA observations, in addition to FIRST \cite[1.4 GHz; VizieR 2014 catalog][]{helfand2015} values. Where no FIRST catalog source was recorded, NVSS \cite[1.4 GHz; VizieR catalog][]{condon1998} flux densities were used. If neither survey was available, an upper limit was used of NVSS 1$\sigma$ 0.45 mJy \cite[][]{condon1998}. The different symbols highlight classes of interest: star+quasar superposition systems are marked with stars, gravitational lenses are marked with diamonds, and multi-component systems (the most likely to be candidate AGN pairs) are marked as triangles. All remaining targets are marked with circles. The arrows used represent lower limits on the spectral indices due to the use of NVSS upper limit flux densities.}
    \label{fig:radiocc}
\end{figure*}

In addition to the above classifications (see Table \ref{tab:sed}), the radio spectra were also characterized using a radio color-color plot (see Figure \ref{fig:radiocc}). Three flux densities measured at three sufficiently separated frequencies were used to calculate two spectral indices: the new 3 GHz and 10 GHz VLA observations, and either the FIRST (1.4 GHz) or the NVSS (1.4 GHz) survey values. These flux density measurements produce spectral indices $\alpha^{3}_{1.4}$ and $\alpha^{10}_{3}$. The quadrants of the radio color-color plot represent classifications.

Figure \ref{fig:radiocc} illustrates the distribution of our target spectral shapes and reveals that the sample is diverse in radio spectral shape. However, the majority of our targets, independent of the potential driver of excess astrometric noise, exhibit a spectral index that is consistent with, slightly flatter, or slightly steeper than that of optically thin synchrotron emission (lower left quadrant). We note that the impact of having sources with time variable behavior or extended emission that is not identically sampled at all frequencies (in this case, the 1.4 GHz survey fluxes) can result in apparent shifts in the source location in this color-color plot. In particular, sources exhibiting complex morphologies may have extended emission at 1.4 GHz that is resolved out in the higher frequency measurements, artificially shifting their location on the color-color plot to the left. We refer the reader to Section \ref{sec:appendixD} for detailed spectral modeling on each individual source.

There are many processes that might explain genuine absorption exhibited by some targets (i.e. synchrotron self-absorption \cite[SSA;][]{dekool1989, tingay2003, odea2021}, or free-free absorption \cite[FFA;][]{kellerman1966, odea1998, bicknell1998}). For example, the environment of the radio target could be a source of absorption \cite[such as in the case of radio jet activity;][]{nyland2020,patil2021,patil2022}. Spectral indices steeper than what is expected from optically thin synchrotron emission may suggest ageing in the system, though further observations are required to fully characterize the underlying mechanism.

The spectral shape classifications were also compared to the radio morphologies investigated in Section \ref{sec:results}. No preferential radio spectral classification is found for any of the radio morphologies investigated in this paper, nor for the sample as a whole. Overall, however, the sample is diverse in radio spectral shape.

\begin{deluxetable*}{cccc}[ht]
\tablenum{7}
\tablecaption{Tapered Image Properties\label{tab:taper}}
\tablewidth{0pt}
\tablehead{
\colhead{Source} & \colhead{$B_{maj}, B_{min}, PA$} &  \colhead{$S_{10 GHz}$} & \colhead{$\sigma_{10 GHz}$}\\
\colhead{[SDSS]} & \colhead{[asec, asec, $^\circ$]} & \colhead{[mJy]} & \colhead{[$\upmu$Jy/bm]}\\
\colhead{(1)} & \colhead{(2)} & \colhead{(3)} & \colhead{(4)}\\
}
\startdata
J011114.41+171328.5 & 0.84, 0.60, 81 & 70.80 $\pm$ 0.22 & 59 \\
J024131.89 - 053139.6 & 0.71, 0.60, 48 & 0.18 $\pm$ 0.05 & 25 \\
J074922.97+225511.8 & 0.66, 0.57, 66 & 66.40 $\pm$ 0.16 & 32 \\
J080009.98+165509.4 & 0.67, 0.58, 70 & 2.23 $\pm$ 0.06 & 33 \\
J081830.46+060138.0 & 0.69, 0.58, 67 & 0.12 $\pm$ 0.04 & 16 \\
J095031.63+432908.4 & 0.69, 0.61, -67 & 1.79 $\pm$ 0.05 & 15 \\
J095122.57+263513.9 & 0.72, 0.59, -69 & 1.08 $\pm$ 0.04 & 14 \\
J104406.33+295900.9 & 0.81, 0.55, 82 & 128.1 $\pm$ 0.24 & 65 \\
J105811.94+345808.7 & 0.76, 0.55, -84 & 0.07 $\pm$ 0.02 & 15 \\
J112818.49+240217.4 & 0.67, 0.59, 78 & 0.13 $\pm$ 0.04 & 16 \\
J121544.36+452912.7 & 0.66, 0.58, 73 & 16.40 $\pm$ 0.06 & 25 \\
J141546.24+112943.4 & 0.79, 0.59, -85 & 0.99 $\pm$ 0.15 & 14 \\
J143333.02+484227.7 & 0.71, 0.59, 86 & 14.11 $\pm$ 0.44 & 35 \\
J155859.13+282429.3 & 0.74, 0.56, -88 & 0.71 $\pm$ 0.03 & 14 \\
J162501.98+430931.6 & 0.71, 0.58, 88 & 0.63 $\pm$ 0.05 & 15 \\
J172308.14+524455.5 & 0.66, 0.59, 41 & 3.01 $\pm$ 0.03 & 16 \\
J173330.80+552030.9 & 0.66, 0.59, 43 & 4.71 $\pm$ 0.03 & 16 \\
J210947.09+065634.7 & 0.73, 0.58, 82 & 1.57 $\pm$ 0.04 & 18 \\
\enddata
\caption{Note - Column 1: Coordinate names in the form of ``hhmmss.ss$\pm$ddmmss.s" based on SDSS DR16Q. Column 2: Restoring beam of uv-tapered imaging: major axis, minor axis, position angle. All imaging done with Briggs weighting and a robust factor of 1. Column 3: 10 GHz uv-tapered image source flux density $\pm$ error. Column 4: 10 GHz uv-tapered image 1-$\sigma$ noise.}
\end{deluxetable*}

\begin{deluxetable*}{ccccccc}[ht]
\tablenum{8}
\tablecaption{Radio Spectral Modeling Properties\label{tab:sed}}
\tablewidth{0pt}
\tablehead{
\colhead{Source} & \colhead{$S_{3 GHz}$} & \colhead{$S_{10 GHz}$} & \colhead{$\alpha_{Power}$} & \colhead{$\alpha_{Curved}$} & \colhead{Class} & \colhead{Spectral Shape}\\
\colhead{[SDSS]} & \colhead{[mJy]} & \colhead{[mJy]} & \colhead{} & \colhead{} & \colhead{} & \colhead{}\\
\colhead{(1)} & \colhead{(2)} & \colhead{(3)} & \colhead{(4)} & \colhead{(5)} & \colhead{(6)} & \colhead{(7)}\\
}
\startdata
J011114.41+171328.5 & 78.7 $\pm$ 0.24 & 70.8 $\pm$ 0.22 & -0.09 $\pm$ 0.01 & -0.30 $\pm$ 0.06 & Multi Component & Flat\\
J024131.89 - 053139.6 & 0.63 $\pm$ 0.05 & 0.18 $\pm$ 0.05 & -1.09 $\pm$ 0.09 & -1.33 $\pm$ 0.01 & Star+quasar & Curved\\
J074922.97+225511.8 & 104 $\pm$ 0.19 & 66.4 $\pm$ 0.16  & -0.37 $\pm$ 0.02 & 0.14 $\pm$ 0.21 & Multi Component & Flat\\
J080009.98+165509.4 & 1.89 $\pm$ 0.04 & 2.23 $\pm$ 0.06 & 0.04 $\pm$ 0.15 & -1.46 $\pm$ 0.47 & Unresolved & Upturned\\
J081830.46+060138.0 & 0.31 $\pm$ 0.04 & 0.12 $\pm$ 0.04 & -1.93 $\pm$ 0.47 & -3.01 $\pm$ 0.01 & Lens & Upturned\\
J095031.63+432908.4 & 2.61 $\pm$ 0.06 & 1.79 $\pm$ 0.05 & 0.08 $\pm$ 0.17 & 0.44 $\pm$ 0.07 & Multi Component & Peaked\\
J095122.57+263513.9 & 0.76 $\pm$ 0.06 & 1.08 $\pm$ 0.04 & 0.07 $\pm$ 0.38 & -2.45 $\pm$ 0.01 & Lens & Upturned\\
J104406.33+295900.9 & 156 $\pm$ 0.24 & 128 $\pm$ 0.24 & -0.06 $\pm$ 0.08 & 3.09 $\pm$ 0.45 & Unresolved & Curved\\
J105811.94+345808.7 & 0.32 $\pm$ 0.03 & 0.07 $\pm$ 0.02 & -0.94 $\pm$ 0.05 & -0.97 $\pm$ 0.01 & Unresolved & Curved\\
J112818.49+240217.4 & 0.26 $\pm$ 0.04 & 0.13 $\pm$ 0.04 & -0.58 $\pm$ 0.15 & - $\pm$ - & Lens & Standard\\
J121544.36+452912.7 & 23.4 $\pm$ 0.05 & 16.4 $\pm$ 0.06 & -0.30 $\pm$ 0.04 & -0.72 $\pm$ 0.08 & Jet & Curved\\
J141546.24+112943.4 & 2.61 $\pm$ 0.18 & 0.99 $\pm$ 0.15 & -1.13 $\pm$ 0.09 & -1.18 $\pm$ 0.15 & Lens & Standard\\
J143333.02+484227.7 & 34.9 $\pm$ 0.22 & 14.1 $\pm$ 0.44 & -0.86 $\pm$ 0.11 & -0.88 $\pm$ 0.14 & Jet & Standard\\
J155859.13+282429.3 & 1.27 $\pm$ 0.04 & 0.71 $\pm$ 0.03 & -0.58 $\pm$ 0.19 & -1.14 $\pm$ 0.23 & Star+quasar & Curved\\
J162501.98+430931.6 & 1.31 $\pm$ 0.08 & 0.63 $\pm$ 0.05 & -0.89 $\pm$ 0.08 & -0.94 $\pm$ 0.06 & Multi Component & Standard\\
J172308.14+524455.5 & 1.11 $\pm$ 0.03 & 3.01 $\pm$ 0.03 & 0.79 $\pm$ 0.11 & -1.29 $\pm$ 0.03 & Star+quasar & Upturned\\
J173330.80+552030.9 & 7.01 $\pm$ 0.03 & 4.71 $\pm$ 0.03 & -0.33 $\pm$ 0.03 & -0.69 $\pm$ 0.26 & Unresolved & Flat\\
J210947.09+065634.7 & 5.58 $\pm$ 0.04 & 1.57 $\pm$ 0.04 & -0.91 $\pm$ 0.15 & 0.46 $\pm$ 0.29 & Star+quasar & Peaked\\
\enddata
\caption{Note - Column 1: Coordinate names in the form of ``hhmmss.ss$\pm$ddmmss.s" based on SDSS DR16Q. Column 2: VLA 3 GHz (S-band) total flux density $\pm$ error. Column 3: VLA 10 GHz (X-band) uv-tapered total flux density $\pm$ error. Column 4: Spectral index from least squares fitting of radio spectral fitting with standard power law $\pm$ error. Column 5: Spectral index from least squares fitting of radio spectral fitting with curved power law $\pm$ error, where possible. Column 6: Classification. Column 7: Spectral shape classification.}
\end{deluxetable*}

\section{Discussion} \label{sec:discussion}

\subsection{Chance Superposition with Stars}
\label{sec:star-quasars}

One possible alternative explanation for the unresolved targets is star+quasar superposition. To assess this possibility, we retrieved the SDSS spectra for the targets in this sample and searched the spectra for $z=0$ absorption lines using Bayesian AGN Decomposition Analysis for SDSS Spectra \citep[BADASS;][]{sexton2021} package.\footnote{https://github.com/remingtonsexton/BADASS3} BADASS is an open-source spectral analysis program designed for deconvolution of AGN and host galaxy spectra via simultaneous fitting for a variety of spectral features, including but not limited to: the AGN power law continuum, individual broad, narrow, and absorption lines, and stellar host templates or (alternatively) the stellar line-of-sight velocity distribution.

For each target in the sample, we began by fitting host stellar population templates to the observed SDSS spectrum around common $z=0$ stellar absorption features (Ca triplet, CaT, rest-frame 8500\,\AA{}, 8544\,\AA{}, and 8664\,\AA{}; Ca H+K lines, rest-frame 3934\,\AA{} and 3969\,\AA{}; NaD, rest-frame 5890\,\AA{} and 5896\,\AA{}, MgIb wavelength; H$\alpha$, rest-frame 6563\,\AA{}; H$\beta$, rest-frame 4861\,\AA{}). We did not incorporate any fitting of emission or absorption lines or the quasar continuum during these initial `quick-look' fits. Each fit was then visually inspected for signs of absorption lines. In cases where absorption lines were identified, we swapped the host stellar template for an AGN power law continuum (to the model the background quasar continuum) and included narrow absorption lines to the fit of the observed spectrum; we centered these absorption lines on the expected $z=0$ rest-frame wavelengths of the lines (see above), left their amplitudes free to vary, and then refit the spectrum.

\begin{figure*}
    \centering
    \begin{minipage}{0.45\textwidth}
        \includegraphics[width=\linewidth]{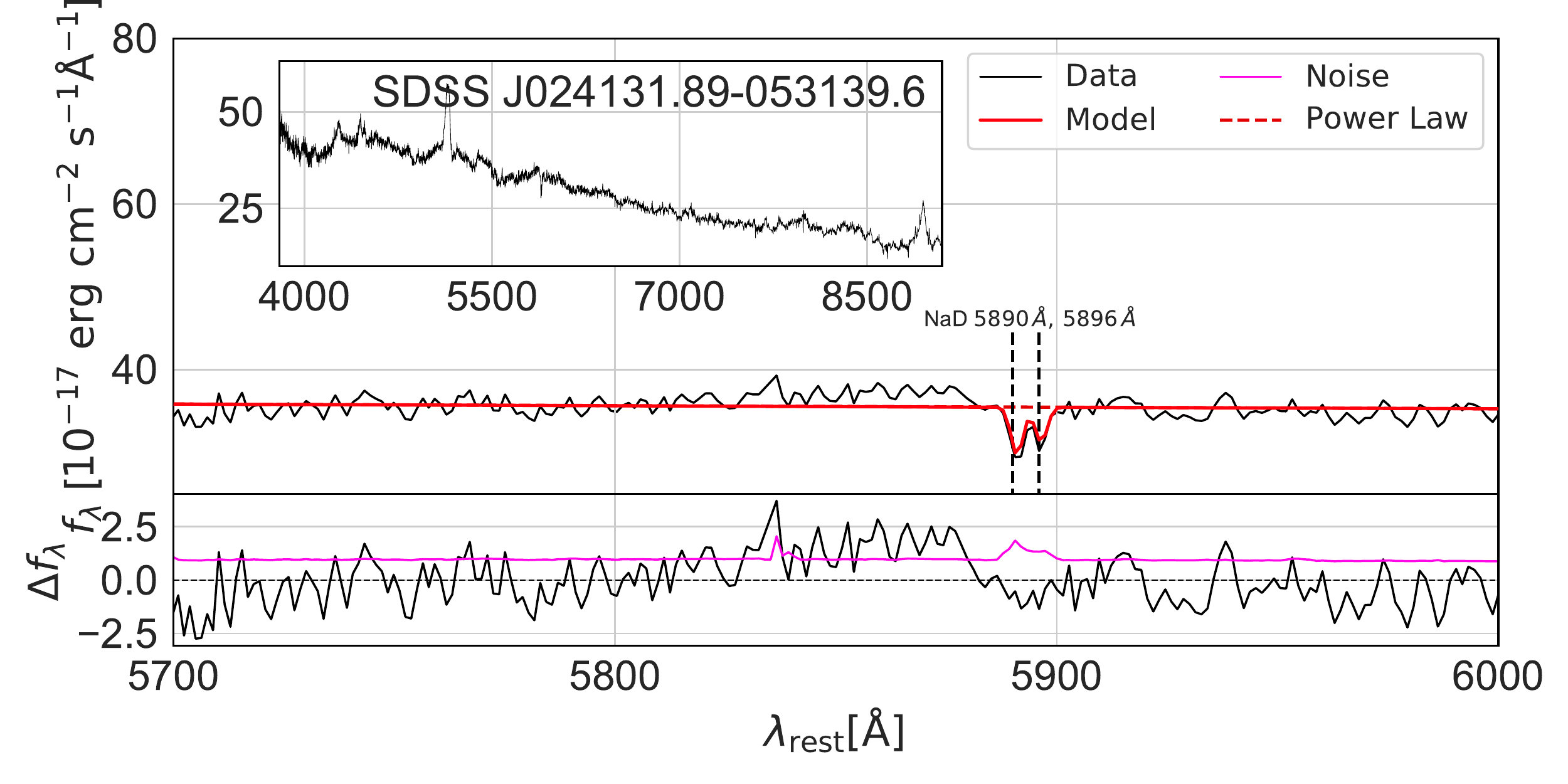}
    \end{minipage}
    \begin{minipage}{0.45\textwidth}
        \includegraphics[width=\linewidth]{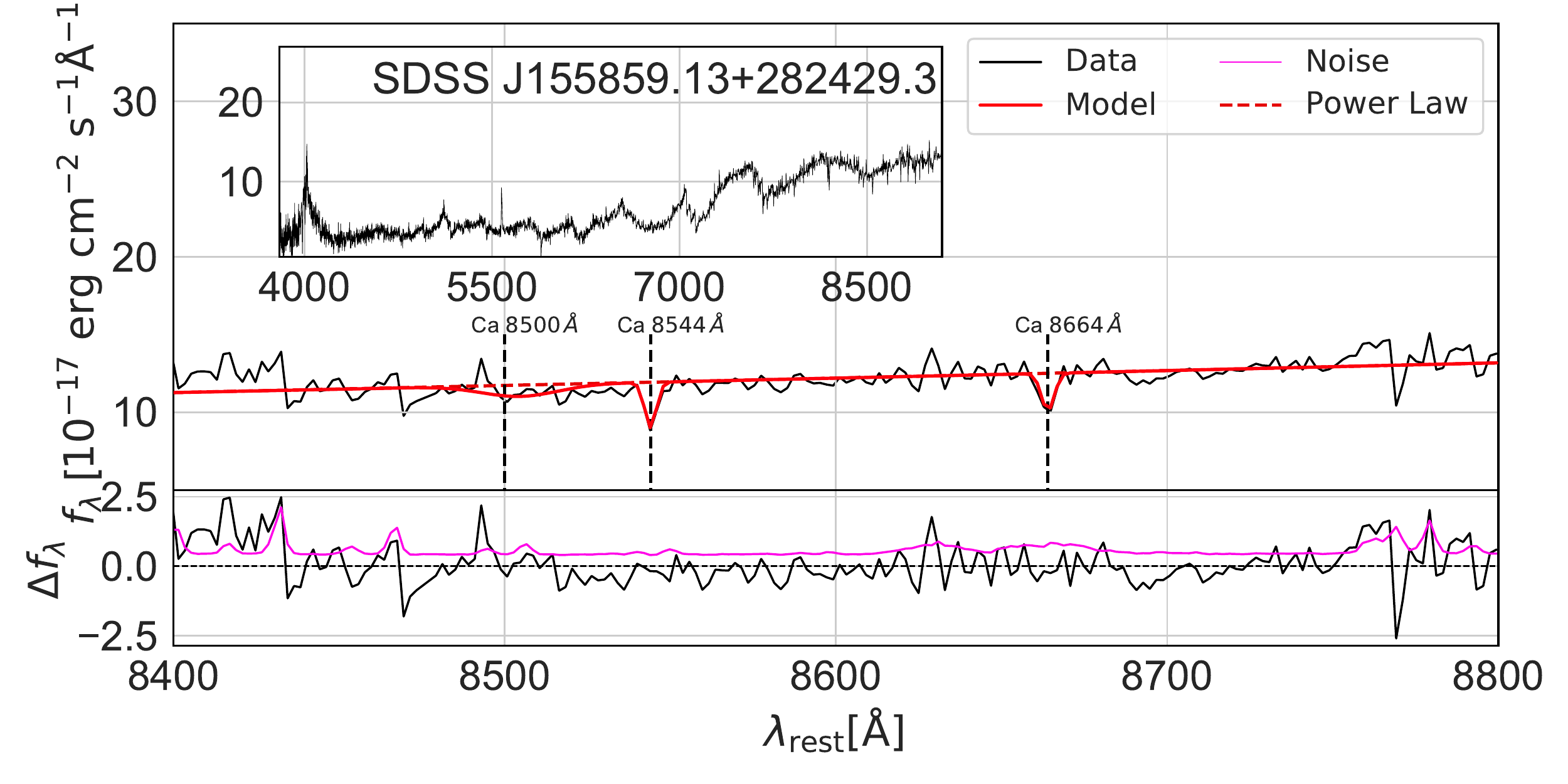}
    \end{minipage}
    
    \begin{minipage}{0.45\textwidth}
        \includegraphics[width=\linewidth]{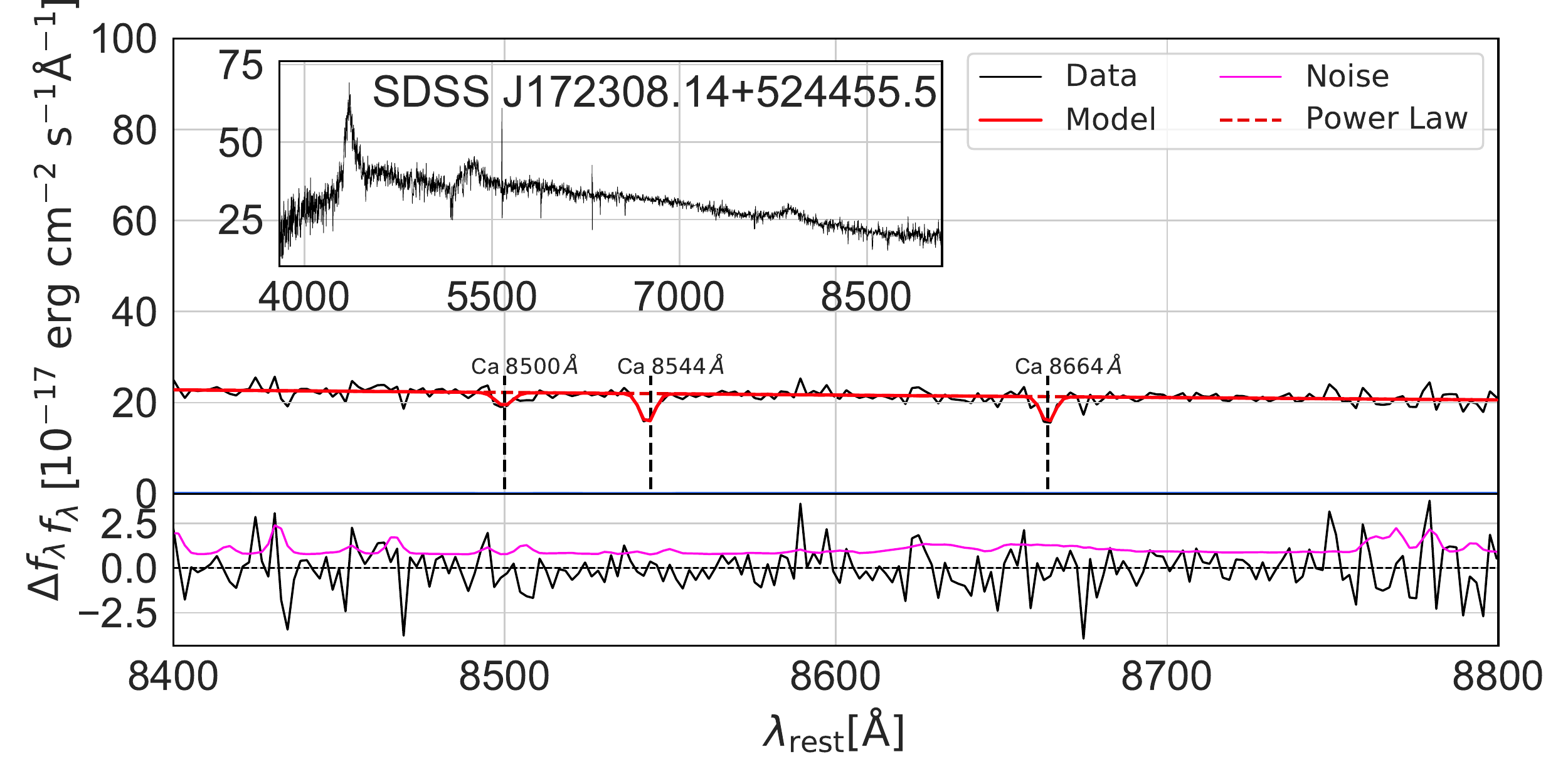}
    \end{minipage}
    \begin{minipage}{0.45\textwidth}
        \includegraphics[width=\linewidth]{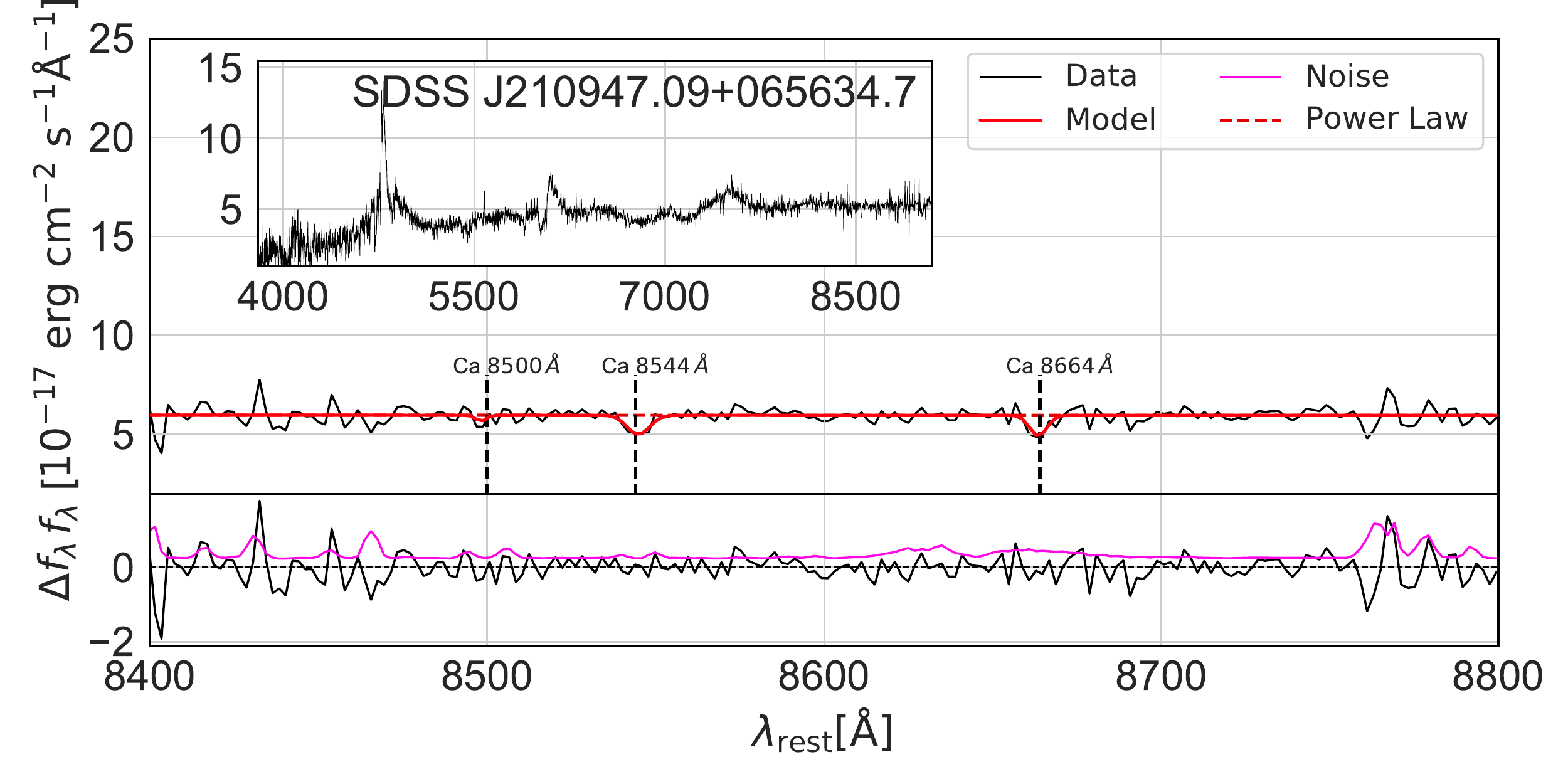}
    \end{minipage}
    
    \caption{SDSS optical spectroscopic absorption line fitting. Here we show the four cases of confirmed foreground star-background quasar pairings (top: J0241-0531, top-middle: J1558+2824, bottom-middle: J1723+5244, bottom: J2109+0656) where we have identified foreground stellar absorption features consistent with $z=0$. Each panel shows (1) the fits to the spectrum across a narrow spectral window straddling the observed absorption lines, (2) the residuals of the fit, and (3) an inset panel showing the full SDSS spectrum. Black lines represent the observed SDSS spectrum (uncorrected for the redshift of the quasar), red lines represent the model (quasar power-law continuum + gaussian absorption lines). We have marked the observed absorption lines with dashed vertical lines and denoted the specific line by name. The full SDSS spectra for these four likely star-quasar pairs are shown in Appendix~\ref{sec:appendixA} along with the SDSS spectra of the rest of the sample.}
    \label{fig:star-quasars}
\end{figure*}

Through this iterative fitting process, we found evidence for CaT in J155859.13+282429.3, J172308.14+524455.5, and J210947.09+065634.7, NaD lines in J024131.89-053139.6, J155859.13+282429.3, J172308.14+524455.5, and J210947.09+065634.7, and H$\alpha$ in J172308.14+524455.5, but we do not find evidence for MgIb, H$\beta$, or Ca H+K lines in any targets. The lack of evidence for MgIb and Ca H+K lines is not particularly surprising; the rest-frame wavelengths for the Ca H+K lines reside in a spectral region of low throughput in SDSS while MgIb is not a particularly strong line relative to other lines. Figure~\ref{fig:star-quasars} displays examples of our absorption line fits to the CaT lines in J155859.13+282429.3, J172308.14+524455.5, and J210947.09+065634.7 and the NaD lines in J024131.89-053139.6, along with inset panels showing the full SDSS spectra. These results demonstrate that 4 systems in this sample comprise quasar-star pairs caught in projection. Assuming no other systems within this sample comprise quasar-star pairs, our sample has a quasar-star contamination rate of $24.2^{+10.4}_{-8.7}$\% (estimated using \textsc{scipy.special.betaincinv}). However, the absence of foreground stellar absorption lines does not necessarily preclude the presence of other quasar-star pairs within this sample; it has been shown even in the local Universe that the continuum of background AGNs can easily dominate the stellar continuum of foreground stars \citep[even when the sources are separated by $\sim6$\arcsec{},][]{pfeifle2023}, and foreground, $z=0$ stellar absorption features can be either diluted or completely elusive. Our quasar-star pair contamination rate is therefore a lower limit for this sample.

\subsection{Dual Quasars versus Lensed Quasars}
\label{sec:dualslenses}

One of the primary goals of this study is to identify candidate dual AGN systems. For the portion of the sample identified as multi component targets, it is necessary to eliminate both star+quasar superpositions and lensed quasars contaminants. After eliminating foreground stellar contaminants, \cite{shen_nature} argued that the abundance of high-redshift sub-arcsecond gravitationally lensed quasars is insufficient to account for most of the resolved pairs in a systematic search. This indicates that, while we can expect some additional lensed interlopers, the majority of our remaining sample will be bonafide AGN pairs. 

Four targets in the pilot sample have been identified as lensed quasars (two are shown in Figure \ref{fig:3xHST}: J1415+1129; left, and J0951+2635; right). In J1415+1129, the unresolved but extended source at 3 GHz resolves into at least four individual components at 10 GHz. Archival HST observations reveal that the target is a lensed quasar, an Einstein Cross type system known colloquially as the Cloverleaf System \cite[][]{chartas2004}. Interestingly, the radio and optical peaks are not aligned, as was similarly observed at 8.4 GHz \cite[][]{kayser1990}, where an offset between the optical and modeled radio centroid was measured. However, CO(7-6) observations display good spatial alignment \cite[][]{alloin1997} and a recent multi-wavelength study reports the discovery of a radio lobe \cite[][]{zhang2022}. 

In J0951+2635, the double lens is resolved at both new radio frequencies. Archival HST observations reveal a similar optical structure. The target was identified as a lensed quasar via spectroscopy \cite[][]{Schechter_1998}, and followed-up by the GLENDAMA \cite[][]{gilmerino2018} and COSMOGRAIL \cite[][]{Sluse_2012} projects. Finally, follow-up HST observations \cite[][]{Jakobsson_2005} confirm the identification.

J0818+0601 was identified by \cite{hutsem2020} as a gravitationally lensed quasar using spectropolarimetric observations. They detect an additional foreground absorption system tentatively identified as the lensing galaxy. Finally, J1128+2402 has been identified by \cite{Inada_2014} as a gravitationally lensed quasar based on the similar spectral energy distributions of the stellar components, in addition to the existence of an extended object in between the measured components.

For the other multi-component and unresolved targets, a lens cannot be ruled out. One possible avenue of further investigation would be follow-up high-resolution integral-field (IFU) spectroscopy of both peaks \cite[][]{Rusu_2019,Lemon_2018}. While a nearly identical optical spectrum does not necessarily preclude a dual AGN morphology (given the general similarity of quasar spectra), similar changes in luminosity over time (requiring multi-epoch IFU) might be indicative of a lensed system \cite[e.g.][]{Ciurlo_2023,lemon2020,lemon2023}. Another method might be to search for possible lensing galaxies in deeper, high angular resolution imaging of these systems \cite[][]{Lemon_2018,lemon2023}.

\subsection{Individual Targets}
\label{sec:individualtargets}

In this section, we present a more in-depth study of the existing ancillary multi-frequency data that exists for each target in the sample, in conjunction with the new VLA observations, radio spectral analysis, and SDSS spectra analysis. Final radio images are presented in Appendix A, Figures \ref{fig:radiounresolved1}, \ref{fig:radiomc1}, and \ref{fig:radioextended1}.

\tab \textbf{SDSS011114.41+171328.5:} exhibits multiple bright components in both radio bands, with a separation of 0.94 arcseconds, or 7.96 kiloparsecs. This target could be tentatively identified as a candidate dual AGN with a relatively flat spectral index. However, we note that the flux ratio between the primary and secondary peaks is quite large, $\sim$ 20, possibly indicative of jet activity.

\tab \textbf{SDSS024131.89-053139.6:} is an unresolved point source at sub-arcsecond scales, with perhaps some slight extension to the NE. The SDSS spectrum for this target reveals evidence of a $z=0$ stellar absorption Na doublet, and has thus been identified as a star+quasar superposition.  

\tab \textbf{SDSS074922.97+225511.8:} exhibits two radio peaks in both VLA bands, with a separation of 0.46 arcseconds, or 3.91 kiloparsecs. It is another candidate dual AGN, and has been previously identified as such by \cite{chen_hst}, with dual-band \textit{HST} observations (see Figure \ref{fig:3xHST}; middle), in addition to Gemini GMOS spectra. \cite{shen_nature} reported a separation of 3.8 kiloparsecs, essentially matching the separation measured from the new VLA observations. Finally, \cite{shen_nature} reports initial follow-up VLBA observations that confirm the existence of and indicate morphological differences in the dual cores. 

\tab \textbf{SDSS081830.46+060138.0}: is a faint, unresolved point source in both VLA bands. It has been identified by \cite{hutsem2020} as a doubly imaged, gravitationally lensed quasar, and \textbf{exhibits an upturned radio spectral shape with a steep spectrum component at low frequencies that may be caused by a previous phase of activity superimposed on more recent, ongoing activity from a compact region driving the flat-spectrum component that emerges at higher frequencies. Alternatively, the spectral shape may be tied to temporal variability and/or lensing effects.} 

\tab \textbf{SDSS095031.63+432908.4:} exhibits a single, extended radio source at 3 GHz that resolves into two component peaks at 10 GHz, with a separation of 0.34 arcseconds, or 2.86 kiloparsecs. We identify this target as another candidate dual AGN. The radio spectrum for this target is peaked, indicating SSA, with a peak frequency of 0.37 GHz, and a curved spectral index of 0.44. 

\tab \textbf{SDSS095122.57+263513.9:} exhibits two radio peaks at both VLA bands. It has been identified as a lensed quasar via spectroscopy \cite[][]{Schechter_1998} and followed-up by the Gravitational Lenses and Dark Matter project \cite[GLENDAMA;][]{gilmerino2018, Sluse_2012}. Follow-up HST \cite[][]{Jakobsson_2005} observations confirm that this system comprises a lensed quasar with a separation of 1.11 arcseconds (see Figure \ref{fig:3xHST}; right). As the two components of this target are resolved in both bands, it is possible to calculate spectral indices for each component, in contrast to measuring the overall system. The northwest target exhibits a spectral index of 0.29, while the southeast target exhibits a spectral index of 0.16. 

\begin{figure}[ht!]
    \centering
     \includegraphics[width=6cm]{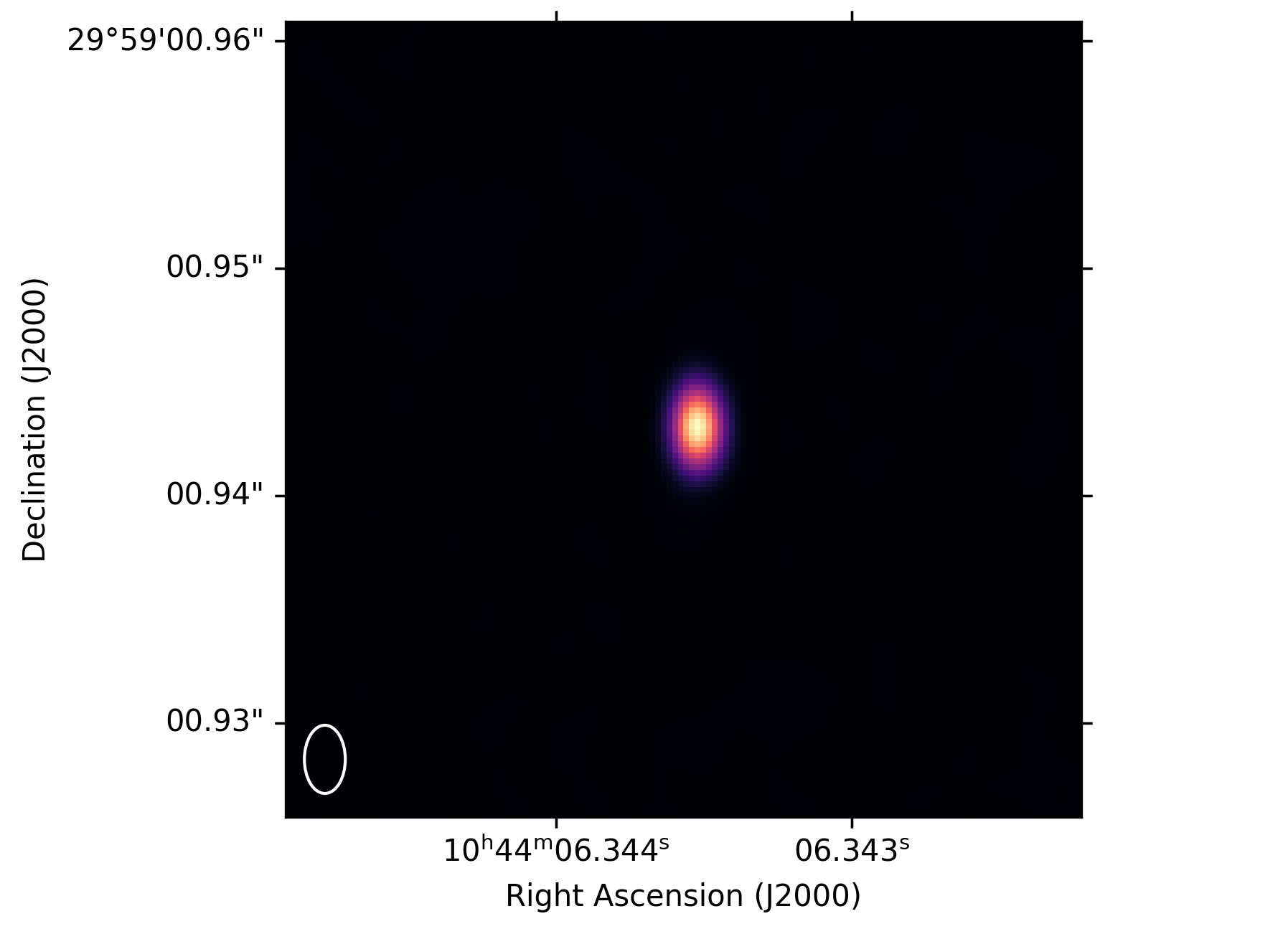}
        \vspace*{1mm}
          \caption{VIPS 5 GHz VLBA observations of SDSS104406.33+295900.9 with a 0.0029$\arcsec{}$ $\times$ 0.0018$\arcsec{}$ beam.}
    \label{fig:VLBA1044}
\end{figure}

\begin{figure*}
    \centering
    \begin{minipage}{0.3\textwidth}
        \includegraphics[width=\linewidth]{141546_HST_3.png}
    \end{minipage}
    \begin{minipage}{0.3\textwidth}
        \includegraphics[width=\linewidth]{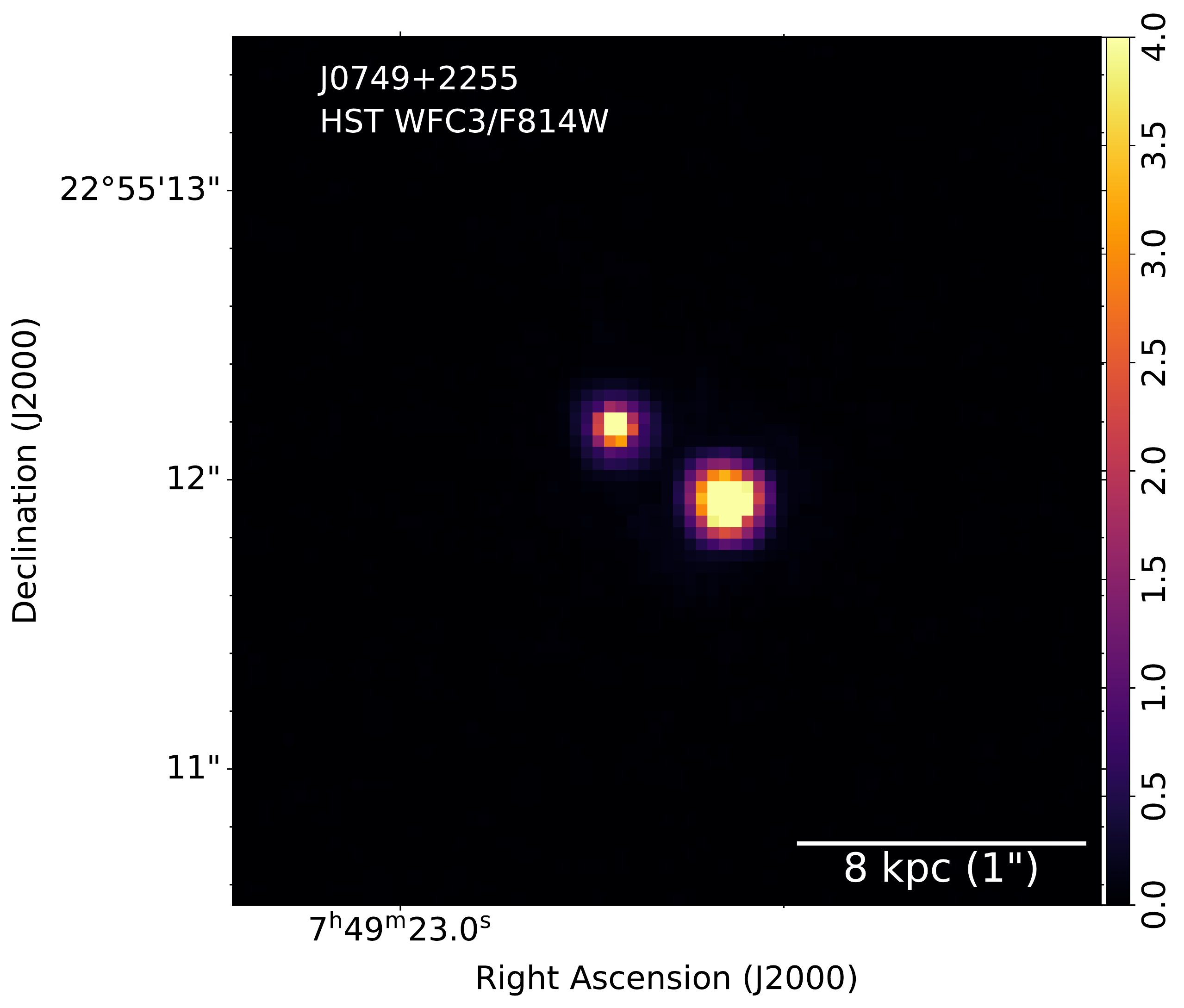}
    \end{minipage}
    \begin{minipage}{0.3\textwidth}
        \includegraphics[width=\linewidth]{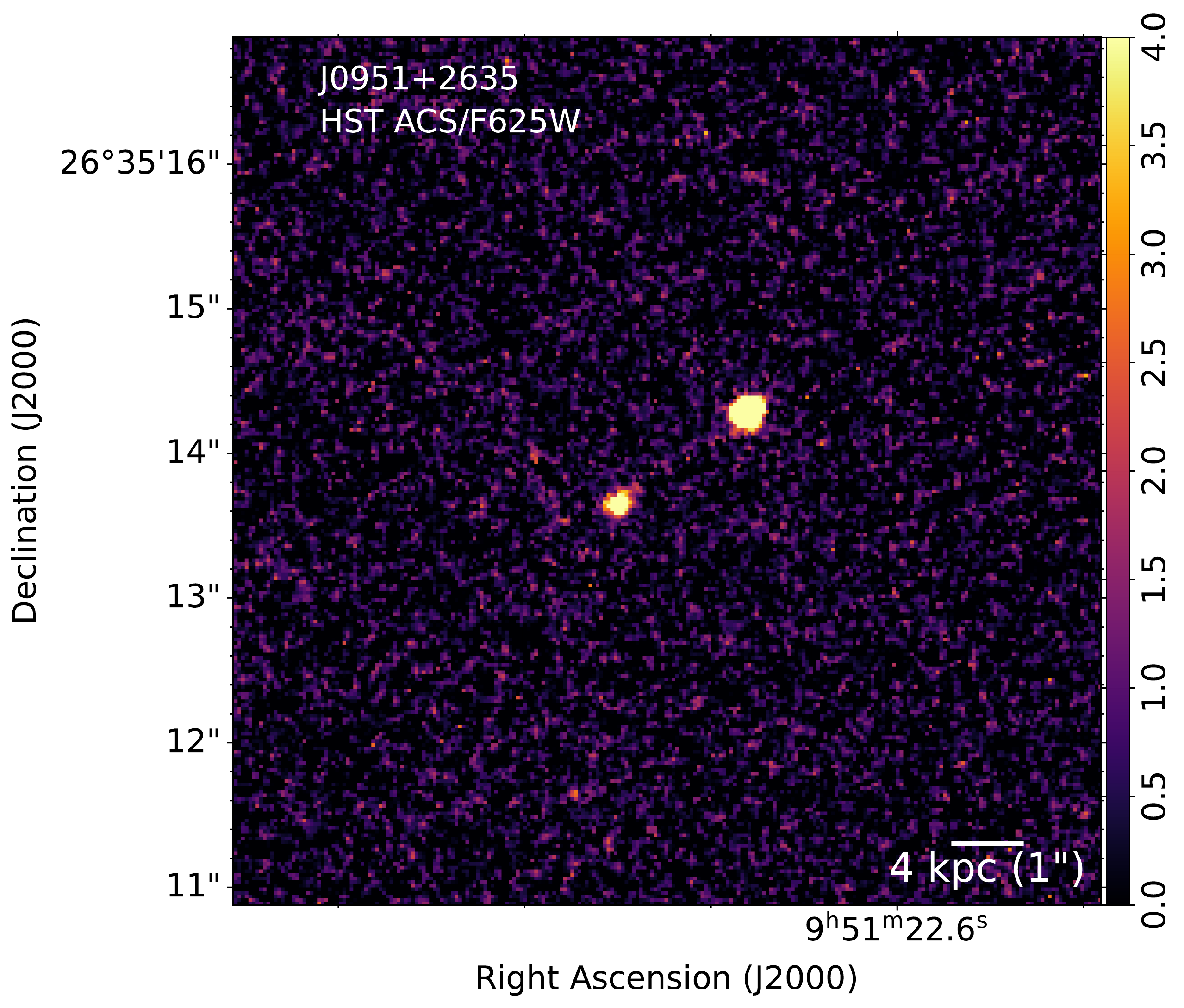}
    \end{minipage}
    
    \caption{Left: HST WFPC2/F336W observations of SDSS141546.24+112943.4, illustrating the lensed quasar, an Einstein Cross known as the Cloverleaf System \cite[][]{chartas2004}. Middle: HST WFC3/F814W observations of SDSSJ074922.97+225511.8, illustrating a likely AGN pair. Right: HST ACS/F625W observations of SDSS095122.57+263513.9, illustrating a known gravitationally-lensed quasar \cite[][]{Schechter_1998}.}
    \label{fig:3xHST}
\end{figure*}

\tab \textbf{SDSS104406.33+295900.9:} is an unresolved point source at sub-arcsecond scales. It is bright in both bands, and has previous VLBA observations from the VLBA Imaging and Polarimetry Sky Survey at 5 GHz \cite[VIPS;][]{helmboldt2007}. Figure \ref{fig:VLBA1044} shows the archival VLBA image, displaying a bright radio source still consistent with an unresolved point source at milliarcsecond scales. The radio spectrum for this target is likely curved, though there are archival flux densities significantly offset from the model that might indicate flux density variability over years-long timescales. Recently, \cite{Wang_2023} found, from new e-MERLIN observations of this target, a significant and large \textit{Gaia}-radio offset.  

\tab \textbf{SDSS112818.49+240217.4:} exhibits two radio peaks at 3 GHz. The eastern peak is not visible at 10 GHz, leaving only the western peak remaining. This target has been identified by \cite{Inada2014} to be a doubly imaged, gravitationally lensed quasar.

\tab \textbf{SDSS121544.36+452912.7:} features a pair of canonical radio jets, between which exists a compact core. The northern jet is diffuse, with a separation from the core of 8.80 arcseconds, or about 74 kiloparsecs. The southern jet exhibits a knot at 7.55 arcseconds from the core, or about 64 kiloparsecs, and then continues on to a diffuse lobe at 11.53 arcseconds, or about 97 kiloparsecs. The spectral index of the target's core is $\sim$ -0.5, while for the northern lobe, it is -0.73, and for the southern lobe, it is -1.15. The target thus follows the canonical model of a radio jet with a flatter core with two steeper radio lobes.

\tab \textbf{SDSS141546.24+112943.4:} is an unresolved point source at 3 GHz that resolves into at least four individual components at 10 GHz. The end-to-end extension of the target is 1.73 arcseconds, or about 14 kiloparsecs. Archival \textit{HST} images reveal that the target is a lensed quasar (see Figure \ref{fig:3xHST}; left); an Einstein Cross known colloquially as the Cloverleaf System \cite[][]{magain1988,chartas2004}. The target's radio spectral shape is best fit to a standard power law, with a spectral index of -1.2, slightly steeper than might be expected for a gravitationally-lensed quasar.

\tab \textbf{SDSS143333.02+484227.7:} exhibits a jet-like extension to the west of the radio core, and a smaller extension to the east. The extension is relatively small-scale: measured from end-to-end, the full length of the target is 1.39 arcseconds, or about 12 kiloparsecs. For this target, there is a secondary faint (magnitude of $\sim$21) \textit{Gaia} detection close to the primary source ($\sim$0.767$\arcsec{}$). The secondary source is $\sim$18\% of the brightness of the primary source. 

\tab \textbf{SDSS155859.13+282429.3:} is an unresolved point source in both bands. The SDSS spectrum for this target reveals evidence for both a Ca triplet and a Na doublet absorption system, and has thus been identified as a star+quasar superposition. This target is the most obvious star-quasar pair; the SDSS spectrum also exhibits prominent TiO absorption bands, which are a common characteristic of M dwarfs \cite[][]{morgan1943}.

\tab \textbf{SDSS162501.98+430931.6:} exhibits a single, extended radio source at 3 GHz, that resolves into two radio components at 10 GHz. This target is another candidate dual AGN, with a separation of 0.52 arcseconds, or 4.52 kiloparsecs. 

\tab \textbf{SDSS172308.14+524455.5:} is an unresolved point source at sub-arcsecond scales, with no clear extension in the fairly bright radio signature. The SDSS spectrum for this target reveals evidence for a $z=0$ Ca triplet, Na doublet, and H$\alpha$ absorption lines, and has thus been identified as a star+quasar superposition. 

\tab \textbf{SDSS210947.09+065634.7:} is an unresolved point source at sub-arcsecond scales. The SDSS spectrum for this target reveals evidence of a $z=0$ Na doublet and a Ca triplet absorption line system, and has thus been identified as a star+quasar superposition. The radio spectrum for this target is curved, with a peak frequency of 1.67 GHz and a curved spectral index of 0.46. Recently, \cite{Wang_2023} found, from new e-MERLIN observations of this target, a significant and large \textit{Gaia}-radio offset. 

\subsection{Understanding the Varstrometry-selected Radio Parameters}
\label{sec:parameters}

In addition to understanding each individual source, it is equally important to understand the sample that the \texttt{varstrometry} method selects for. In order to characterize the overall target sample, a control sample was prepared. The control sample was generated alongside the target sample, and systems identified as star+quasar superpositions were identified by their SDSS spectra and were eliminated from the target sample. 

The control sample was generated from a cross-match of the SDSS DR16Q and \textit{Gaia} EDR3 catalogs, to within 1.5$\arcsec{}$. The sample \textit{Gaia} G-magnitude (G $<$ 20) and redshift (z $>$ 0.5) cuts were applied. In this case, the important \textit{Gaia} parameter, the \texttt{astrometric\_excess\_noise\_significance} was set to $<$ 4, in order to generate a control sample of sources that do not display significant astrometric jitter. Finally, it was required that each control source have a VLASS signature so that overall control and target sample radio parameters could be compared. The control sample was then spatially matched to the target sample, so as to ensure that the controls covered the same portion of the sky as the targets to minimize systematics arising from the \textit{Gaia} scanning law. The number of controls per target object was maximized and required to be the same for each target. Additionally, no controls were shared between targets, resulting in a control sample of about 120 sources. 

The controlled parameters, \textit{Gaia} G-magnitude and redshift, show probability values calculated using the Anderson-Darling, Kolmogorov-Smirnov, and Kuiper's tests that are $\mathrm{>} 0.05$, indicating that the null hypothesis (in this case, that both the target and control samples arise from the same parent population) cannot be ruled out. The same is true for the \textit{Gaia} G-magnitude signal-to-noise ratio, the peak radio luminosities, the total radio luminosities, the ratio of peak-to-total radio luminosity, and radio spectral indices. For the controlled parameter AENS, the p-values are $\mathrm{<} 0.05$, rejecting the null hypothesis, and indicating that the two samples vary by more than the parameter of interest. 

\begin{figure*}[ht!]
\centering
\begin{minipage}{0.32\textwidth}
    \includegraphics[width=\linewidth]{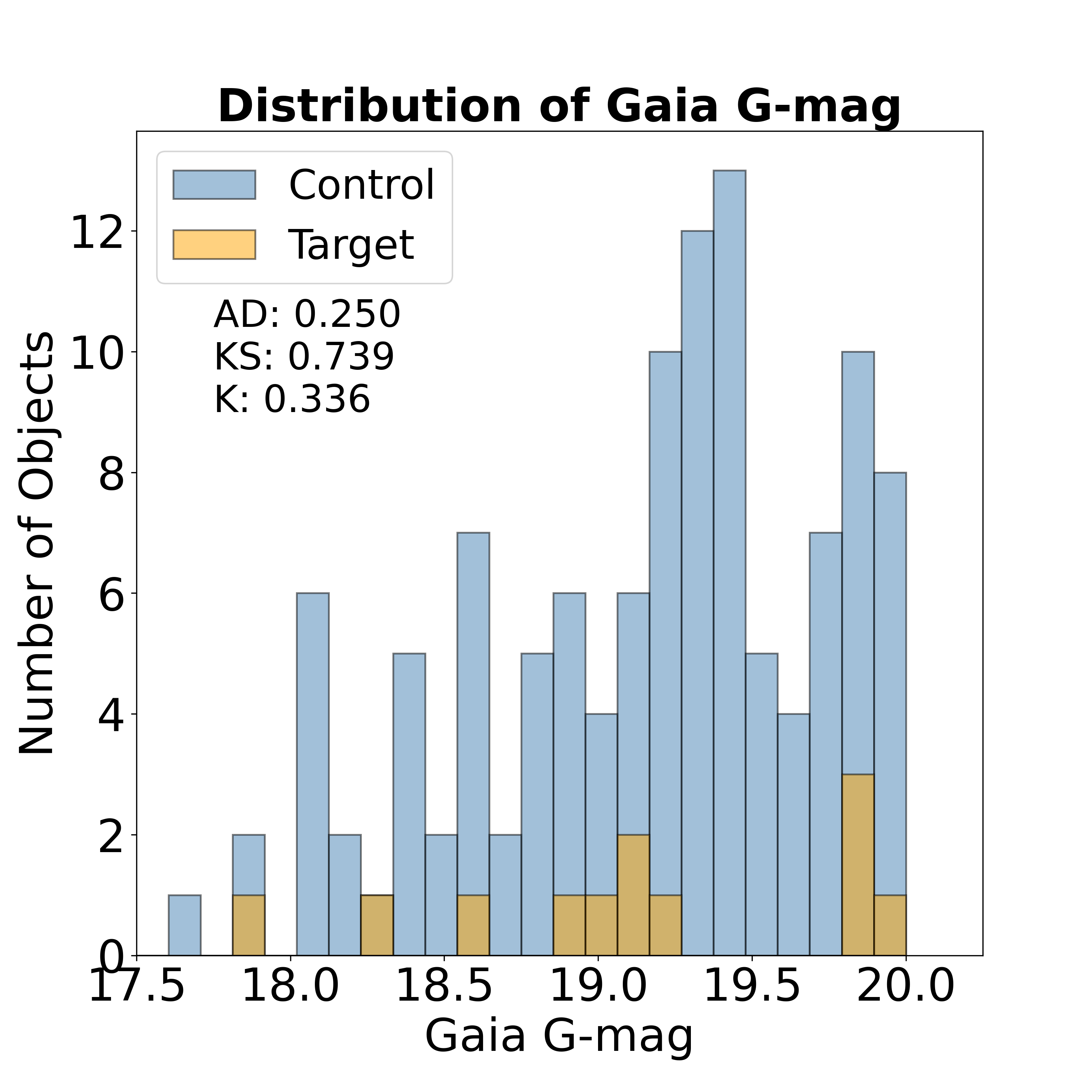}
\end{minipage}
\begin{minipage}{0.32\textwidth}
    \includegraphics[width=\linewidth]{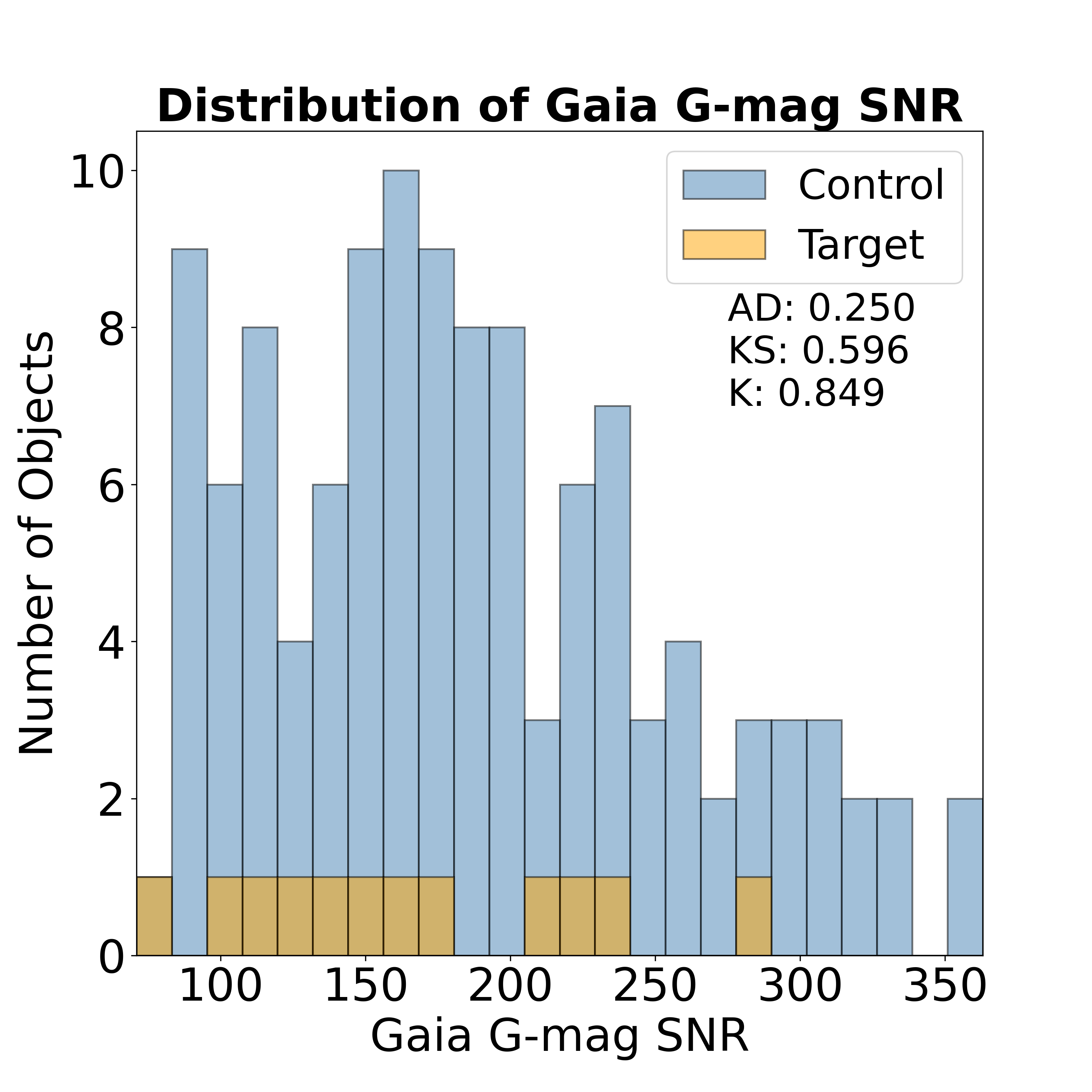}
\end{minipage}
\begin{minipage}{0.32\textwidth}
    \includegraphics[width=\linewidth]{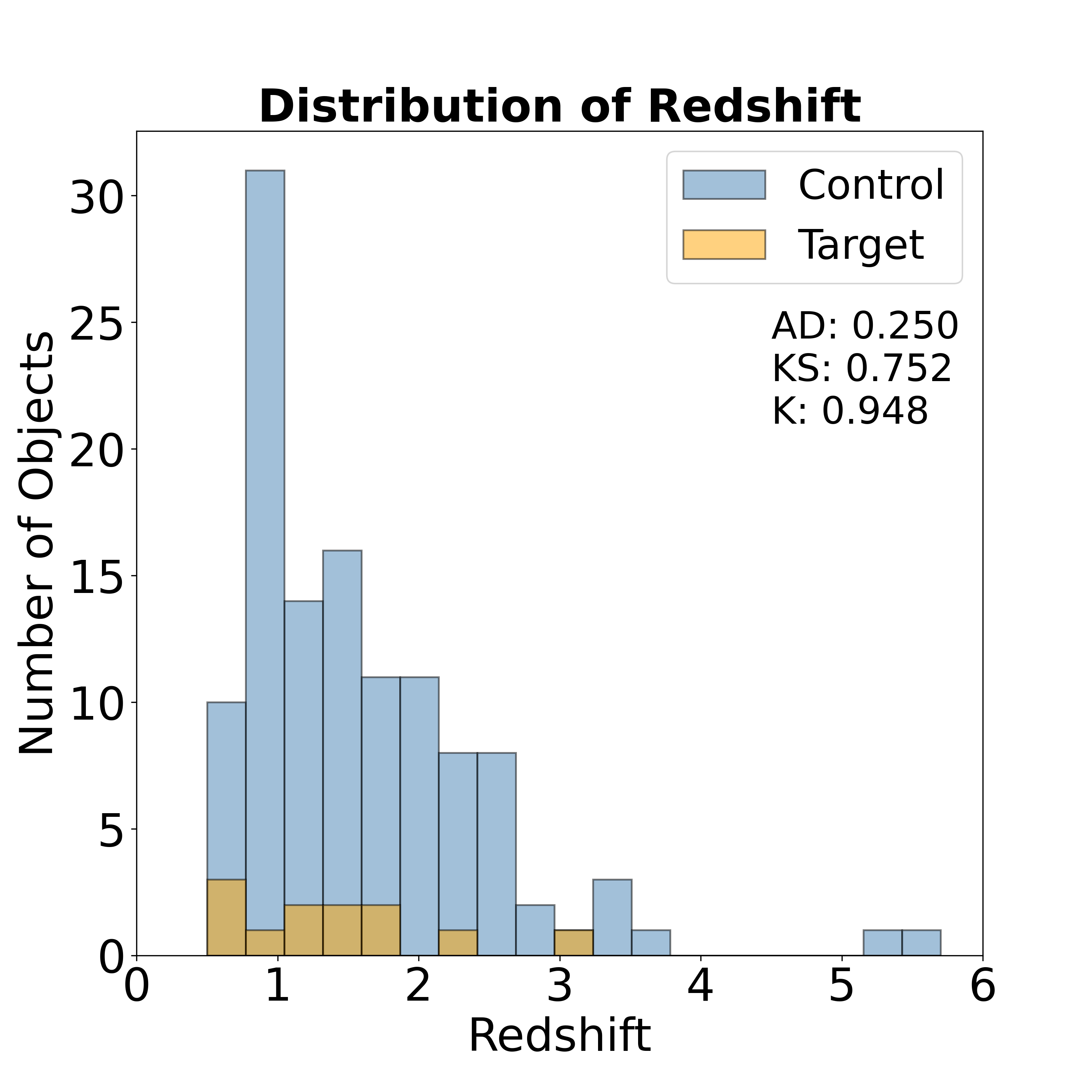}
\end{minipage}

\vspace{-0.4cm}

\begin{minipage}{0.32\textwidth}
    \includegraphics[width=\linewidth]{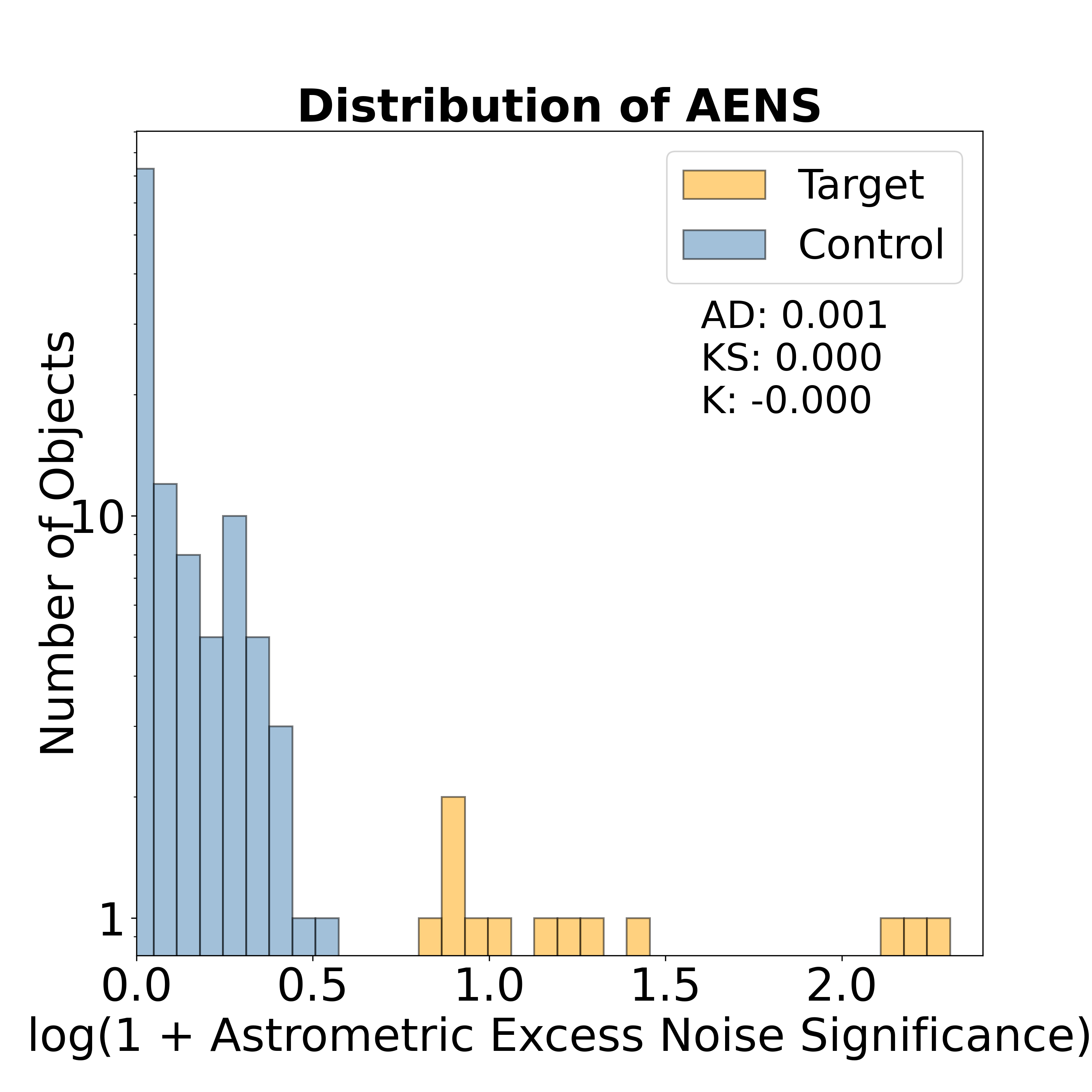}
\end{minipage}
\begin{minipage}{0.32\textwidth}
    \includegraphics[width=\linewidth]{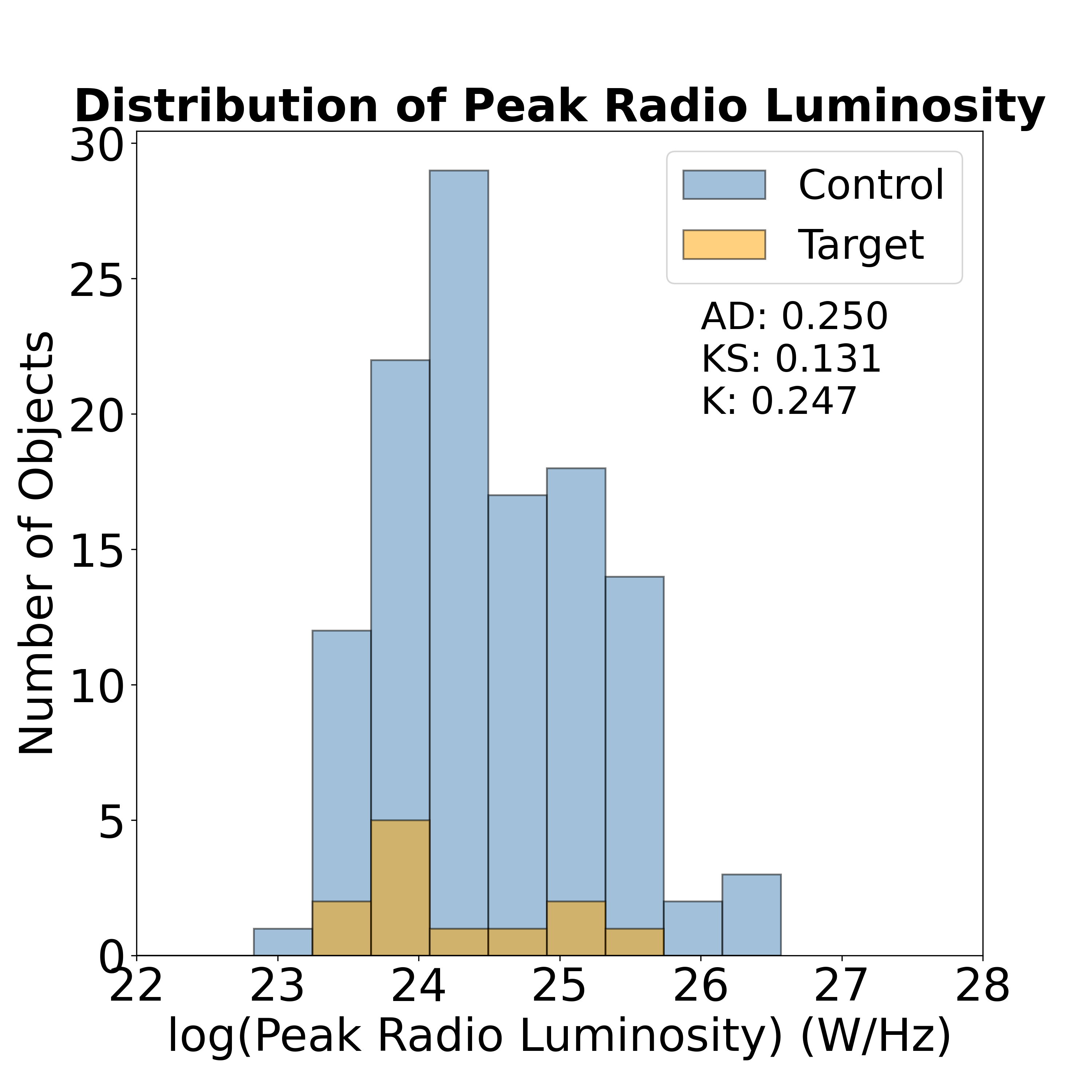}
\end{minipage}
\begin{minipage}{0.32\textwidth}
    \includegraphics[width=\linewidth]{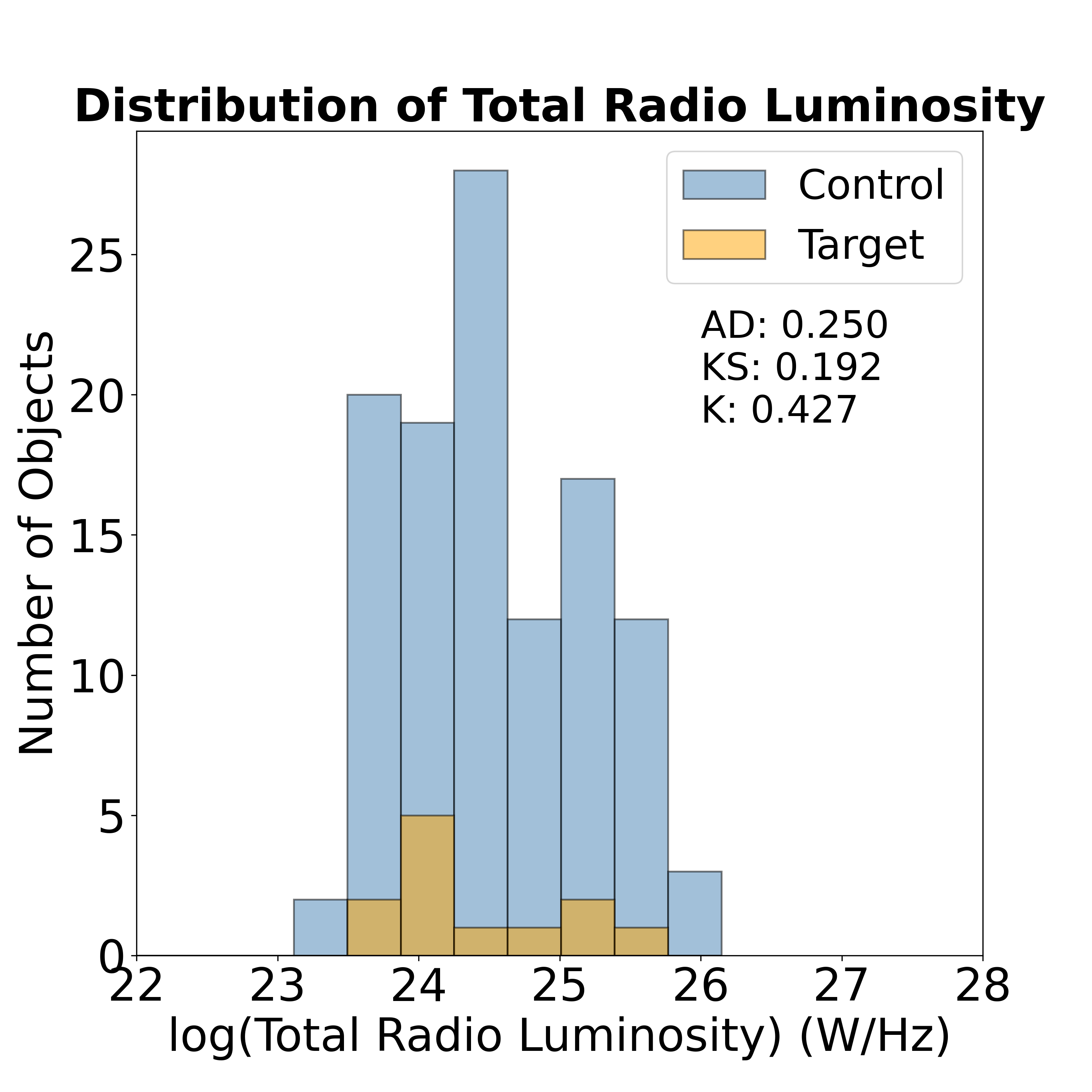}
\end{minipage}

\vspace{-0.4cm}

\begin{minipage}{0.32\textwidth}
    \includegraphics[width=\linewidth]{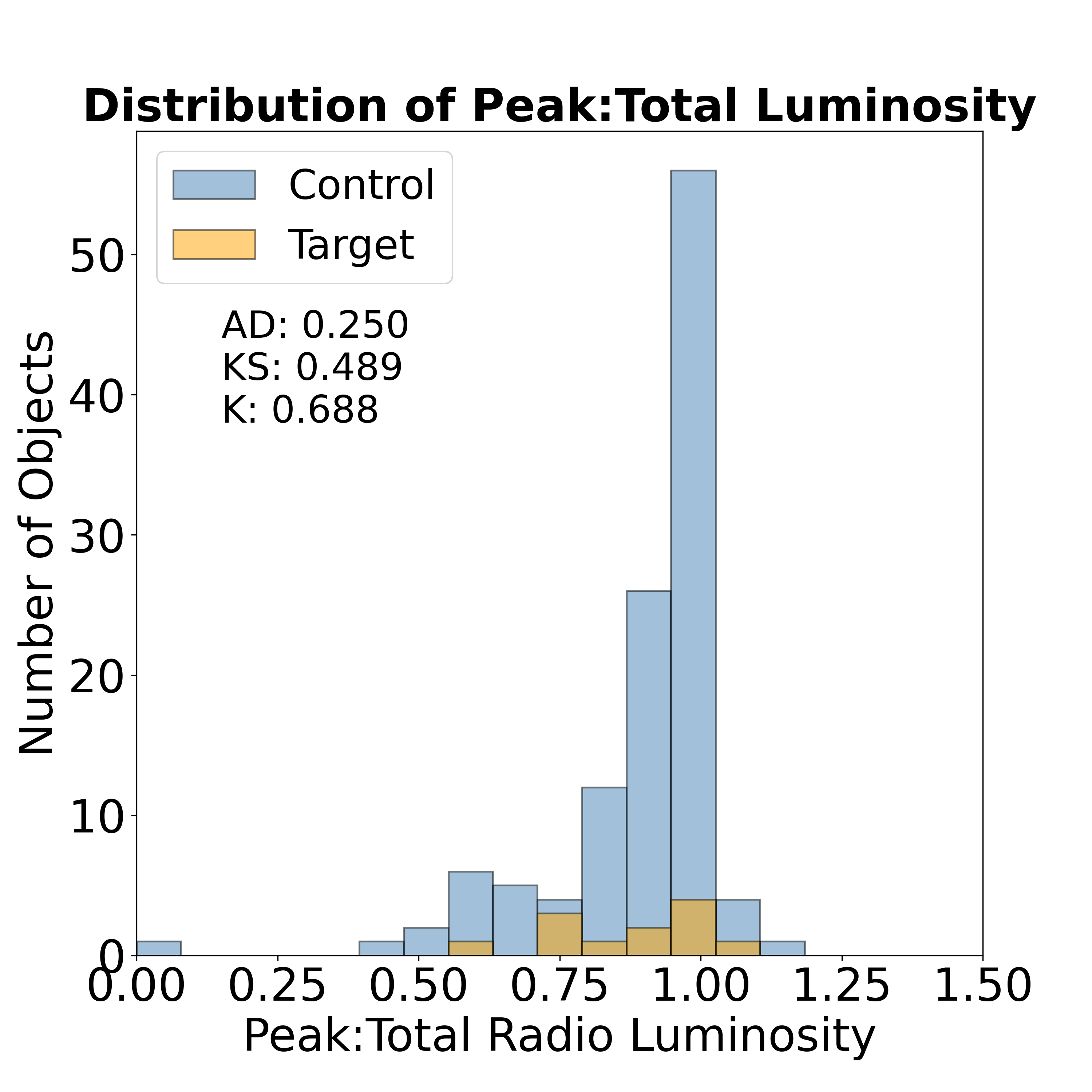}
\end{minipage}
\begin{minipage}{0.32\textwidth}
    \includegraphics[width=\linewidth]{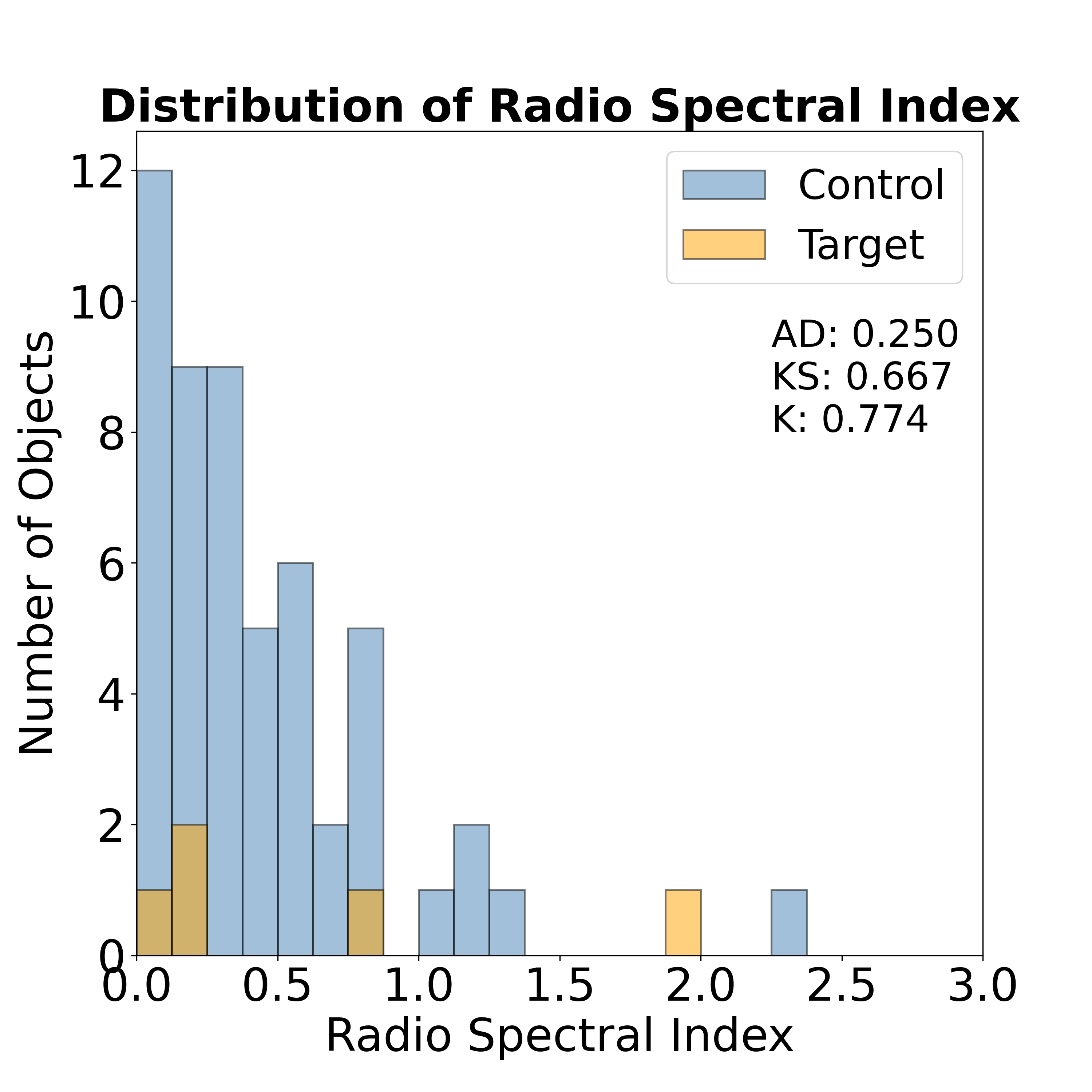}
\end{minipage}

\caption{These histograms compare the control sample of 120 targets to a target sample of 12 targets (all with AENS $>5$). The peak and total radio luminosities are those measured from VLASS. The spectral index is calculated between measurements from FIRST (at 1.4 GHz) and measurements from VLASS (at 3 GHz). The purpose of these plots is to compare the control and target samples for the important radio parameters. \textit{Gaia} G-band magnitude, G-band magnitude SNR, and target redshift were controlled for in the creation of the control sample, and the p-values listed on the plots show the lack of disagreement between the two samples. Note: In Panel 4, the logarithmic values of AENS + 1 are used. In Panels 5 and 6, the logarithmic values of the peak radio luminosity are used.}
\label{fig:histograms}
\end{figure*}

Distributions are shown in Figure \ref{fig:histograms}, illustrating the offset in the AENS. The controlled parameters exhibit the relationship driven by the sample selection parameters, but the important radio parameters (peak radio luminosity, total radio luminosity, ratio of peak-to-total radio luminosity, and radio spectral index) illustrate that the target sample is not radio-loud when compared to the matched control sample, likely eliminating the possibility of blazar jet activity driving a large percentage of the high AEN.

\section{Summary and Conclusions} \label{sec:summary}

We present deep, targeted VLA observations of 18 \textit{Gaia}-unresolved quasars identified as astrometrically-variable. The high redshifts (0.8 $\leq$ z $\leq$ 2.9) of these quasars probe an interesting observational gap in the current population of confirmed and candidate kiloparsec-scale dual AGN near cosmic noon. This pilot study, which combines the varstrometry method \cite[][]{hwang_initial} with high-resolution radio observations, is the first in a series of studies designed to fully understand the radio+varstrometry sample. Our conclusions are as follows:

\begin{itemize}
    \item The new VLA observations reveal eight targets identified as multi-component (candidate dual AGN) or gravitational lenses, or $\sim$44\% of the overall sample. Four of the eight are currently identified as gravitational lenses. The result is comparable to that of \cite{chen_hst}. Optical/infrared spectroscopy and imaging will be required to filter out other contaminants by identifying stellar features, modeling lensed quasar systems, and identifying lensing galaxies.  
    \item Three of the 18 targets have been identified as systems exhibiting jet activity or otherwise extended emission. This population includes J1415+1129, which has been identified from existing HST observations as a lensed quasar.
    \item A total of four of the 18 targets have been identified as gravitationally lensed quasars. We note that, while the focus of this search was multiple AGN systems, gravitationally lensed systems are equally interesting, and are the focus of many ongoing studies, including in the high resolution radio regime \cite[][]{spingola2019, casadio2021}. Additionally, gravitational lenses remain a focus of those mining \textit{Gaia}'s rich data set in order to create gravitational lens catalogs \cite[][]{lemon2023,Lemon_2018}. Follow-up high resolution optical observations will be required to identify gravitational lenses with smaller separations.
    \item Nine of the 18 targets have unresolved radio signatures. After careful analysis of the SDSS spectra, four of these nine were identified as superpositions of foreground stars and background quasars, and thus were eliminated as candidate dual AGN. J0818+0601 was additionally identified as a gravitationally lensed quasar. There remain many astrophysical explanations for the excess astrometric noise exhibited by the remaining unresolved targets. Follow-up multiwavelength observations will be required to observe any existing structure at smaller scales and to confirm or reject all scenarios. For the AGN pair scenario, this includes high-resolution optical/infrared imaging with HST or JWST to identify the two quasars and radio interferometric observations at smaller scales to identify two compact cores with the VLBA or the next-generation Very Large Array \cite[ngVLA;][]{selina2018,nyland2018science}.
    \item It is also possible for the unresolved targets to exhibit high astrometric variability caused by the properties of the host galaxy. If the host structure is significantly extended, as in the case of tidal features, this could contribute to the excess astrometric noise \cite[][]{makarov2022,makarov2023}. Despite the quality of the DECaLS imaging, due to the high redshifts of the targets, it is not possible to constrain the host galaxies with currently available optical observations from SDSS and DECaLS. Quasar host galaxy characterization will require follow-up optical/infrared imaging with higher angular resolution.
    \item We note that the excess astrometric noise could also be attributable to \textit{Gaia} systematics, though this is unlikely given the careful sample selection. Future \textit{Gaia} data releases will provide increasingly better overall astrometric measurements. Additionally, \textit{Gaia}'s pipeline continues to develop better treatment of extended sources, and continues to increase the reliability of the astrometric excess noise parameter, increasing the number of AGN pairs identifiable in future data releases.
    \item An in-depth study of the overall sample illustrated that the targets sample is not particularly radio loud in comparison to a matched control sample. This likely eliminates the possibility of blazar jet activity as the main driver of the high excess astrometric noise.
    \item Finally, a thorough review of the radio spectral shapes of each target reveals that the majority of the targets, no matter their spectral shape classification, exhibit a spectral index that is consistent with that of optically thin synchrotron emission. Targets with curved radio spectral shapes would benefit from follow-up simultaneous observations with broad frequency coverage above and below the peak frequency.
\end{itemize}

Overall, we demonstrate the potential of using radio+varstrometry to systematically discover genuine kiloparsec-scale dual quasar candidates, including at redshifts greater than $z\sim 0.8$. Additionally, these results will assist in refining the targeting strategy to facilitate a more efficient identification of dual and binary AGN specifics by providing radio constraints that can be applied to a larger, more systematic search of existing radio surveys. 

\clearpage
\begin{acknowledgements}
We thank the anonymous referee for the many helpful suggestions that have significantly improved the paper. We thank Cameron Lemon, Phil Cigan, and Pallavi Patil for their guidance and their many useful comments and suggestions. E. S. gratefully acknowledges support from the National Radio Astronomy Observatory's Student Observing Support fellowship. R. W. P. gratefully acknowledges support through an appointment to the NASA Postdoctoral Program at Goddard Space Flight Center, administered by ORAU through a contract with NASA. 
This work has made use of data from the European Space Agency (ESA) mission Gaia (https://www.cosmos.esa.int/gaia), processed by the Gaia Data Processing and Analysis Consortium (DPAC, https://www.cosmos.esa.int/web/gaia/dpac/consortium). Funding for the DPAC has been provided by national institutions, in particular the institutions participating in the Gaia Multilateral Agreement.
This research made use of Astropy, a community-developed core Python package for Astronomy (\cite{astropy}), $\mathrm{TOPCAT}$ (\cite{topcat}), the Common Astronomy Software Application (\cite{casanew2022}), and the Python Blob Detector and Source Finder (\cite{pybdsf}). 
Funding for the Sloan Digital Sky Survey IV has been provided by the Alfred P. Sloan Foundation, the U.S.
Department of Energy Office of Science, and the Participating Institutions.
 The National Radio Astronomy Observatory is a facility of the National Science Foundation operated under cooperative agreement by Associated Universities, Inc. Basic research in radio astronomy at the U.S. Naval Research Laboratory is supported by 6.1 Base Funding. This work made use of data from the VLA Low-band Ionosphere and Transient Experiment (VLITE). Construction and installation of VLITE was supported by the NRL Sustainment, Restoration, and Maintenance funding.

\end{acknowledgements}

\vspace{5mm}
\facilities{\textit{Gaia}, VLA (NRAO), Sloan, HST, VLBA}

\software{astropy (\cite{astropy}), CASA (\cite{casanew2022}), PyBDSF (\cite{pybdsf}), $\mathrm{TOPCAT}$ (\cite{topcat})}

\bibliography{citation.bib} 

\bibliographystyle{aasjournal}
\nolinenumbers

\section{Appendix A: Individual Target VLA Observations}
\label{sec:appendixA}

\begin{figure*}[!htb]
\gridline{\fig{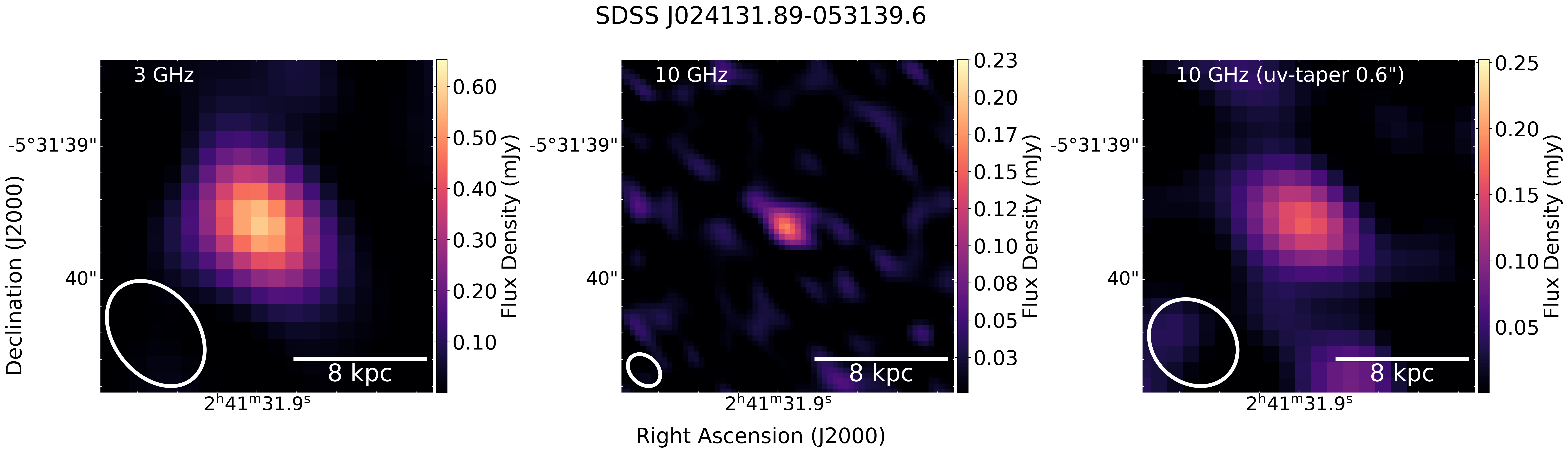}{\textwidth}{(a) \textbf{SDSSJ024131.89-053139.6:} Left: VLA 3 GHz observations with a 0.87$\arcsec{}$ $\times$ 0.63$\arcsec{}$ beam. Middle: VLA 10 GHz observations with a 0.19$\arcsec{}$ $\times$ 0.17$\arcsec{}$ beam. Right: VLA 10 GHz uv-tapered observations with a 0.71$\arcsec{}$ $\times$ 0.60$\arcsec{}$ beam. Scale is 7.50 kiloparsecs per arcsecond. \textbf{This target has been identified as star+quasar superposition.}}}
\gridline{\fig{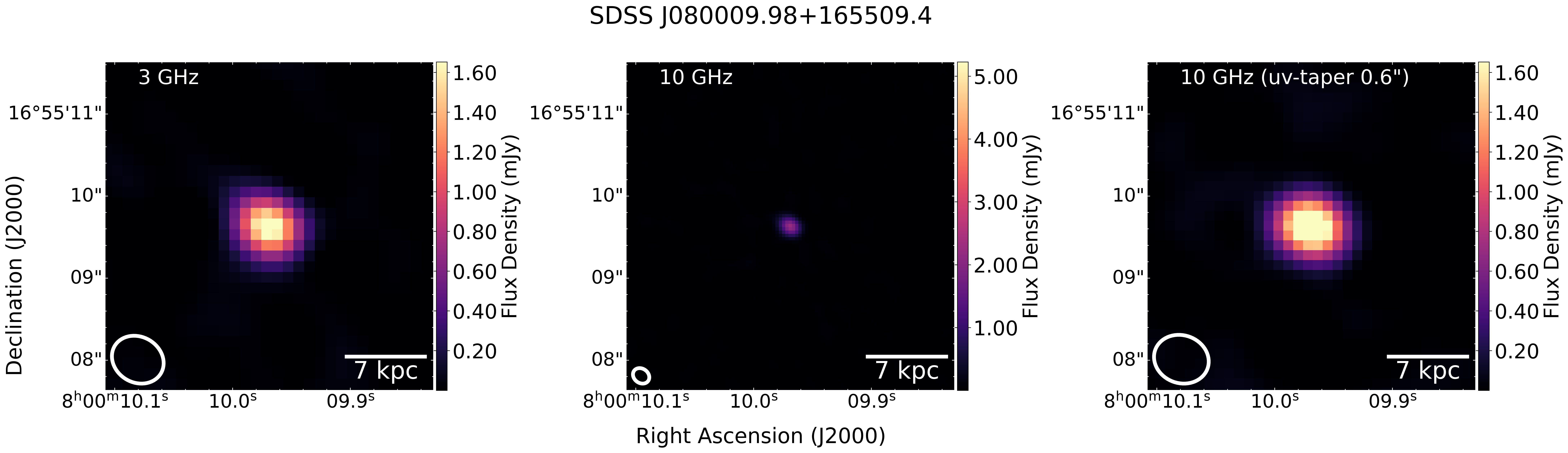}{\textwidth}{(b) \textbf{SDSSJ080009.98+165509.4:} Left: VLA 3 GHz observations with a 0.65$\arcsec{}$ $\times$ 0.55$\arcsec{}$ beam. Middle: VLA 10 GHz observations with a 0.23$\arcsec{}$ $\times$ 0.17$\arcsec{}$ beam. Right: VLA 10 GHz uv-tapered observations with a 0.67$\arcsec{}$ $\times$ 0.58$\arcsec{}$ beam. Scale is 7.14 kiloparsecs per arcsecond.}}
\gridline{\fig{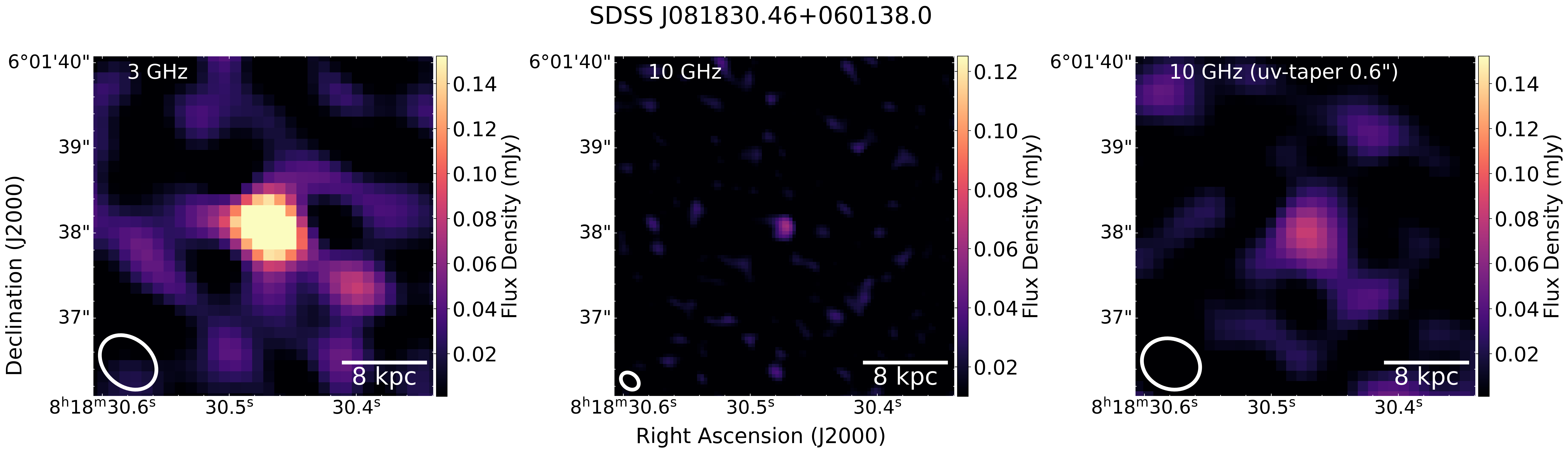}{\textwidth}{(c) \textbf{SDSSJ081830.46+060138.0:} Left: VLA 3 GHz observations with a 0.73$\arcsec{}$ $\times$ 0.56$\arcsec{}$ beam. Middle: VLA 10 GHz observations with a 0.21$\arcsec{}$ $\times$ 0.18$\arcsec{}$ beam. Right: VLA 10 GHz uv-tapered observations with a 0.69$\arcsec{}$ $\times$ 0.58$\arcsec{}$ beam. Scale is 8.20 kiloparsecs per arcsecond. \textbf{This target has been identified as a gravitationally lensed quasar.}}}
\caption{New VLA observations for all targets identified as unresolved.} \label{fig:radiounresolved1}
\end{figure*}
\setcounter{figure}{7}

\begin{figure*}
\gridline{\fig{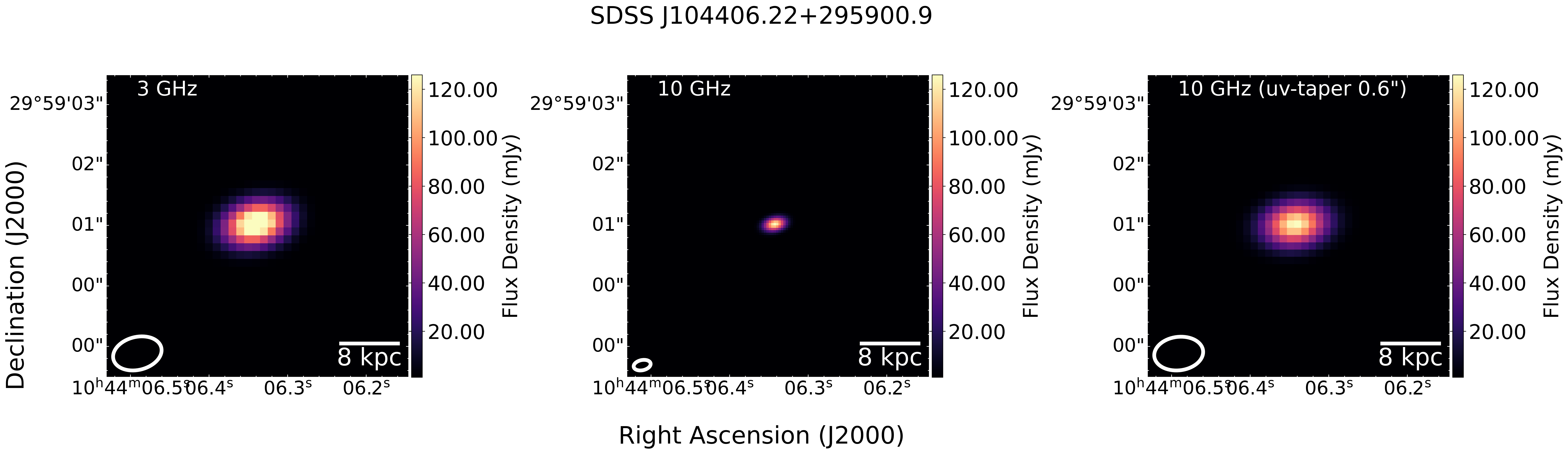}{\textwidth}{(d) \textbf{SDSSJ104406.33+295900.9:} Left: VLA 3 GHz observations with a 0.82$\arcsec{}$ $\times$ 0.54$\arcsec{}$ beam. Middle: VLA 10 GHz observations with a 0.18$\arcsec{}$ $\times$ 0.15$\arcsec{}$ beam. Right: VLA 10 GHz uv-tapered observations with a 0.81$\arcsec{}$ $\times$ 0.55$\arcsec{}$ beam. Scale is 7.77 kiloparsecs per arcsecond. \textbf{This target exhibits possible time-dependent flux density variability over years-long timescales.}}} 
\gridline{\fig{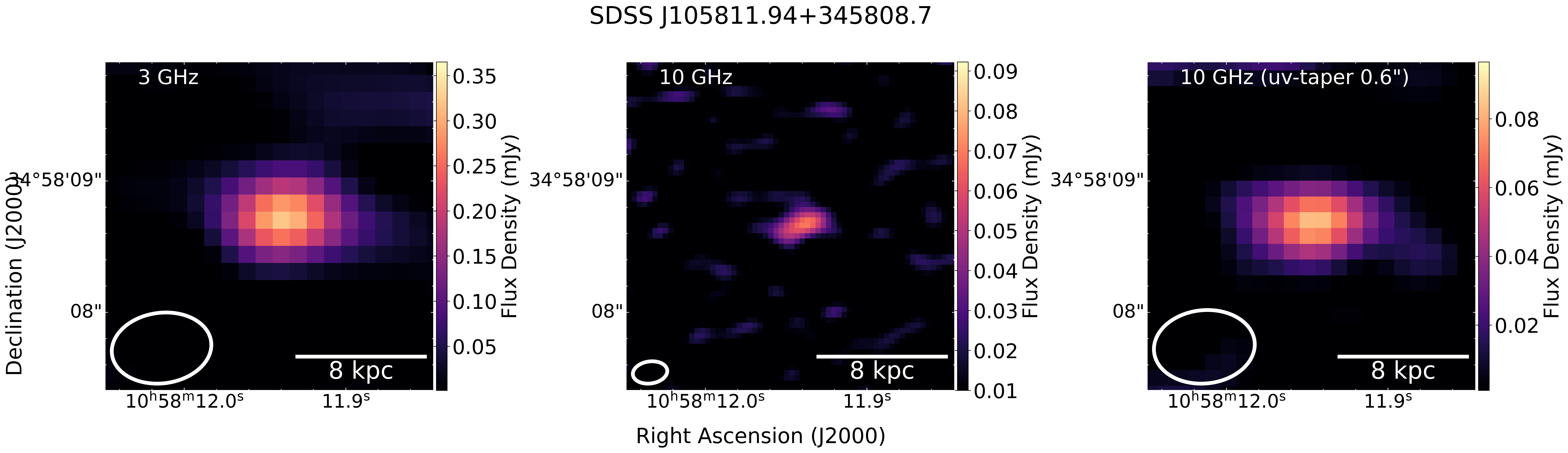}{\textwidth}{(e) \textbf{SDSSJ105811.94+240217.4:} Left: VLA 3 GHz observations with a 0.76$\arcsec{}$ $\times$ 0.54$\arcsec{}$ beam. Middle: VLA 10 GHz observations with a 0.12$\arcsec{}$ $\times$ 0.16$\arcsec{}$ beam. Right: VLA 10 GHz uv-tapered observations with a 0.76$\arcsec{}$ $\times$ 0.55$\arcsec{}$ beam. Scale is 8.17 kiloparsecs per arcsecond.}}
\gridline{\fig{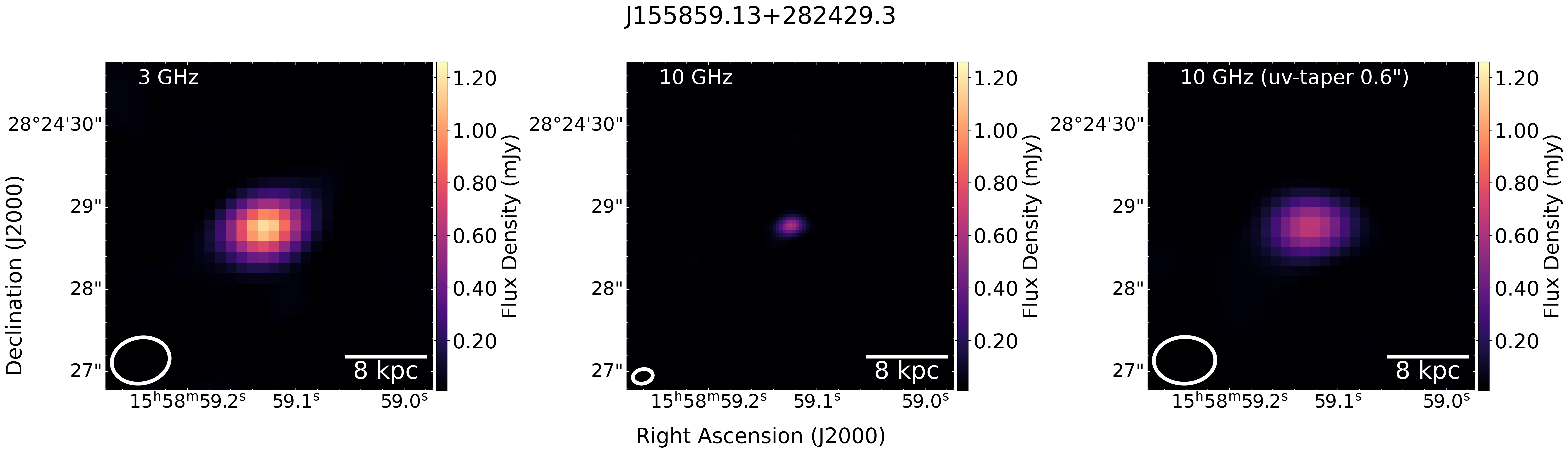}{\textwidth}{(f) \textbf{SDSSJ155859.13+282429.3:} Left: VLA 3 GHz observations with a 0.71$\arcsec{}$ $\times$ 0.56$\arcsec{}$ beam. Middle: VLA 10 GHz observations with a 0.31$\arcsec{}$ $\times$ 0.18$\arcsec{}$ beam. Right: VLA 10 GHz uv-tapered observations with a 0.74$\arcsec{}$ $\times$ 0.56$\arcsec{}$ beam. Scale is 8.21 kiloparsecs per arcsecond. \textbf{This target has been identified as star+quasar superposition.}}} 
\caption{Continued from previous page. New VLA observations for all targets identified as unresolved.} \label{fig:radiounresolved2}
\end{figure*}
\setcounter{figure}{7}

\begin{figure*}
\gridline{\fig{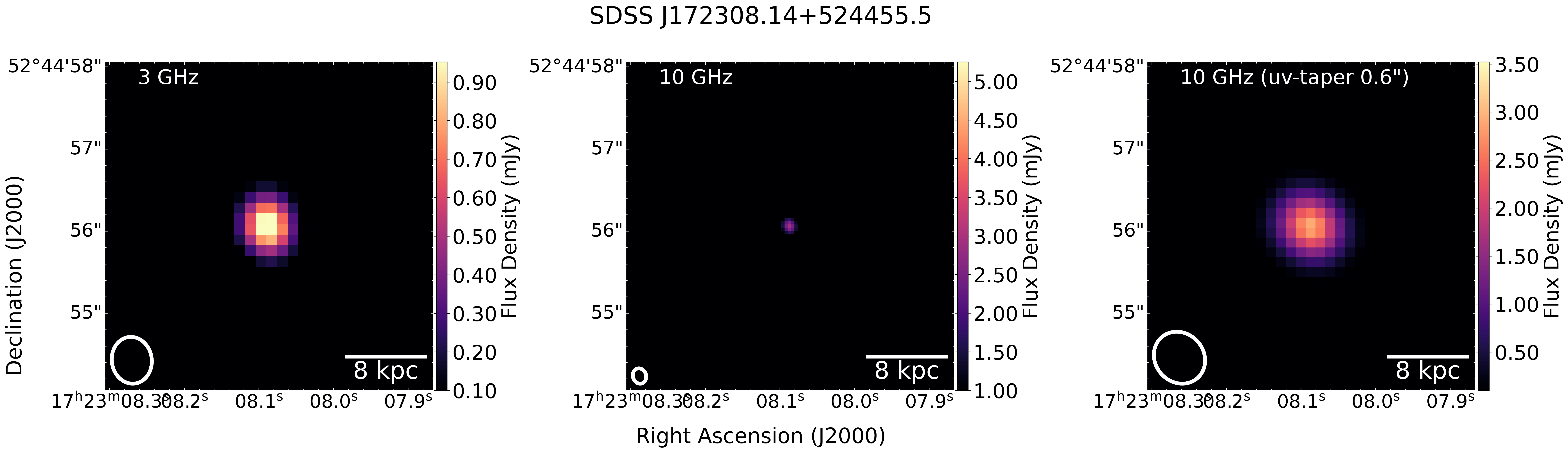}{\textwidth}{(g) \textbf{SDSSJ172308.14+524455.5:} Left: VLA 3 GHz observations with a 0.57$\arcsec{}$ $\times$ 0.49$\arcsec{}$ beam. Middle: VLA 10 GHz observations with a 0.28$\arcsec{}$ $\times$ 0.16$\arcsec{}$ beam. Right: VLA 10 GHz uv-tapered observations with a 0.66$\arcsec{}$ $\times$ 0.59$\arcsec{}$ beam. Scale is 8.44 kiloparsecs per arcsecond. \textbf{This target has been identified as star+quasar superposition.}}} 
\gridline{\fig{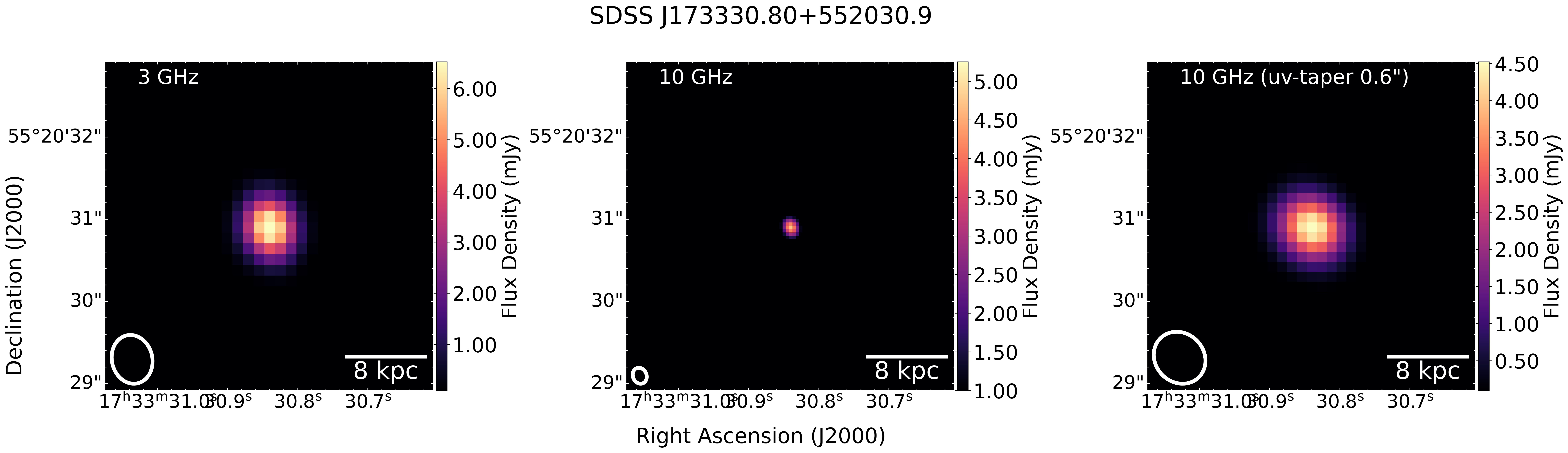}{\textwidth}{(h) \textbf{SDSSJ173330.80+552030.9:} Left: VLA 3 GHz observations with a 0.59$\arcsec{}$ $\times$ 0.49$\arcsec{}$ beam. Middle: VLA 10 GHz observations with a 0.19$\arcsec{}$ $\times$ 0.17$\arcsec{}$ beam. Right: VLA 10 GHz uv-tapered observations with a 0.66$\arcsec{}$ $\times$ 0.59$\arcsec{}$ beam. Scale is 8.29 kiloparsecs per arcsecond.}}
\gridline{\fig{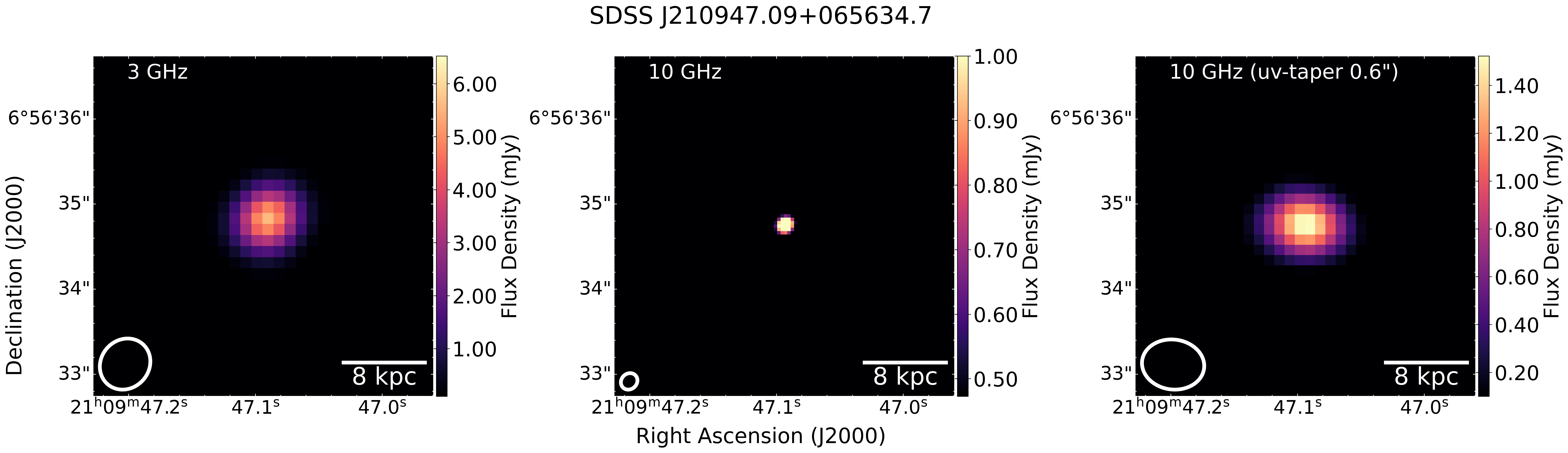}{\textwidth}{(i) \textbf{SDSSJ210947.09+065634.7:} Left: VLA 3 GHz observations with a 0.63$\arcsec{}$ $\times$ 0.57$\arcsec{}$ beam. Middle: VLA 10 GHz observations with a 0.26$\arcsec{}$ $\times$ 0.16$\arcsec{}$ beam. Right: VLA 10 GHz uv-tapered observations with a 0.73$\arcsec{}$ $\times$ 0.58$\arcsec{}$ beam. Scale is 7.77 kiloparsecs per arcsecond. \textbf{This target has been identified as star+quasar superposition.}}}
\caption{Continued from previous page. New VLA observations for all targets identified as unresolved.} \label{fig:radiounresolved3}
\end{figure*}

\clearpage
\setcounter{figure}{8}

\begin{figure*}
\gridline{\fig{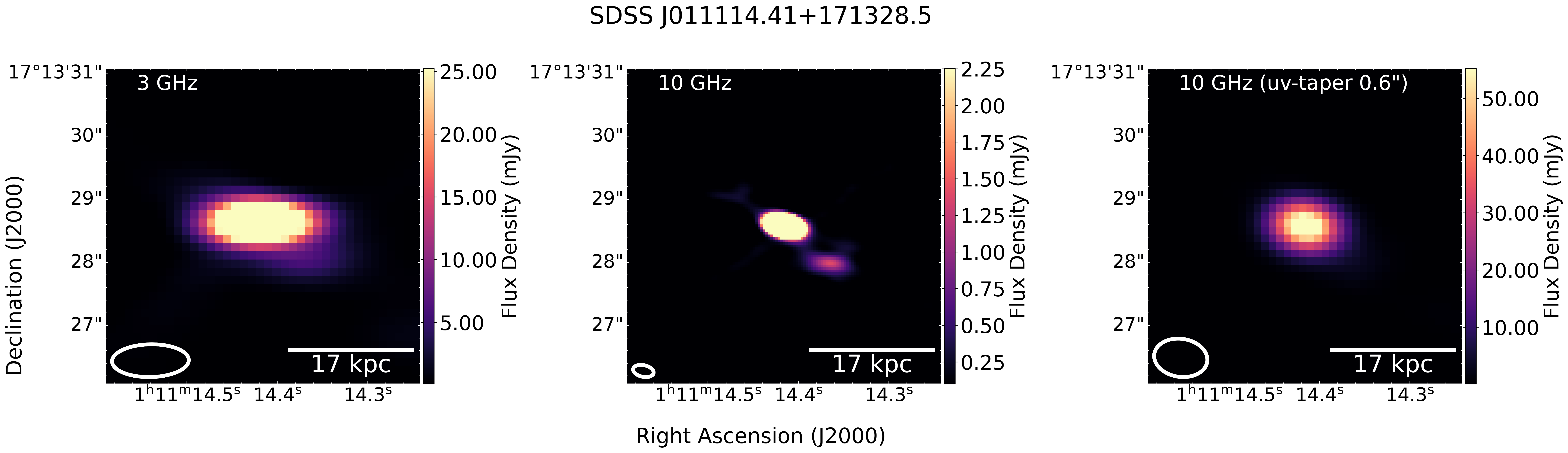}{\textwidth}{(a) \textbf{SDSSJ011114.41+171328.5:} Left: VLA 3 GHz observations with a 1.22$\arcsec{}$ $\times$ 0.52$\arcsec{}$ beam. Middle: VLA 10 GHz observations with a 0.27$\arcsec{}$ $\times$ 0.21$\arcsec{}$ beam. Right: VLA 10 GHz uv-tapered observations with a 0.84$\arcsec{}$ $\times$ 0.60$\arcsec{}$ beam. Scale is 8.26 kiloparsecs per arcsecond. \textbf{This target is a candidate dual AGN.}}} 
\gridline{\fig{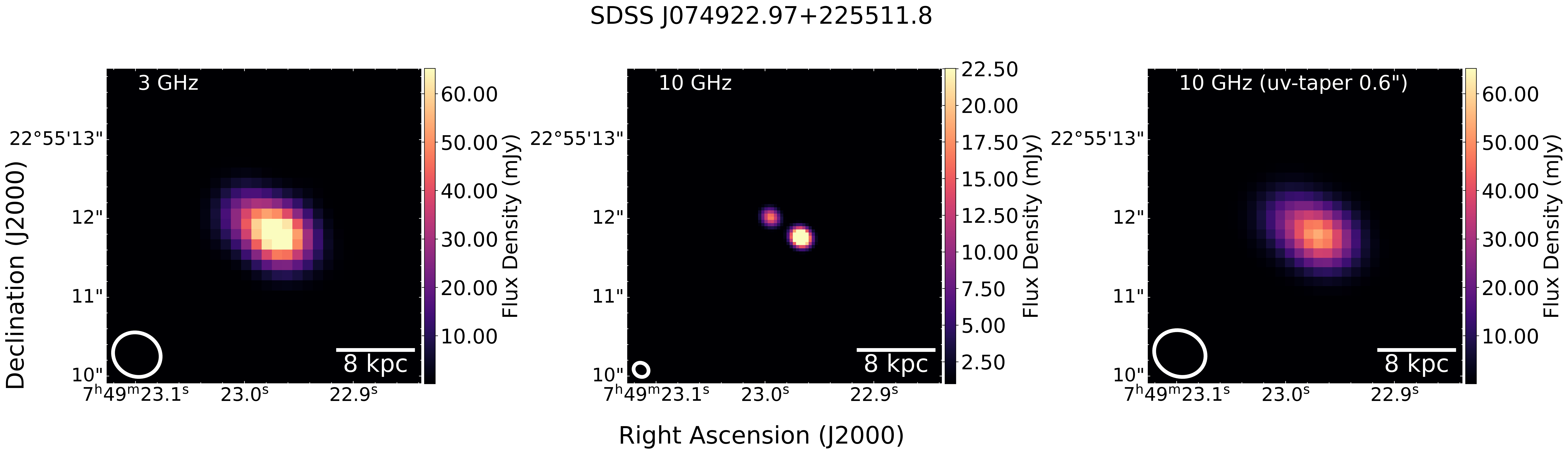}{\textwidth}{(b) \textbf{SDSSJ074922.97+225511.8:} Left: VLA 3 GHz observations with a 0.62$\arcsec{}$ $\times$ 0.55$\arcsec{}$ beam. Middle: VLA 10 GHz observations with a 0.24$\arcsec{}$ $\times$ 0.17$\arcsec{}$ beam. Right: VLA 10 GHz uv-tapered observations with a 0.66$\arcsec{}$ $\times$ 0.57$\arcsec{}$ beam. Scale is 8.32 kiloparsecs per arcsecond. \textbf{This target is a confirmed dual AGN.}}}
\gridline{\fig{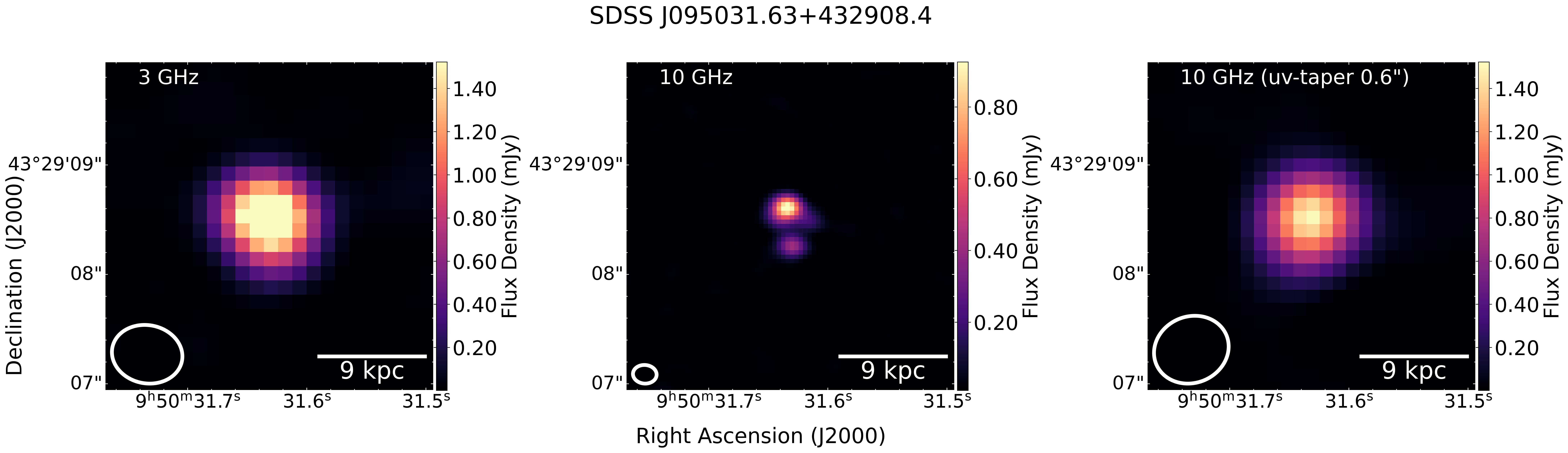}{\textwidth}{(c) \textbf{SDSSJ095031.63+432908.4:} Left: VLA 3 GHz observations with a 0.65$\arcsec{}$ $\times$ 0.53$\arcsec{}$ beam. Middle: VLA 10 GHz observations with a 0.21$\arcsec{}$ $\times$ 0.17$\arcsec{}$ beam. Right: VLA 10 GHz uv-tapered observations with a 0.69$\arcsec{}$ $\times$ 0.61$\arcsec{}$ beam. Scale is 8.46 kiloparsecs per arcsecond. \textbf{This target has been identified as a candidate dual AGN.}}} 
\caption{New VLA observations for all targets identified as having multiple components.} \label{fig:radiomc1}
\end{figure*}
\setcounter{figure}{8}

\begin{figure*}
\gridline{\fig{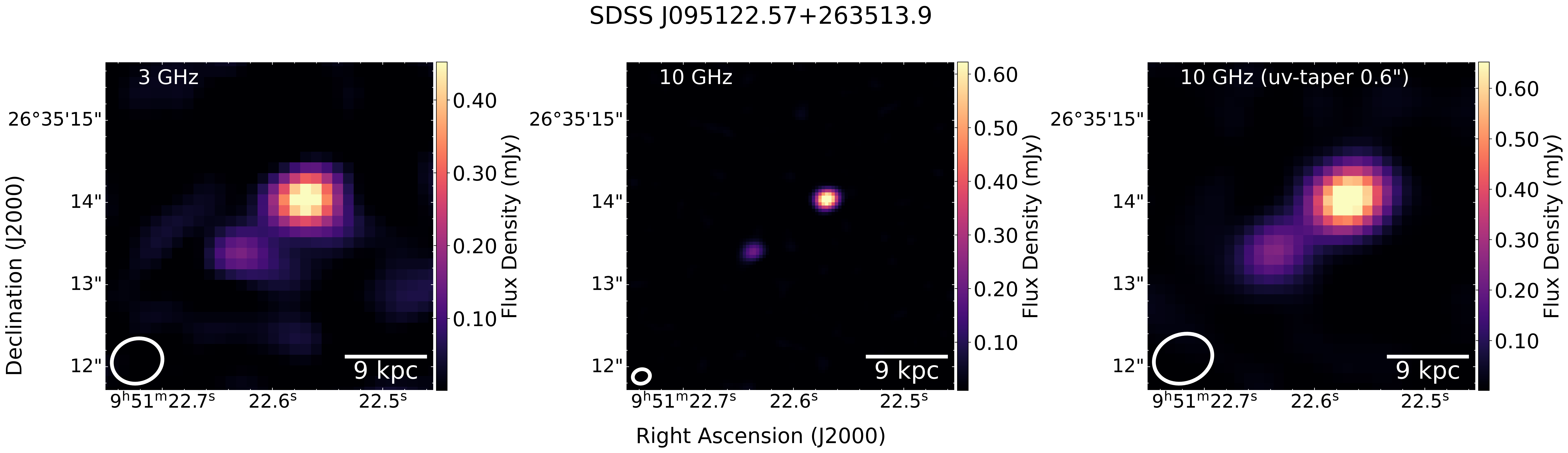}{\textwidth}{(d) \textbf{SDSSJ095122.57+263513.9:} Left: VLA 3 GHz observations with a 0.62$\arcsec{}$ $\times$ 0.54$\arcsec{}$ beam. Middle: VLA 10 GHz observations with a 0.21$\arcsec{}$ $\times$ 0.17$\arcsec{}$ beam. Right: VLA 10 GHz uv-tapered observations with a 0.72$\arcsec{}$ $\times$ 0.59$\arcsec{}$ beam. Scale is 8.23 kiloparsecs per arcsecond. \textbf{This target has been identified as a gravitationally lensed quasar.}}} 
\gridline{\fig{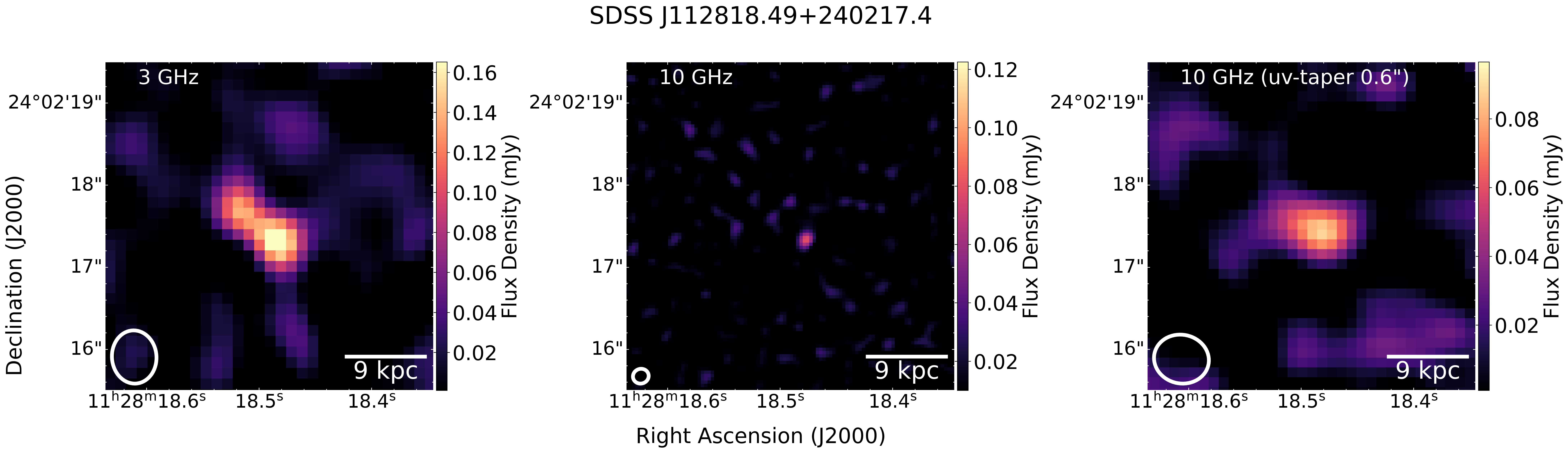}{\textwidth}{(e) \textbf{SDSSJ112818.49+240217.4:} Left: VLA 3 GHz observations with a 0.65$\arcsec{}$ $\times$ 0.54$\arcsec{}$ beam. Middle: VLA 10 GHz observations with a 0.25$\arcsec{}$ $\times$ 0.17$\arcsec{}$ beam. Right: VLA 10 GHz uv-tapered observations with a 0.67$\arcsec{}$ $\times$ 0.59$\arcsec{}$ beam. Scale is 8.47 kiloparsecs per arcsecond. \textbf{This target has been identified as a gravitationally lensed quasar.}}}
\gridline{\fig{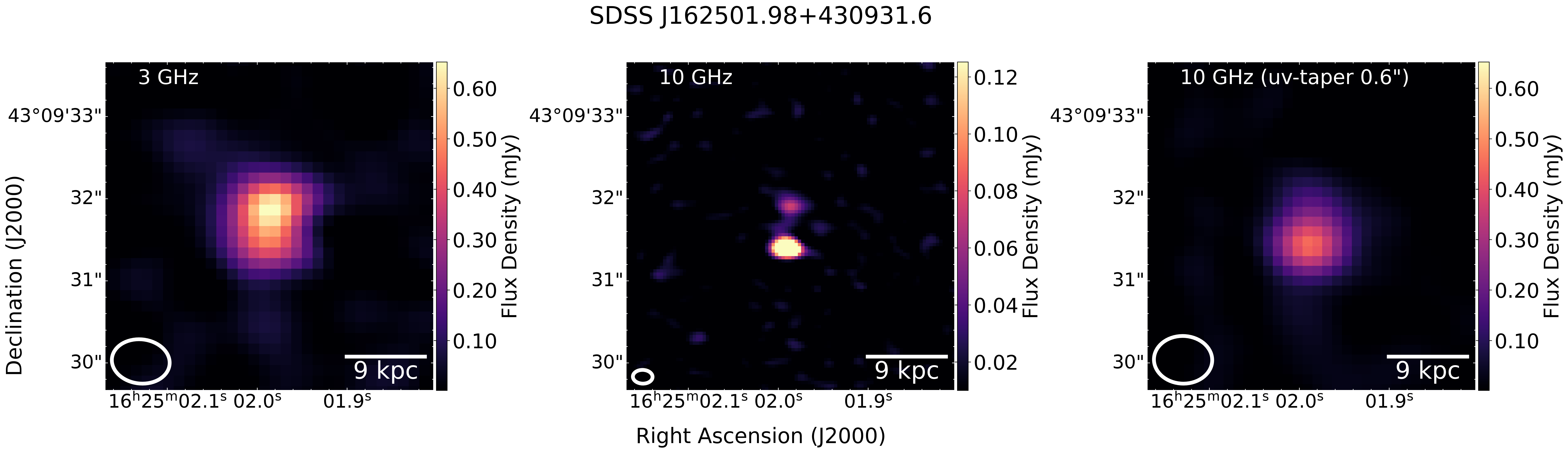}{\textwidth}{(f) \textbf{SDSSJ162501.98+430931.6:} Left: VLA 3 GHz observations with a 0.71$\arcsec{}$ $\times$ 0.54$\arcsec{}$ beam. Middle: VLA 10 GHz observations with a 0.33$\arcsec{}$ $\times$ 0.18$\arcsec{}$ beam. Right: VLA 10 GHz uv-tapered observations with a 0.71$\arcsec{}$ $\times$ 0.58$\arcsec{}$ beam. Scale is 8.47 kiloparsecs per arcsecond. \textbf{This target has been identified as a candidate dual AGN.}}}
\caption{Continued from previous page. New VLA observations for all targets identified as having multiple components.} \label{fig:radiomc2}
\end{figure*}

\clearpage
\setcounter{figure}{9}

\begin{figure*}
\gridline{\fig{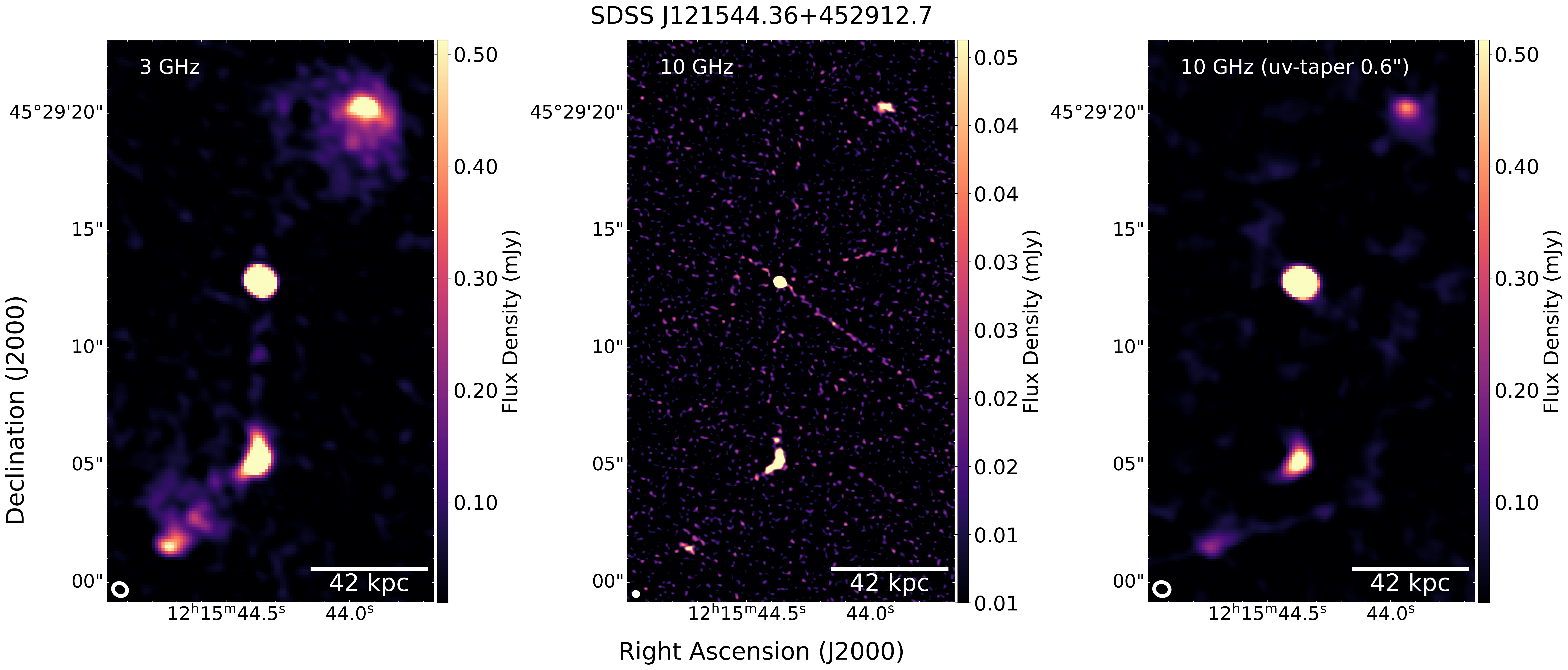}{\textwidth}{(a) \textbf{SDSSJ121544.36+452912.7:} Left: VLA 3 GHz observations with a 0.62$\arcsec{}$ $\times$ 0.53$\arcsec{}$ beam. Middle: VLA 10 GHz observations with a 0.23$\arcsec{}$ $\times$ 0.16$\arcsec{}$beam. Right: VLA 10 GHz uv-tapered observations with a 0.66$\arcsec{}$ $\times$ 0.58$\arcsec{}$ beam. Scale is 8.17 kiloparsecs per arcsecond. \textbf{This target exhibits a pair of jets, featuring a knot in the southern jet.}}} 
\gridline{\fig{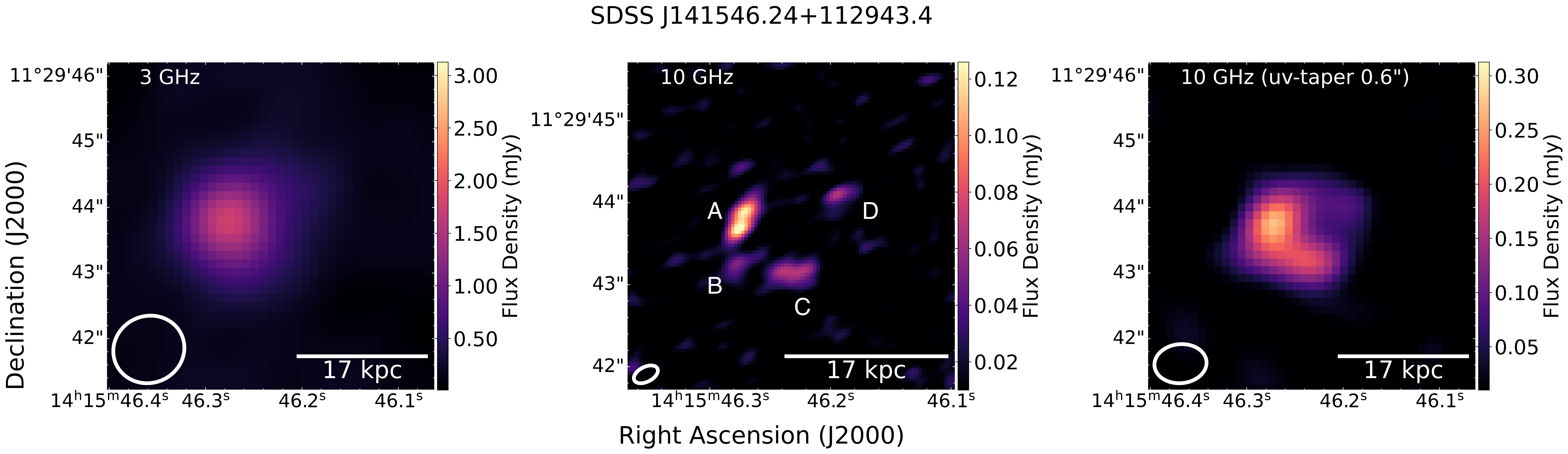}{\textwidth}{(b) \textbf{SDSSJ141546.24+112943.4:} Left: VLA 3 GHz observations with a 1.09$\arcsec{}$ $\times$ 1.02$\arcsec{}$ beam. Middle: VLA 10 GHz observations with a 0.19$\arcsec{}$ $\times$ 0.16$\arcsec{}$ beam. Right: VLA 10 GHz uv-tapered observations with a 0.79$\arcsec{}$ $\times$ 0.59$\arcsec{}$ beam. Scale is 8.07 kiloparsecs per arcsecond. \textbf{This is an extended target that has been identified as a gravitationally lensed quasar known as the Cloverleaf System.}}}
\gridline{\fig{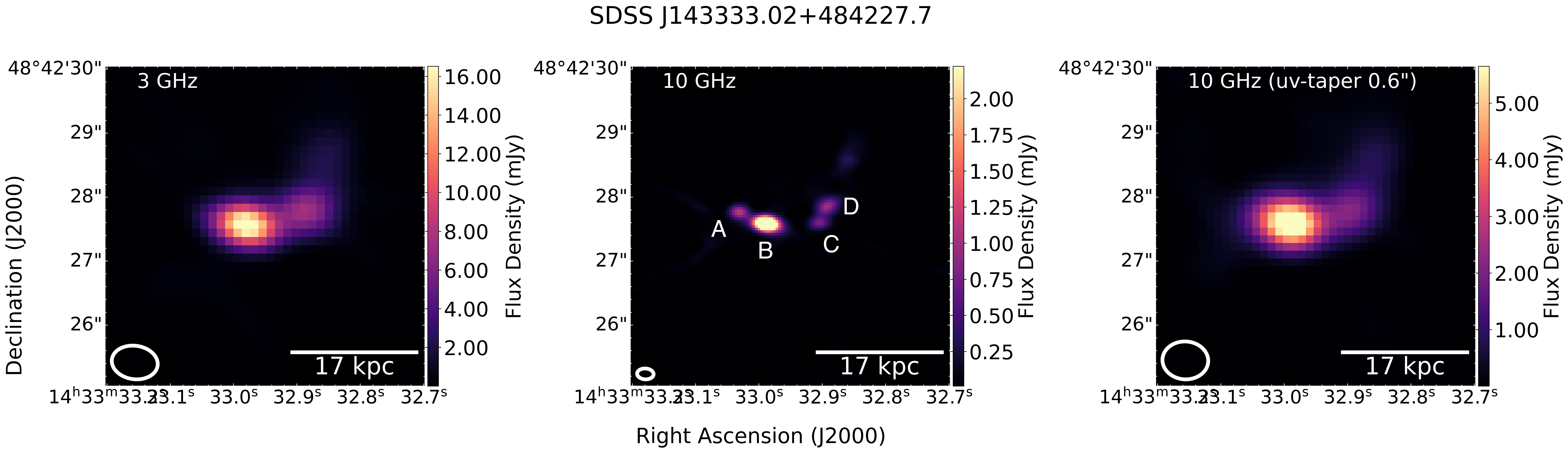}{\textwidth}{(c) \textbf{SDSSJ143333.02+484227.7:} Left: VLA 3 GHz observations with a 0.72$\arcsec{}$ $\times$ 0.52$\arcsec{}$ beam. Middle: VLA 10 GHz observations with a 0.21$\arcsec{}$ $\times$ 0.17$\arcsec{}$ beam. Right: VLA 10 GHz uv-tapered observations with a 0.71$\arcsec{}$ $\times$ 0.59$\arcsec{}$ beam. Scale is 8.37 kiloparsecs per arcsecond. \textbf{This extended target exhibits jet activity.}}} 
\caption{New VLA observations for all targets identified as extended.} \label{fig:radioextended1}
\end{figure*}

\clearpage

\section{Appendix B: Individual Target Optical Spectra}

\setcounter{figure}{10}

\begin{center}
  \begin{longtable}{cc}
\begin{minipage}[t]{0.45\linewidth}
  \includegraphics[width=\linewidth]{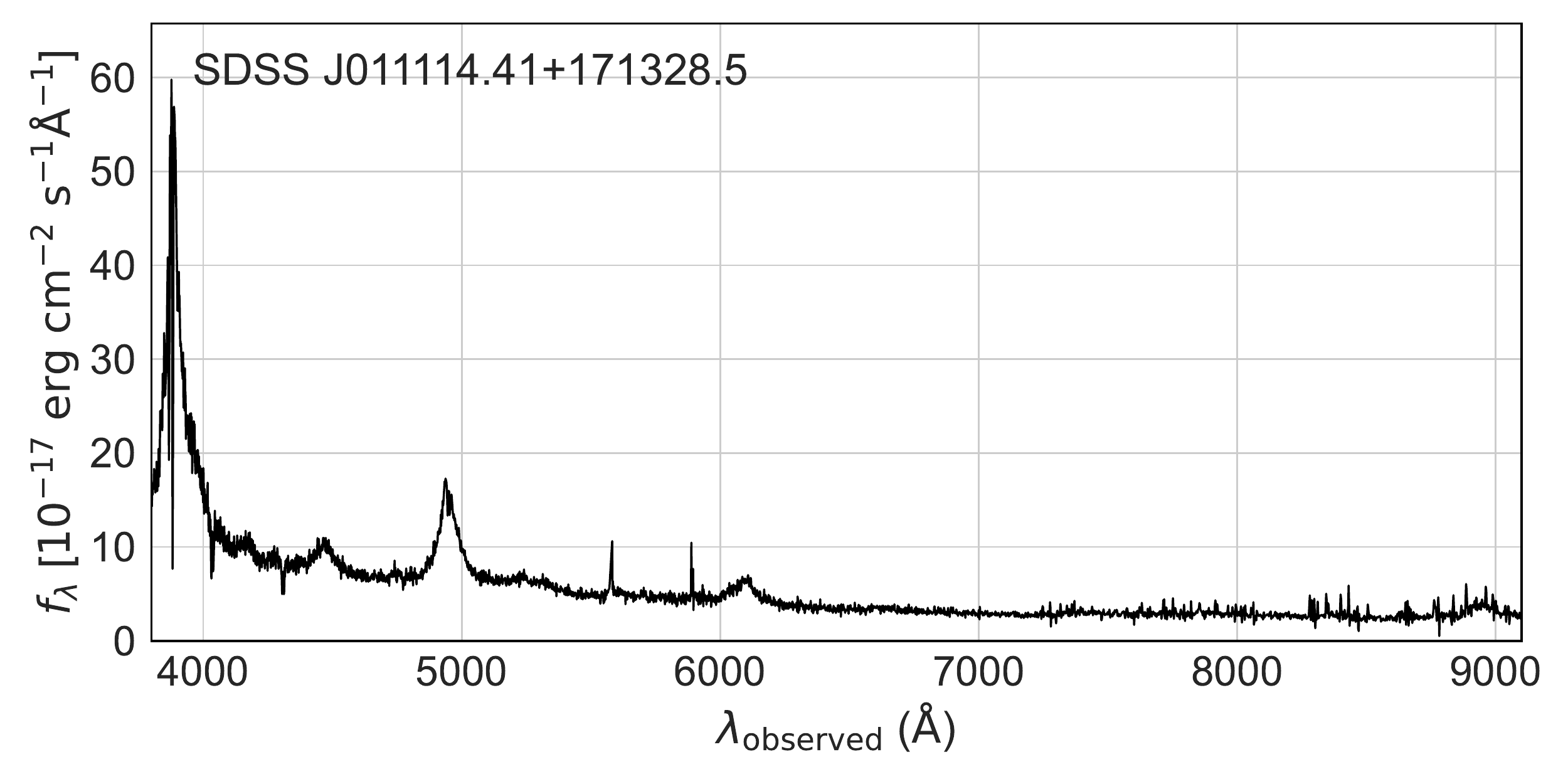}
\end{minipage} &
\begin{minipage}[t]{0.45\linewidth}
  \includegraphics[width=\linewidth]{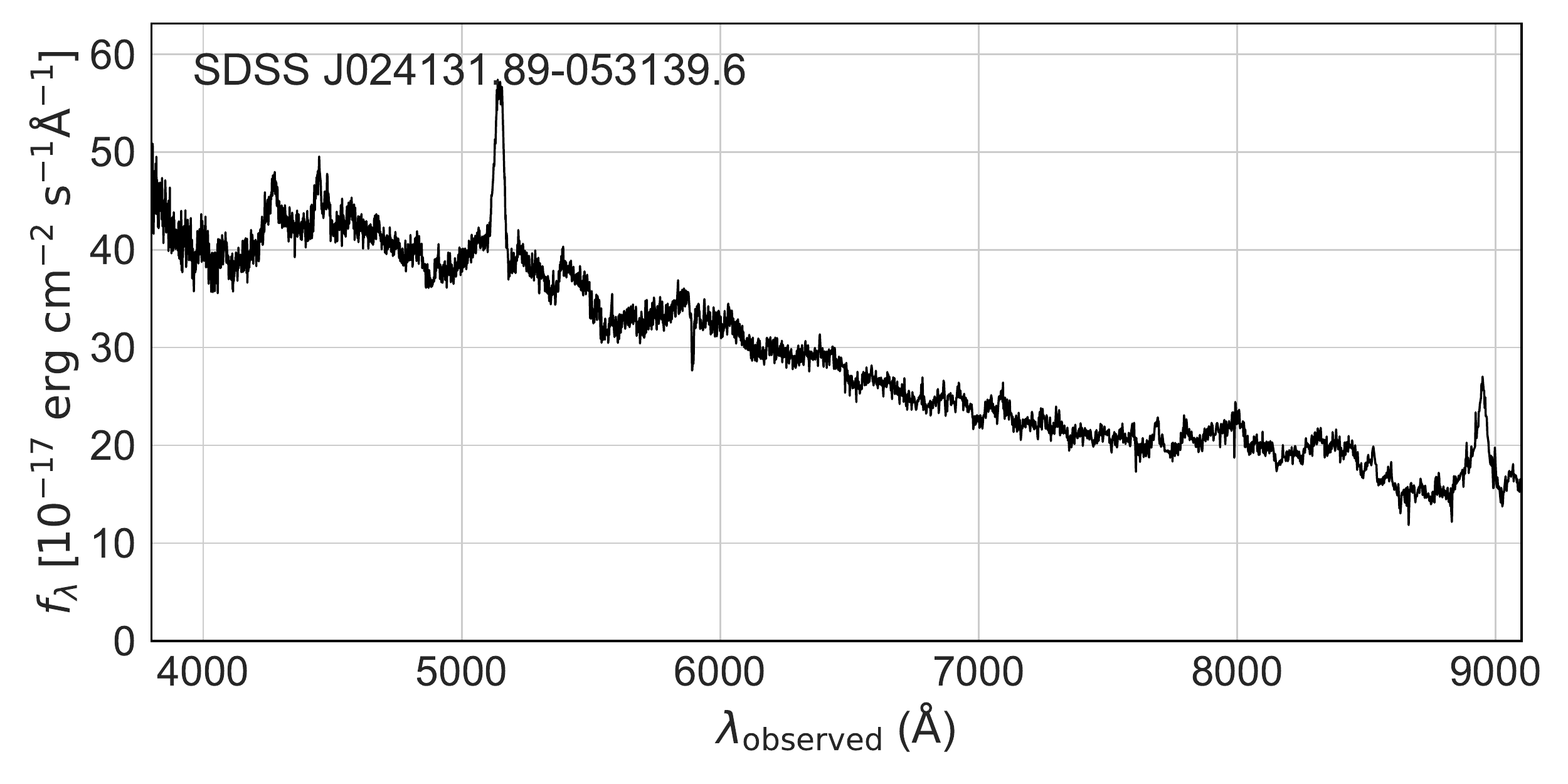}
\end{minipage} \\
\begin{minipage}[t]{0.45\linewidth}
  \includegraphics[width=\linewidth]{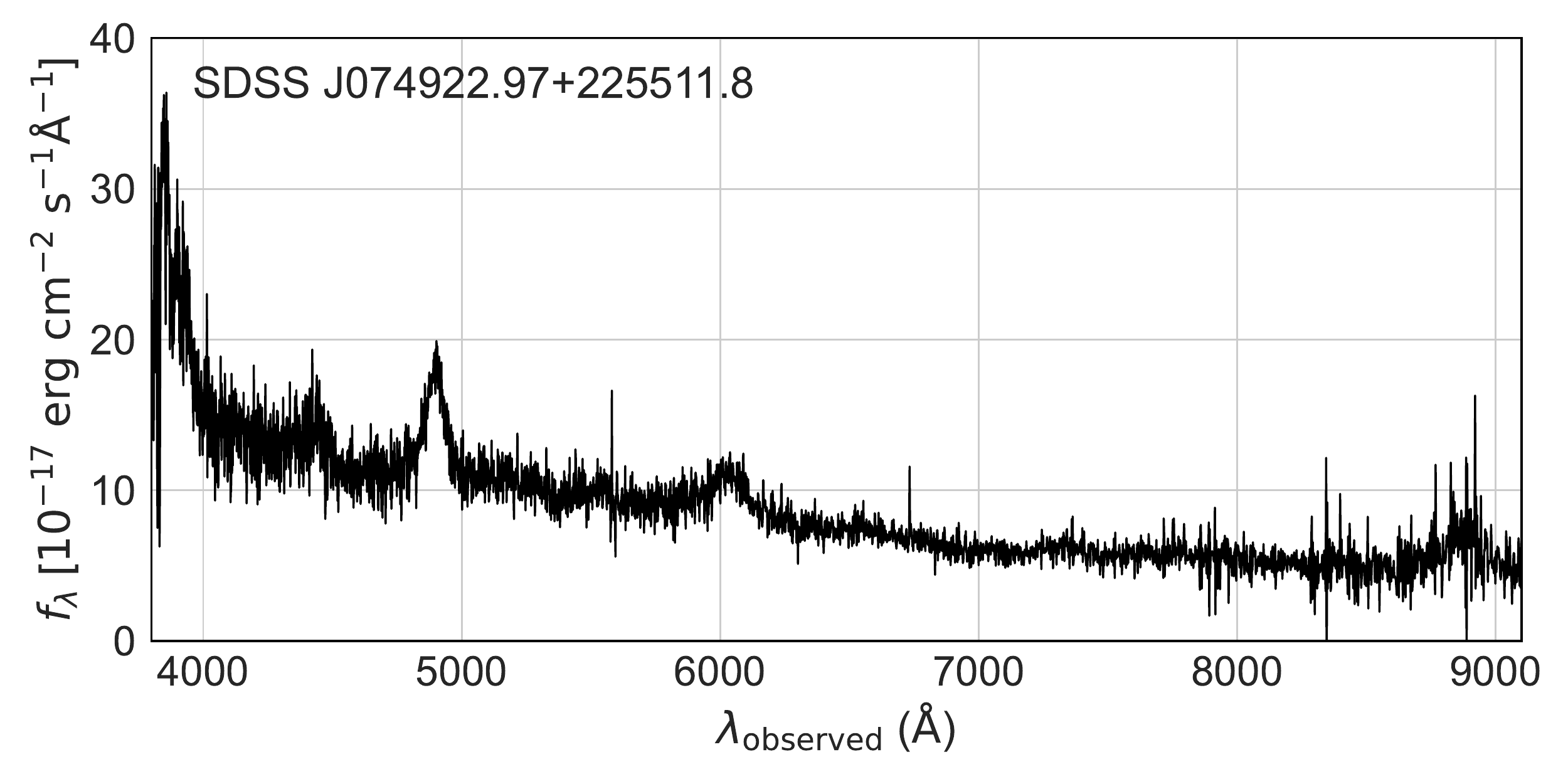}
\end{minipage} &
\begin{minipage}[t]{0.45\linewidth}
  \includegraphics[width=\linewidth]{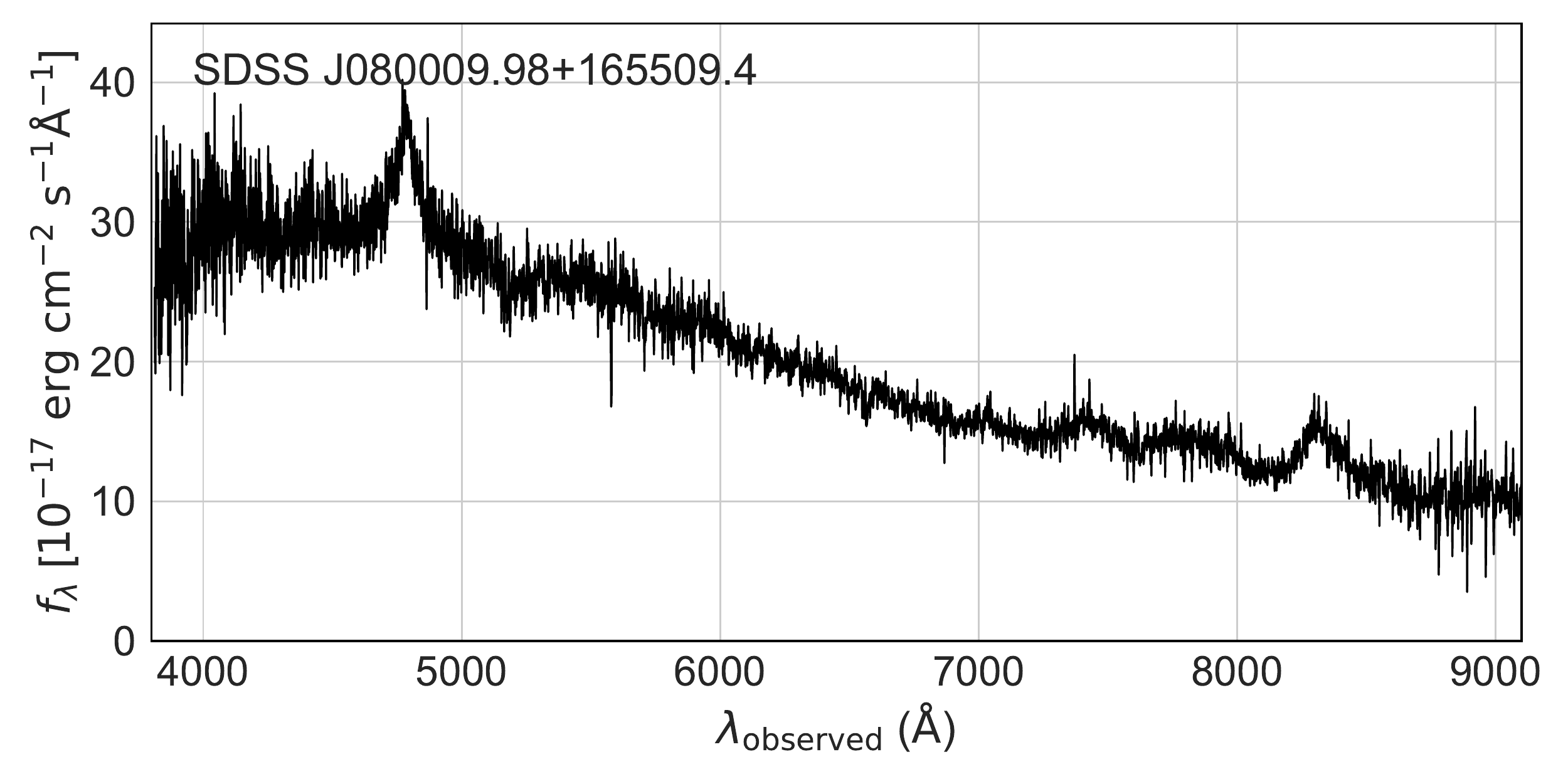}
\end{minipage} \\
\begin{minipage}[t]{0.45\linewidth}
  \includegraphics[width=\linewidth]{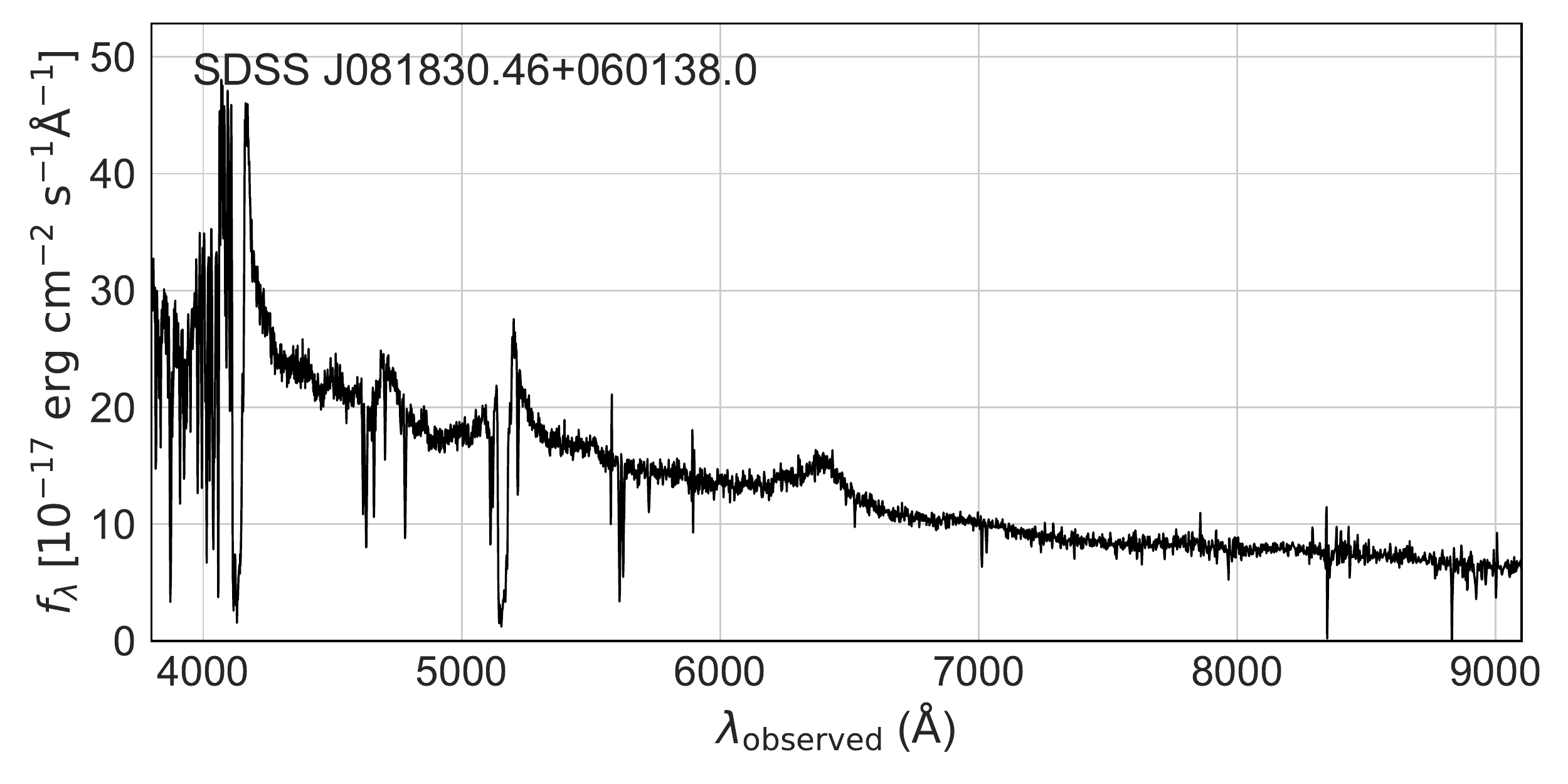}
\end{minipage} &
\begin{minipage}[t]{0.45\linewidth}
  \includegraphics[width=\linewidth]{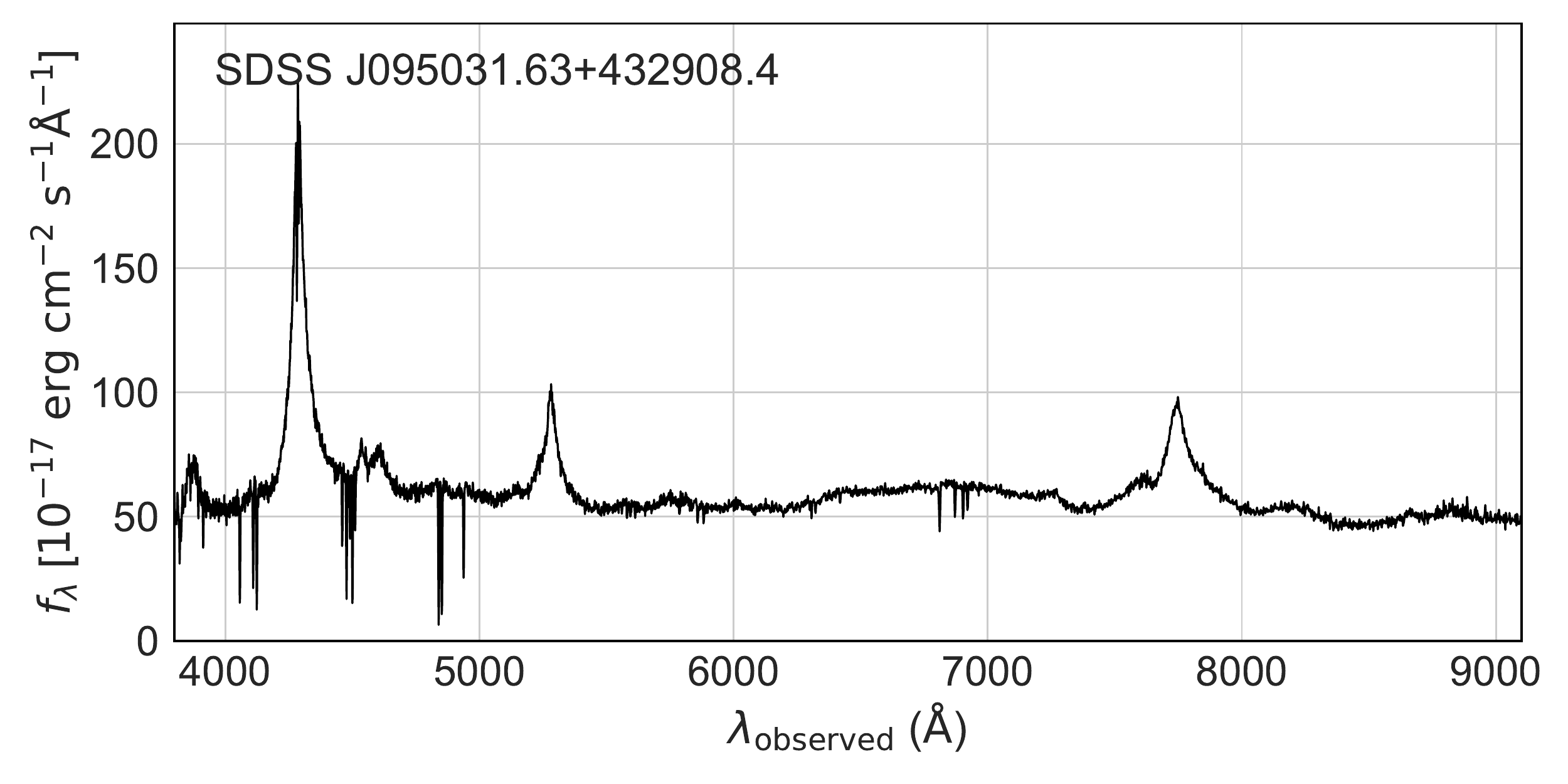}
\end{minipage} \\
\begin{minipage}[t]{0.45\linewidth}
  \includegraphics[width=\linewidth]{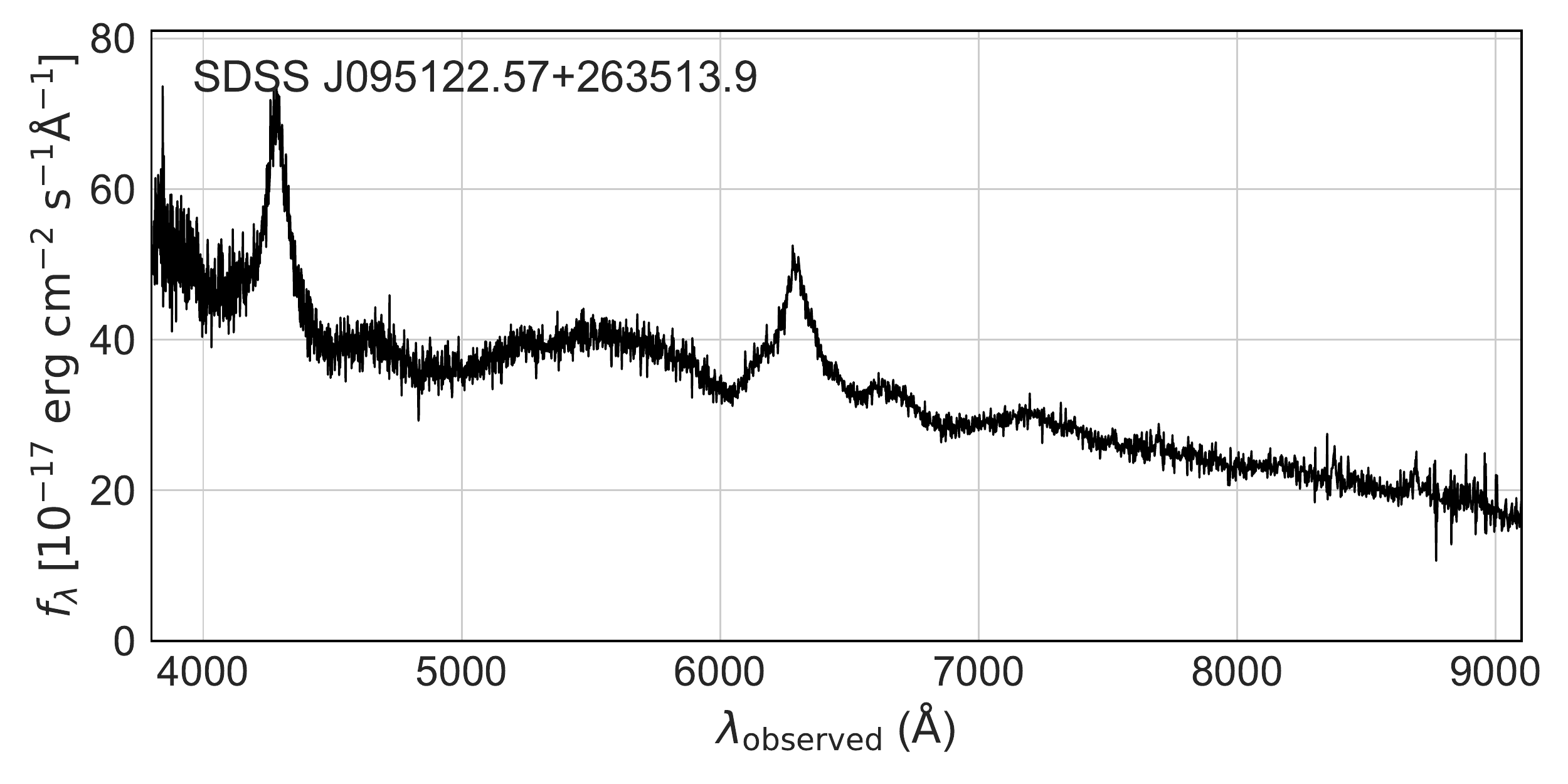}
\end{minipage} &
\begin{minipage}[t]{0.45\linewidth}
  \includegraphics[width=\linewidth]{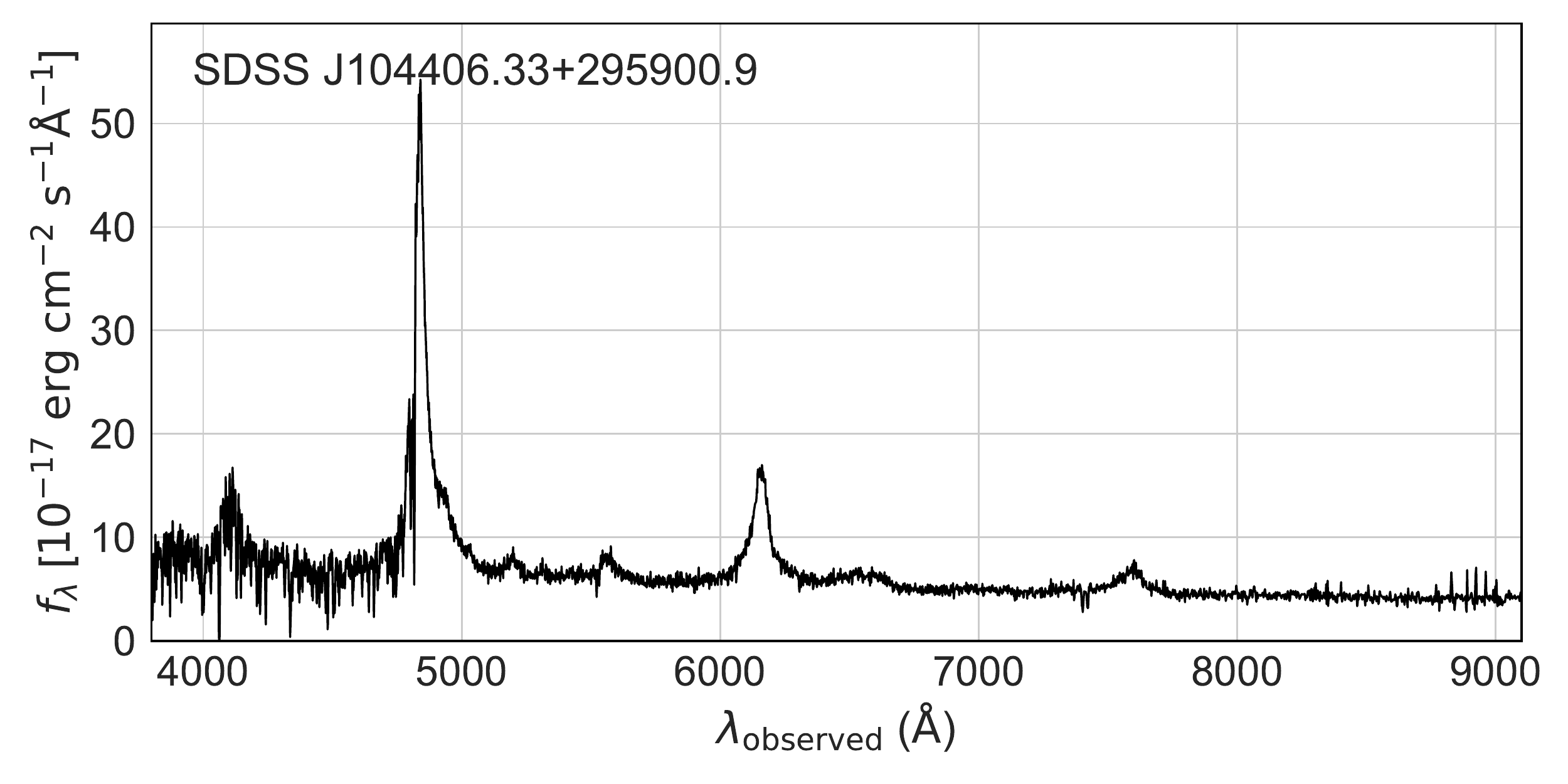}
\end{minipage} \\
\begin{minipage}[t]{0.45\linewidth}
  \includegraphics[width=\linewidth]{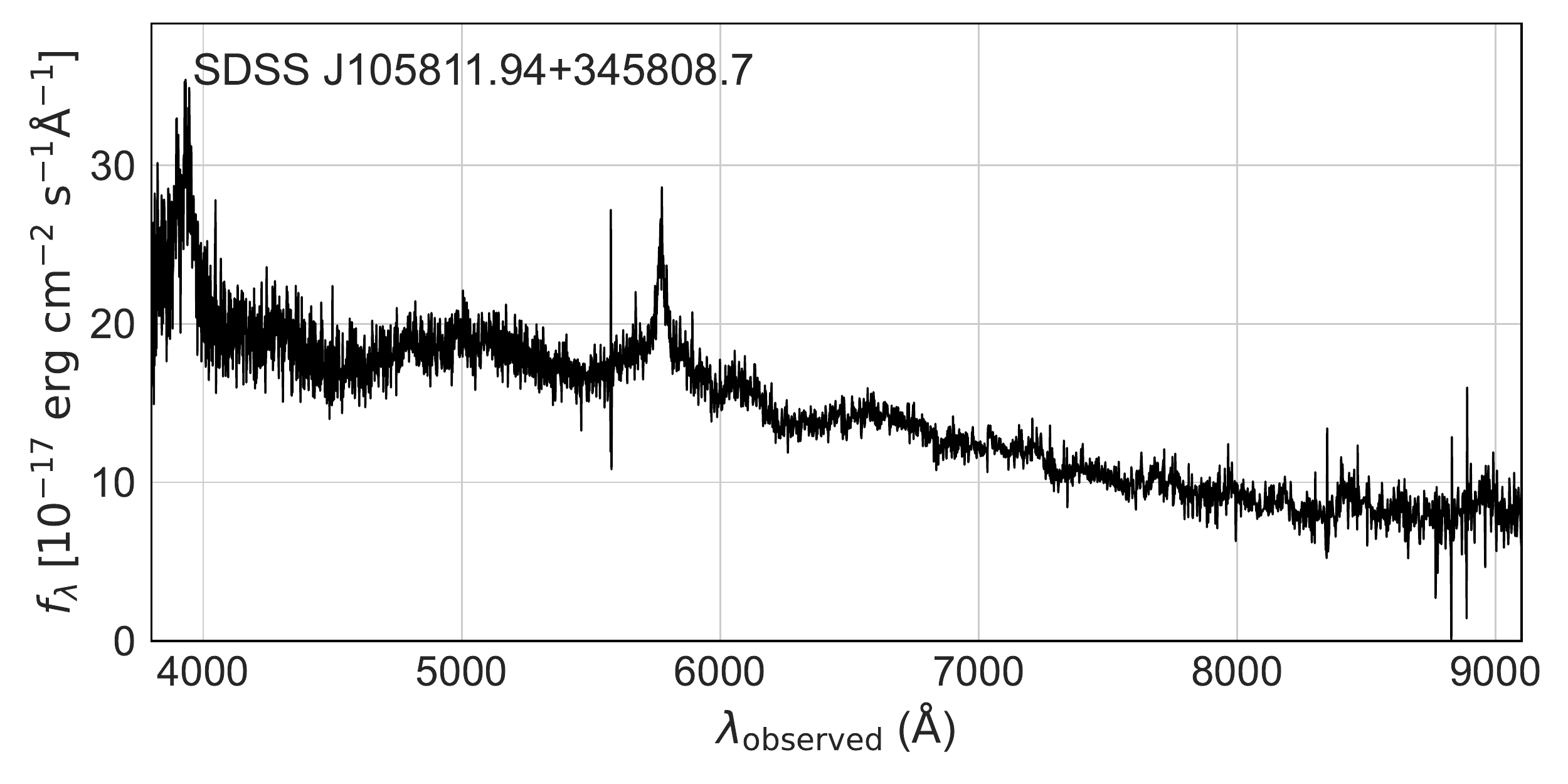}
\end{minipage} &
\begin{minipage}[t]{0.45\linewidth}
  \includegraphics[width=\linewidth]{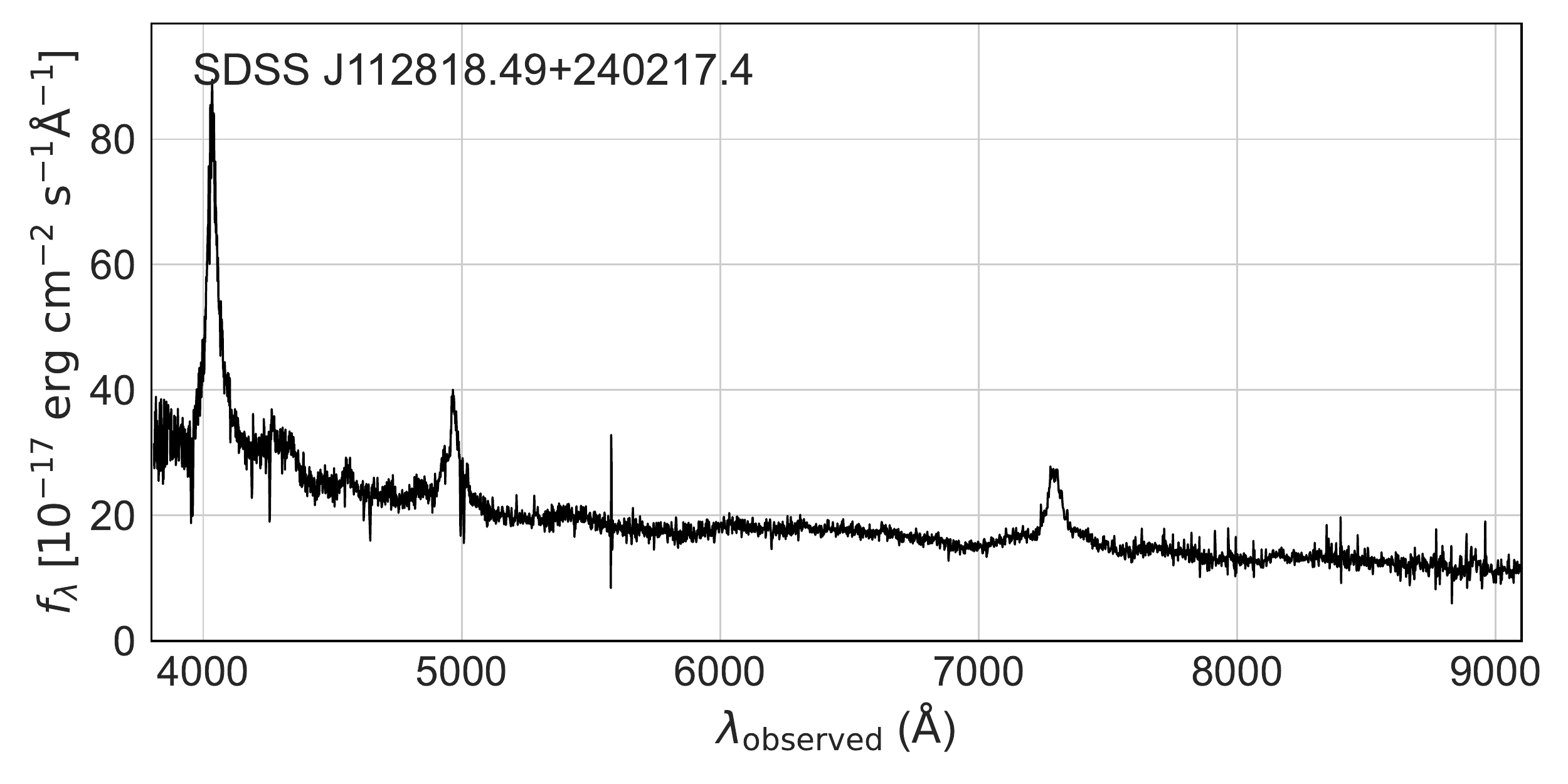}
\end{minipage} \\
\begin{minipage}[t]{0.45\linewidth}
  \includegraphics[width=\linewidth]{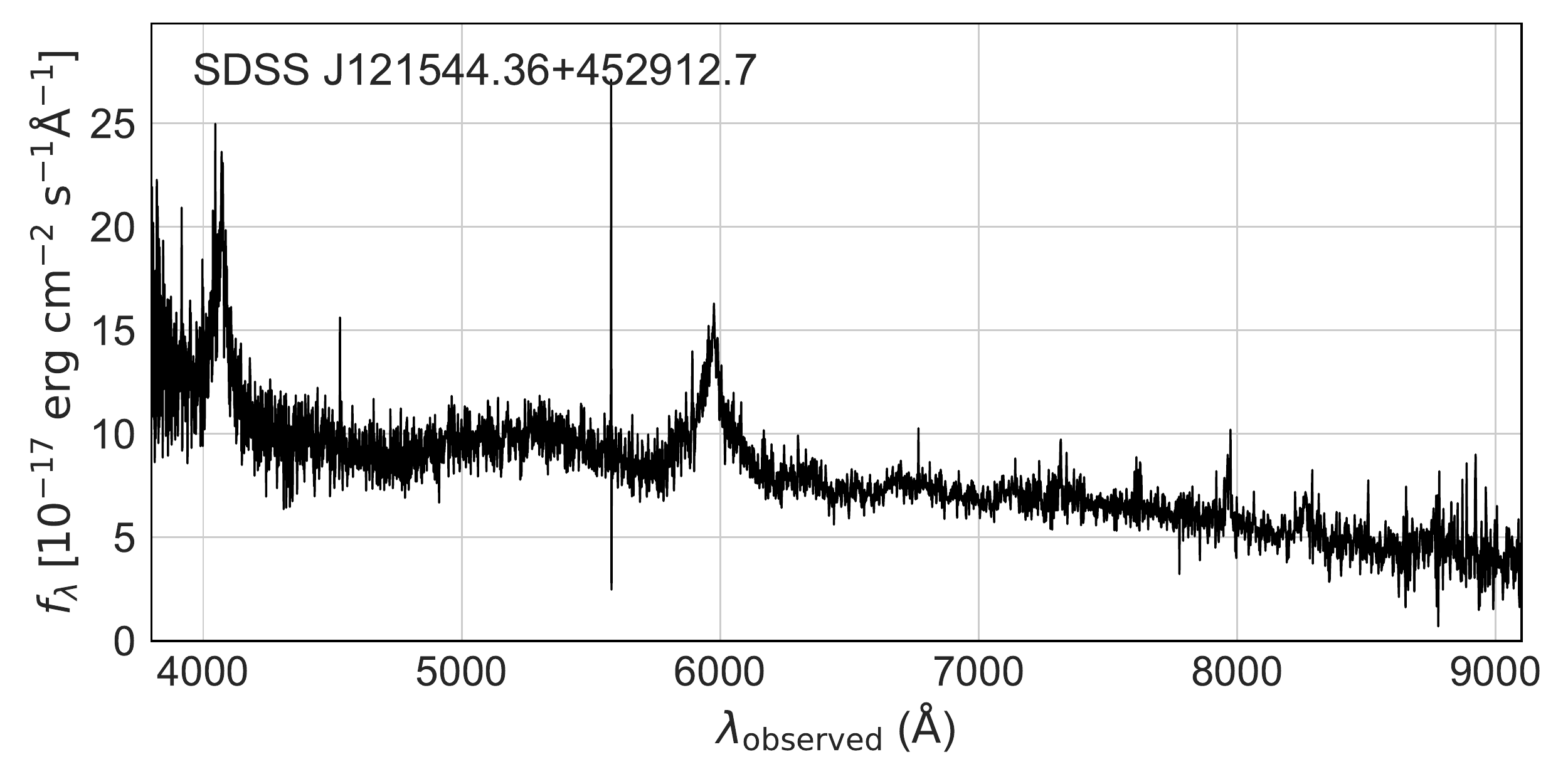}
\end{minipage} &
\begin{minipage}[t]{0.45\linewidth}
  \includegraphics[width=\linewidth]{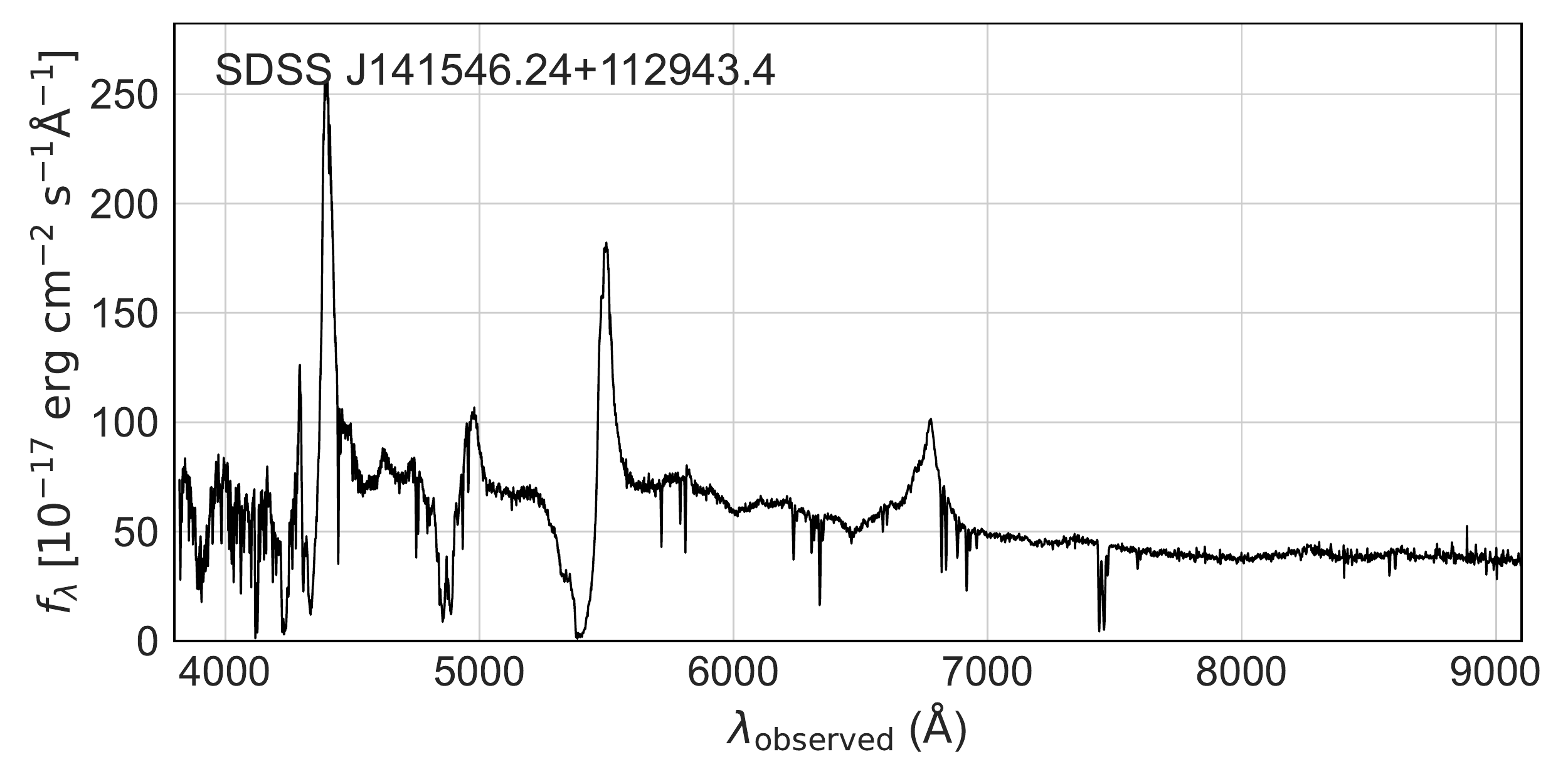}
\end{minipage} \\
\begin{minipage}[t]{0.45\linewidth}
  \includegraphics[width=\linewidth]{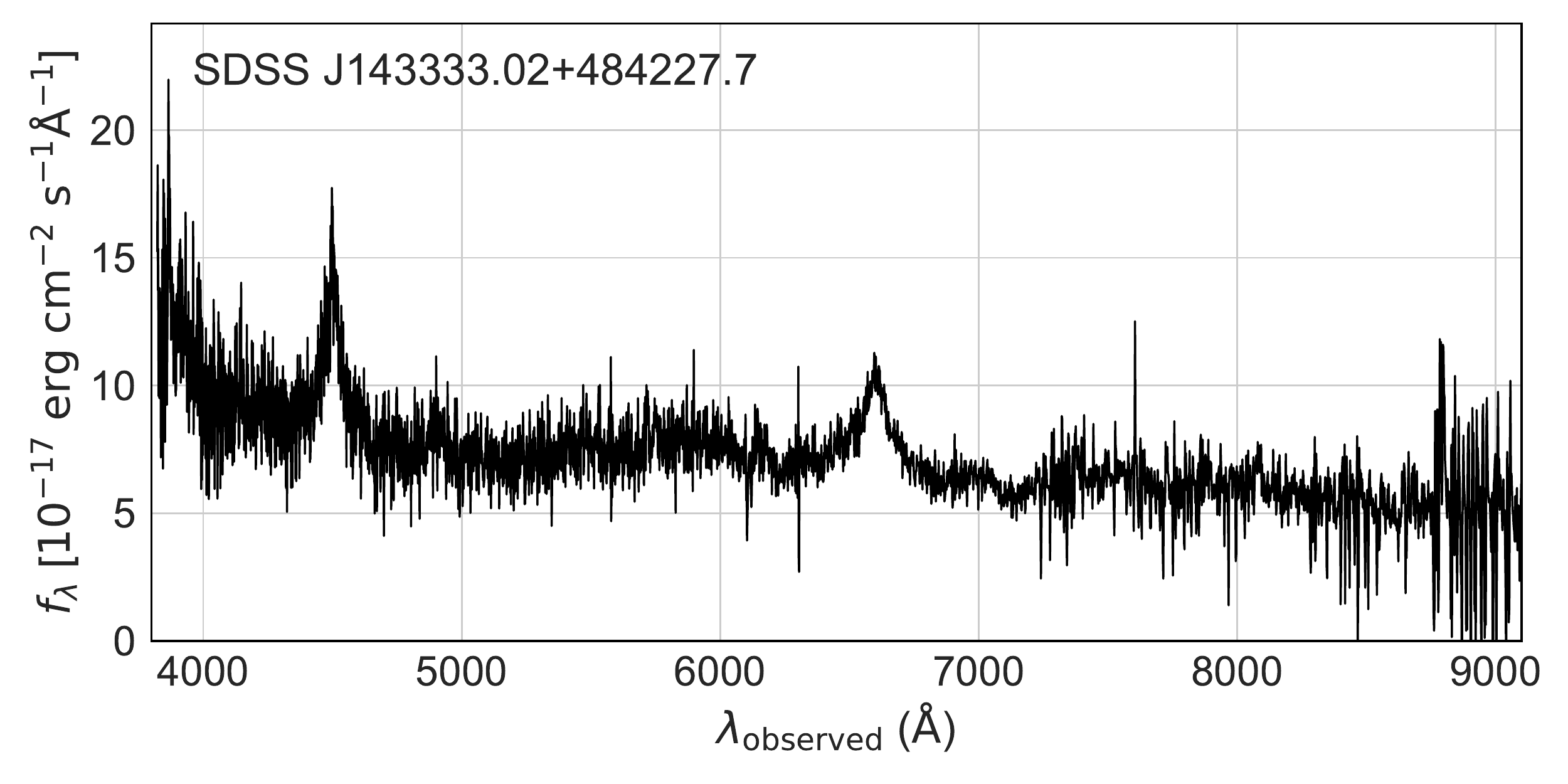}
\end{minipage} &
\begin{minipage}[t]{0.45\linewidth}
  \includegraphics[width=\linewidth]{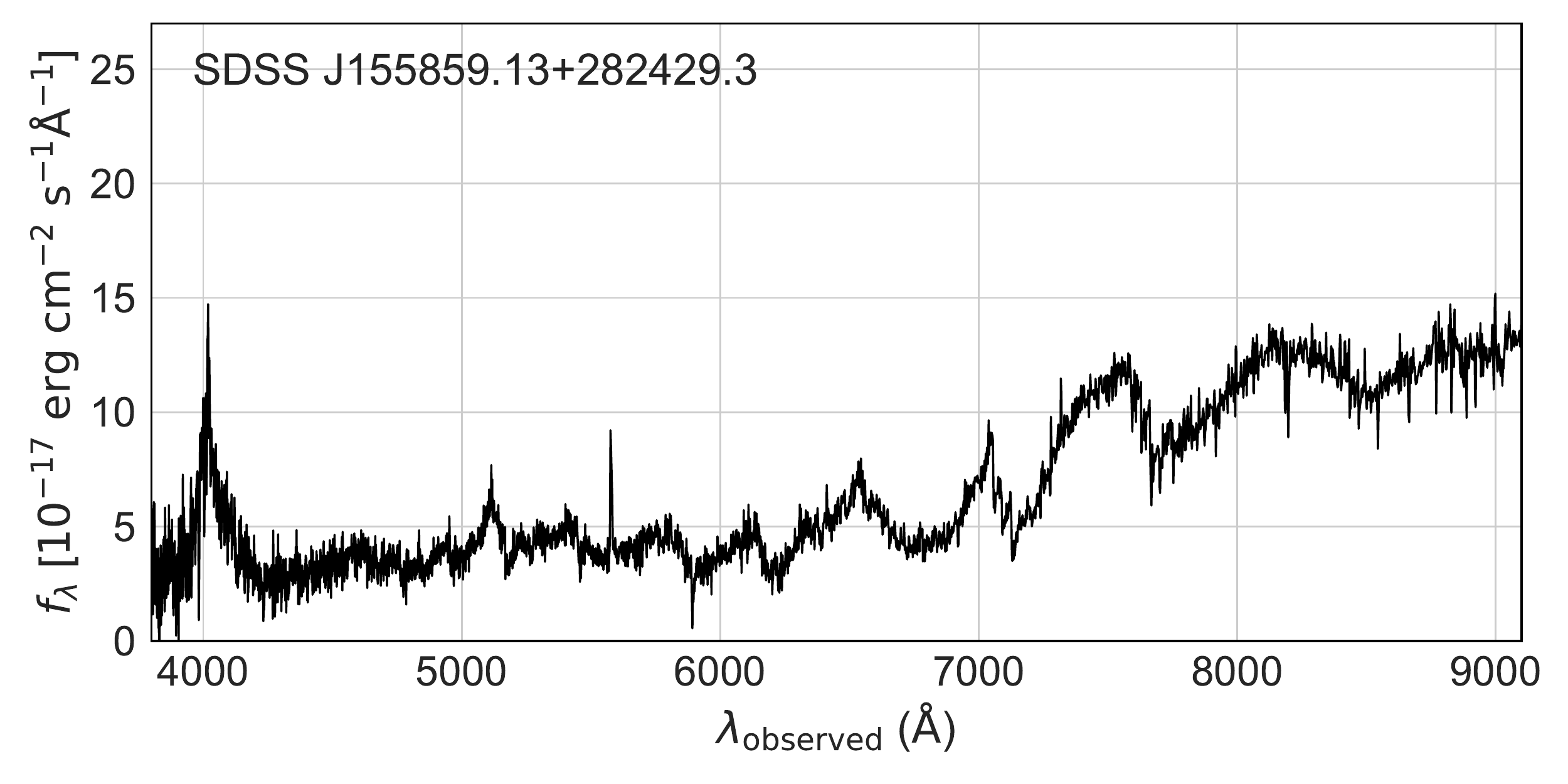}
\end{minipage} \\
\begin{minipage}[t]{0.45\linewidth}
  \includegraphics[width=\linewidth]{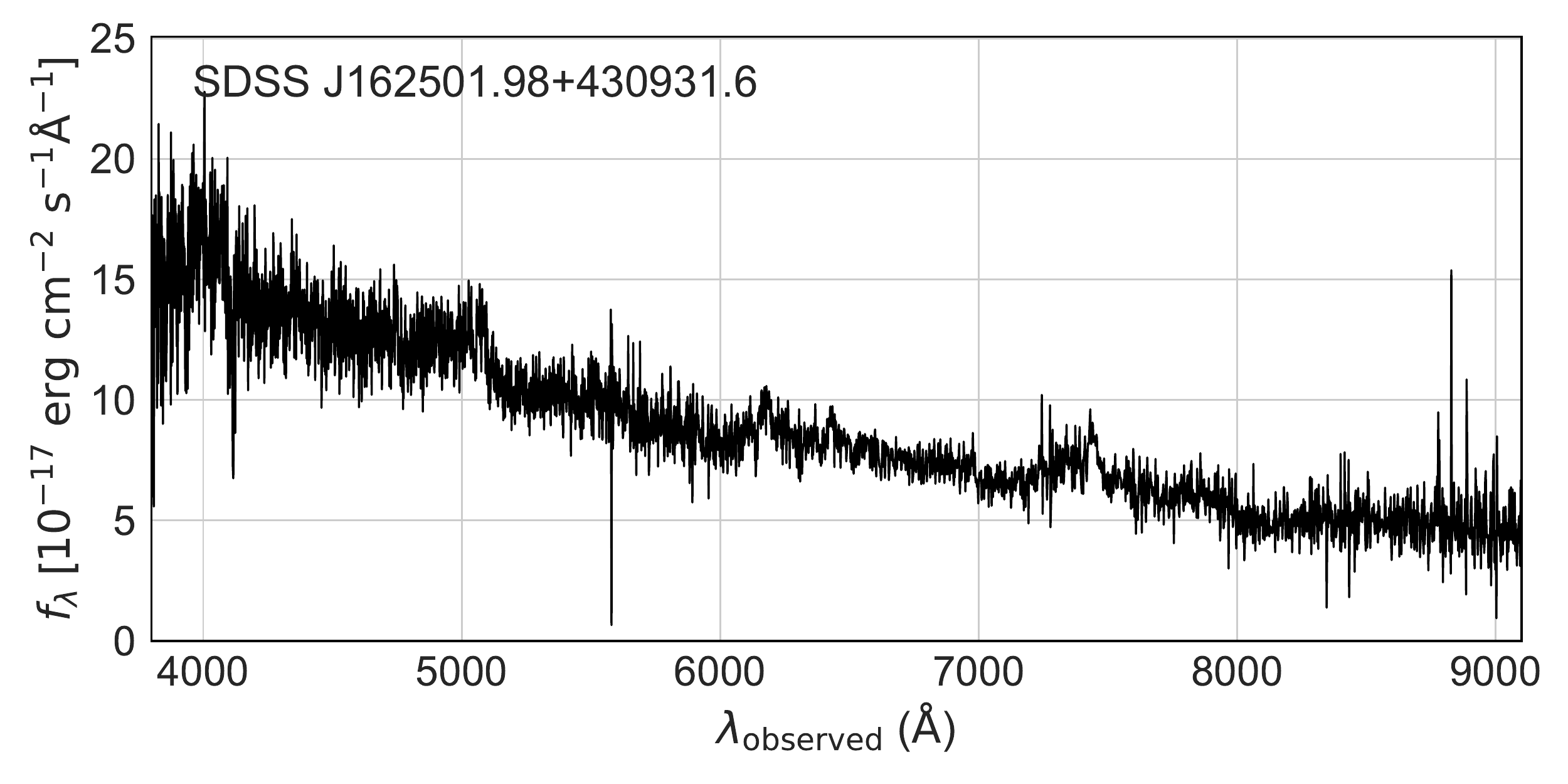}
\end{minipage} &
\begin{minipage}[t]{0.45\linewidth}
  \includegraphics[width=\linewidth]{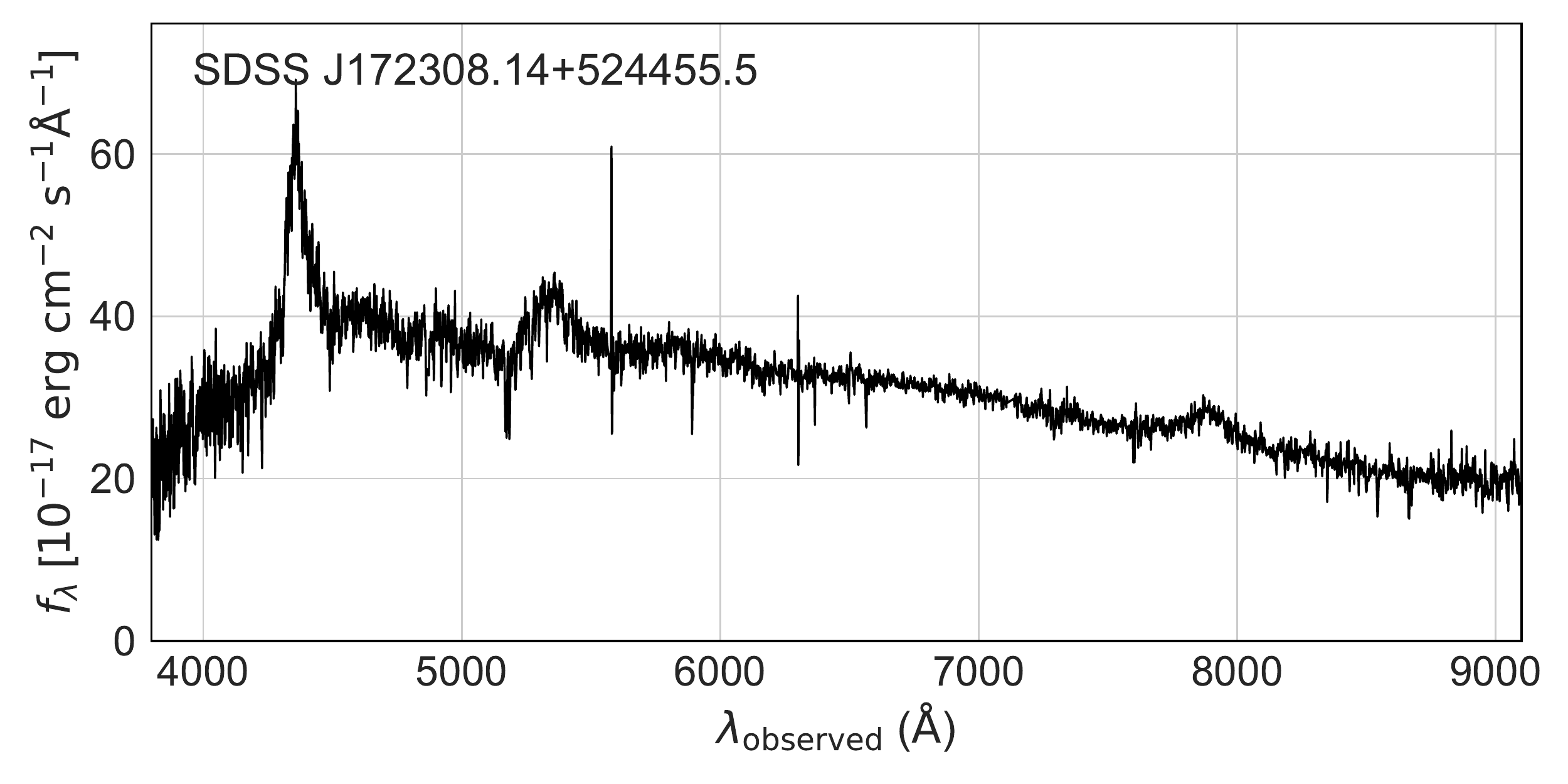}
\end{minipage} \\
\begin{minipage}[t]{0.45\linewidth}
  \includegraphics[width=\linewidth]{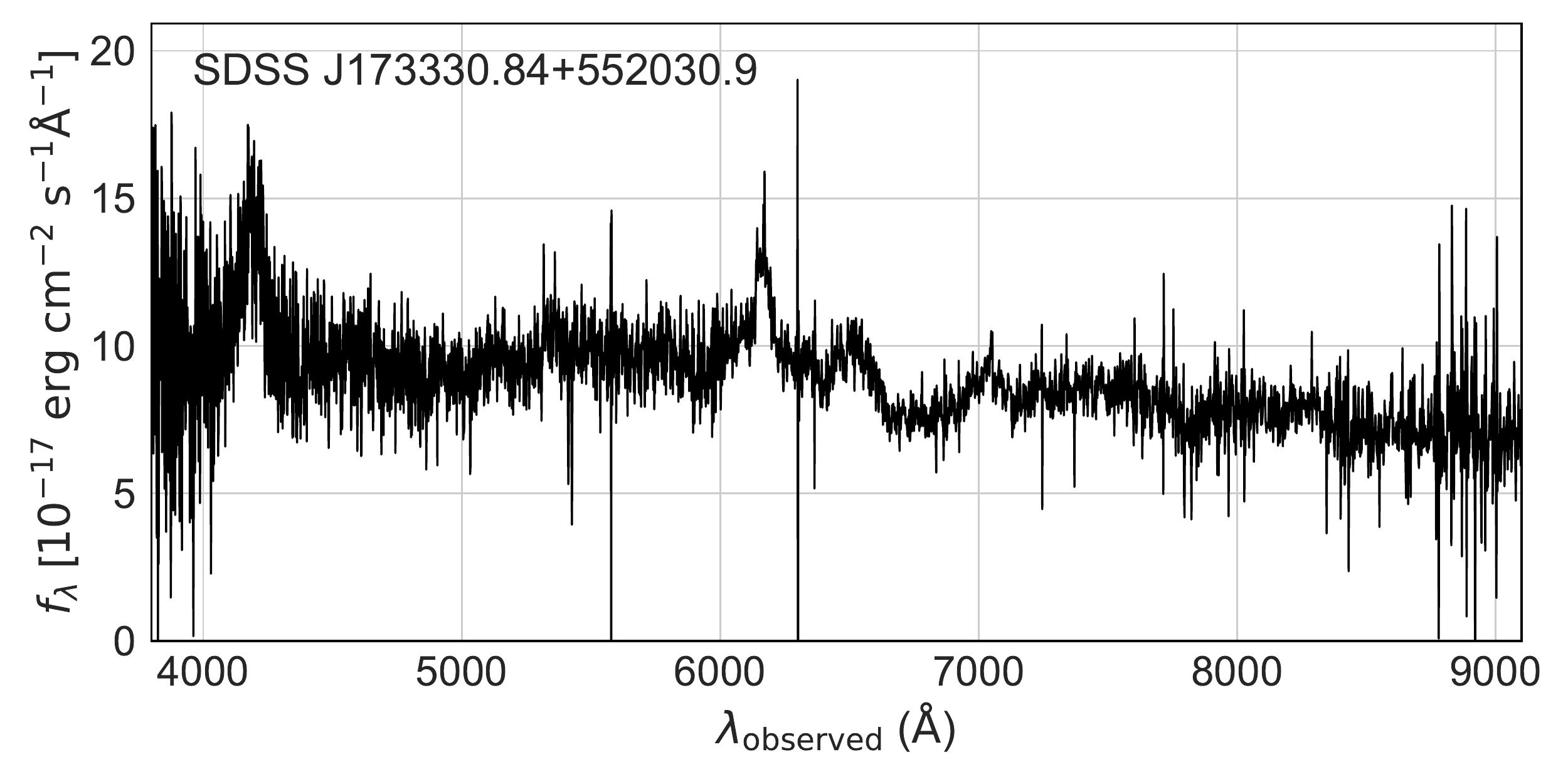}
\end{minipage} &
\begin{minipage}[t]{0.45\linewidth}
  \includegraphics[width=\linewidth]{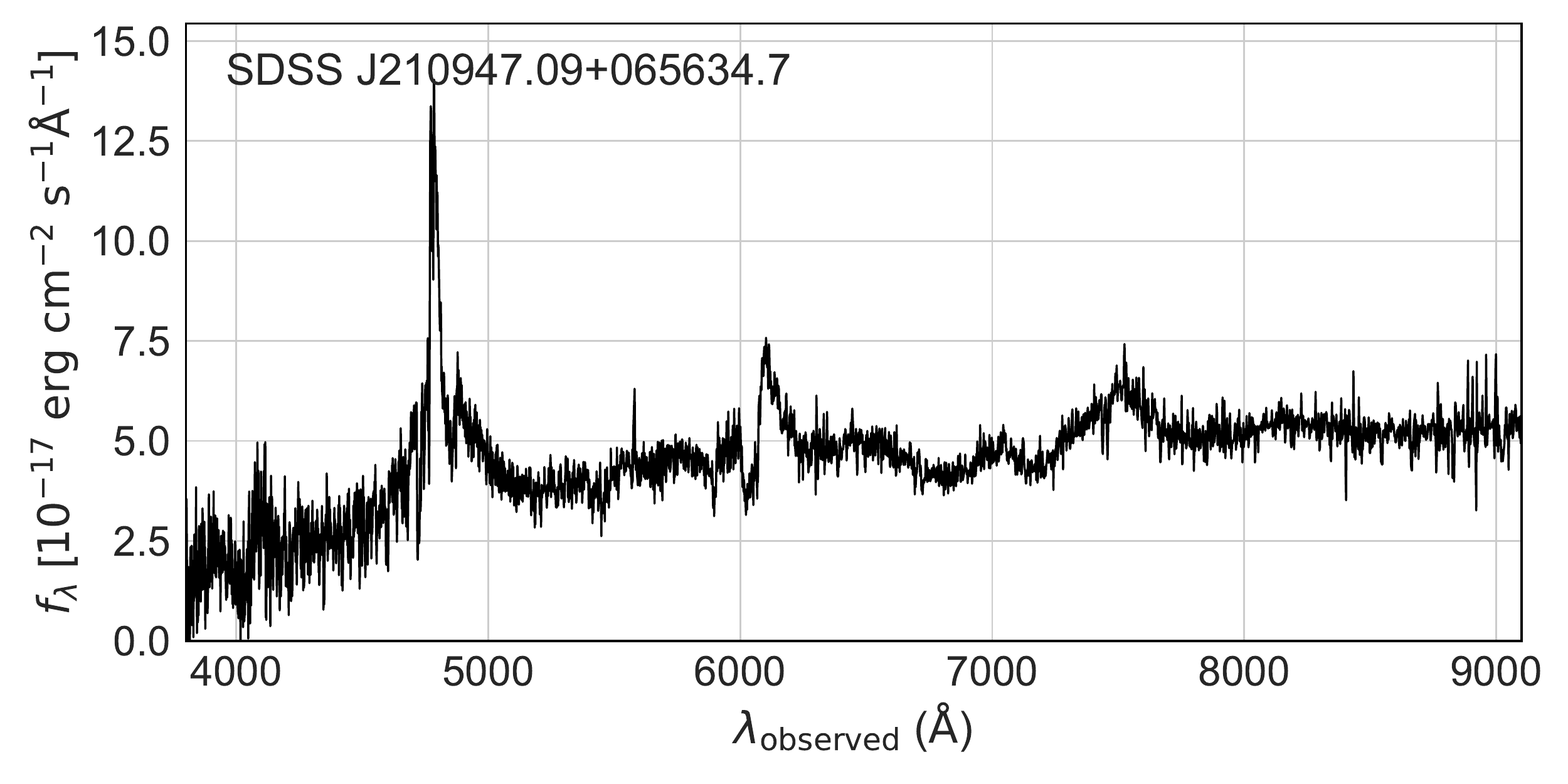}
\end{minipage} \\
  \end{longtable}
  \captionof{figure}{Observed SDSS spectra for the entire sample of 18 targets. These panels display the observed SDSS spectrum ($\sim$4000$\AA$ to $\sim$9000$\AA$) available for each system in the sample; these spectra are not redshift corrected. The designation for each system is listed in the top left corner of each panel. As noted in Section \ref{sec:star-quasars}, four of these systems likely comprise a foreground star and background quasar caught in projection (J0241-0531, J1558+2824, J1723+5244, J2109+0656).}
  \addtocounter{table}{-1}%
 \label{fig:restofspec}
\end{center}

\clearpage
\section{Appendix C: Individual Target Optical Imaging}

\setcounter{figure}{11}

\begin{center}
  \begin{longtable}{ccc}
      \begin{minipage}[t]{0.25\linewidth}
  \includegraphics[width=\linewidth]{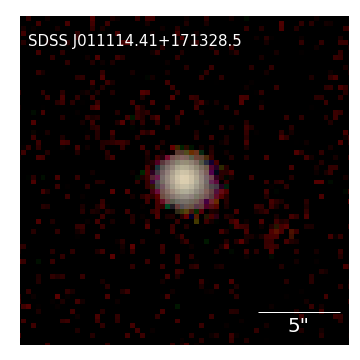}
\end{minipage} &
\begin{minipage}[t]{0.25\linewidth}
  \includegraphics[width=\linewidth]{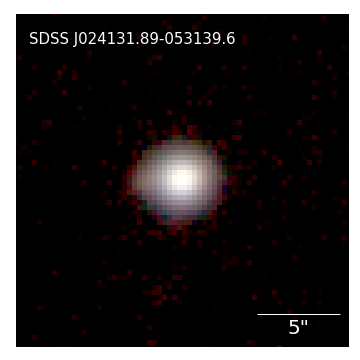}
\end{minipage} &
\begin{minipage}[t]{0.25\linewidth}
  \includegraphics[width=\linewidth]{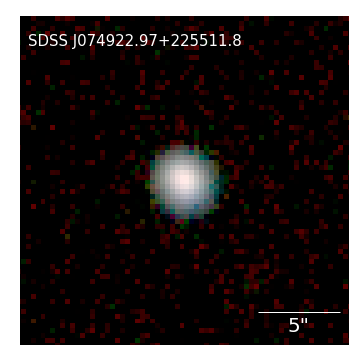}
\end{minipage} \\
\begin{minipage}[t]{0.25\linewidth}
  \includegraphics[width=\linewidth]{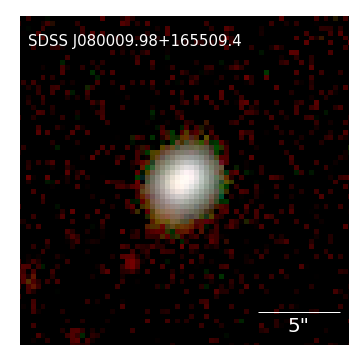}
\end{minipage} &
\begin{minipage}[t]{0.25\linewidth}
  \includegraphics[width=\linewidth]{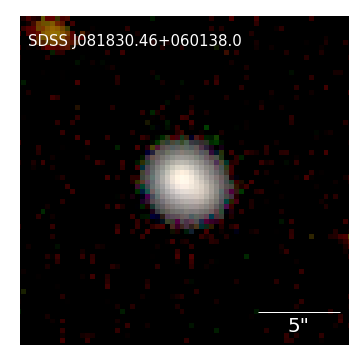}
\end{minipage} &
\begin{minipage}[t]{0.25\linewidth}
  \includegraphics[width=\linewidth]{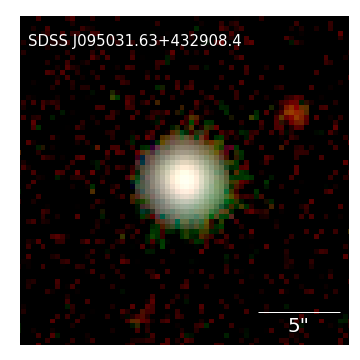}
\end{minipage} \\
\begin{minipage}[t]{0.25\linewidth}
  \includegraphics[width=\linewidth]{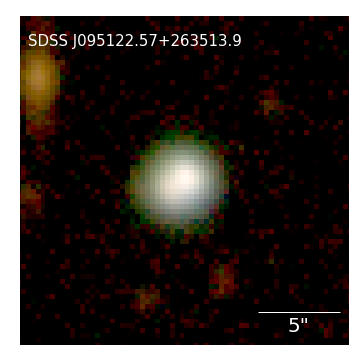}
\end{minipage} &
\begin{minipage}[t]{0.25\linewidth}
  \includegraphics[width=\linewidth]{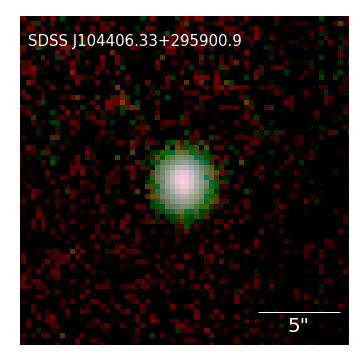}
\end{minipage} &
\begin{minipage}[t]{0.25\linewidth}
  \includegraphics[width=\linewidth]{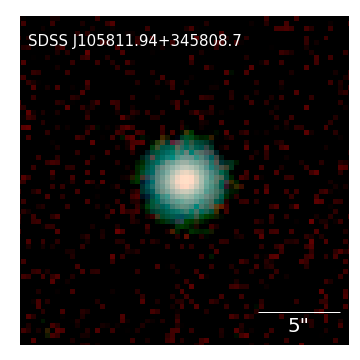}
\end{minipage} \\
\begin{minipage}[t]{0.25\linewidth}
  \includegraphics[width=\linewidth]{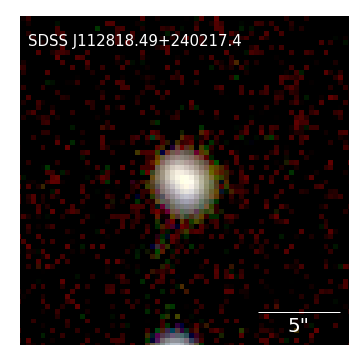}
\end{minipage} &
\begin{minipage}[t]{0.25\linewidth}
  \includegraphics[width=\linewidth]{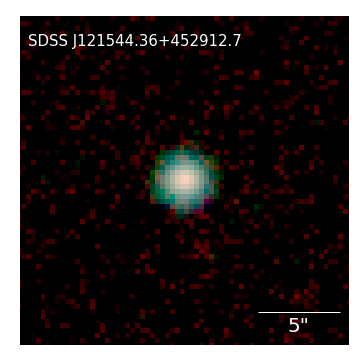}
\end{minipage} &
\begin{minipage}[t]{0.25\linewidth}
  \includegraphics[width=\linewidth]{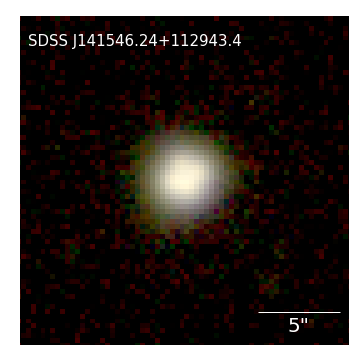}
\end{minipage} \\
\begin{minipage}[t]{0.25\linewidth}
  \includegraphics[width=\linewidth]{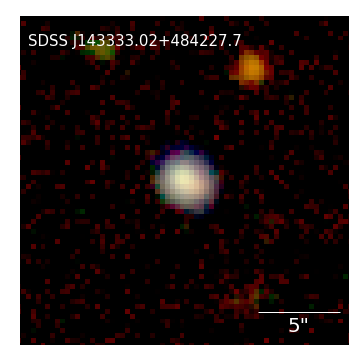}
\end{minipage} &
\begin{minipage}[t]{0.25\linewidth}
  \includegraphics[width=\linewidth]{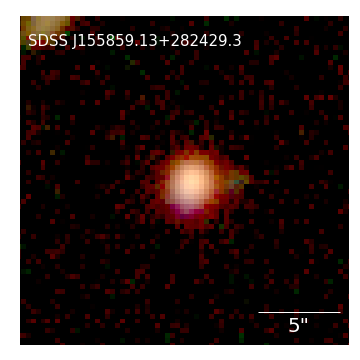}
\end{minipage} &
\begin{minipage}[t]{0.25\linewidth}
  \includegraphics[width=\linewidth]{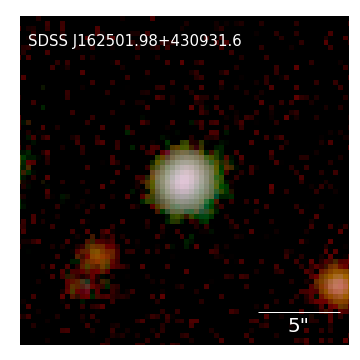}
\end{minipage} \\
\begin{minipage}[t]{0.25\linewidth}
  \includegraphics[width=\linewidth]{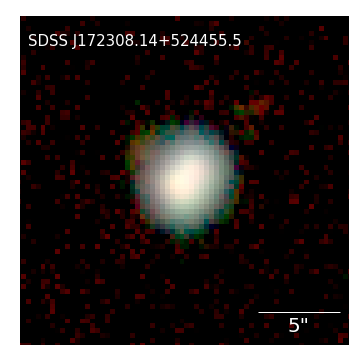}
\end{minipage} &
\begin{minipage}[t]{0.25\linewidth}
  \includegraphics[width=\linewidth]{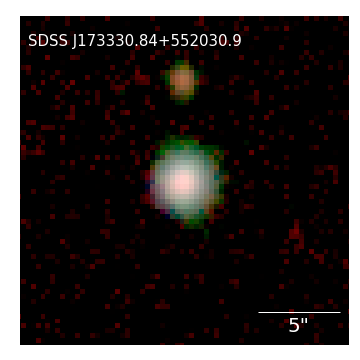}
\end{minipage} &
\begin{minipage}[t]{0.25\linewidth}
  \includegraphics[width=\linewidth]{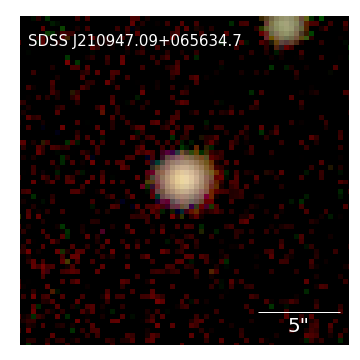}
\end{minipage} \\
  \end{longtable}%
  \captionof{figure}{Optical tricolor $ugz$ images drawn from the DeCaLs Legacy Viewer for the full sample. The SDSS designation for each object is listed in the top left corner of each panel, while the scale bar in the bottom right corner indicates 5''. Notice in several cases that the sources appear elongated rather than point-like, which may potentially point to underlying complex host morphologies and/or a multiplicity of sources, neither of which are not resolvable in these images.}%
  \addtocounter{table}{-1}%
 \label{fig:decals}
\end{center}

\clearpage

\section{Appendix D: Radio Spectral Modeling}
\label{sec:appendixD}
\setcounter{figure}{12}

\begin{center}
  \begin{longtable}{cc}
\begin{minipage}[t]{0.45\linewidth}
  \centering
  \includegraphics[width=0.95\linewidth]{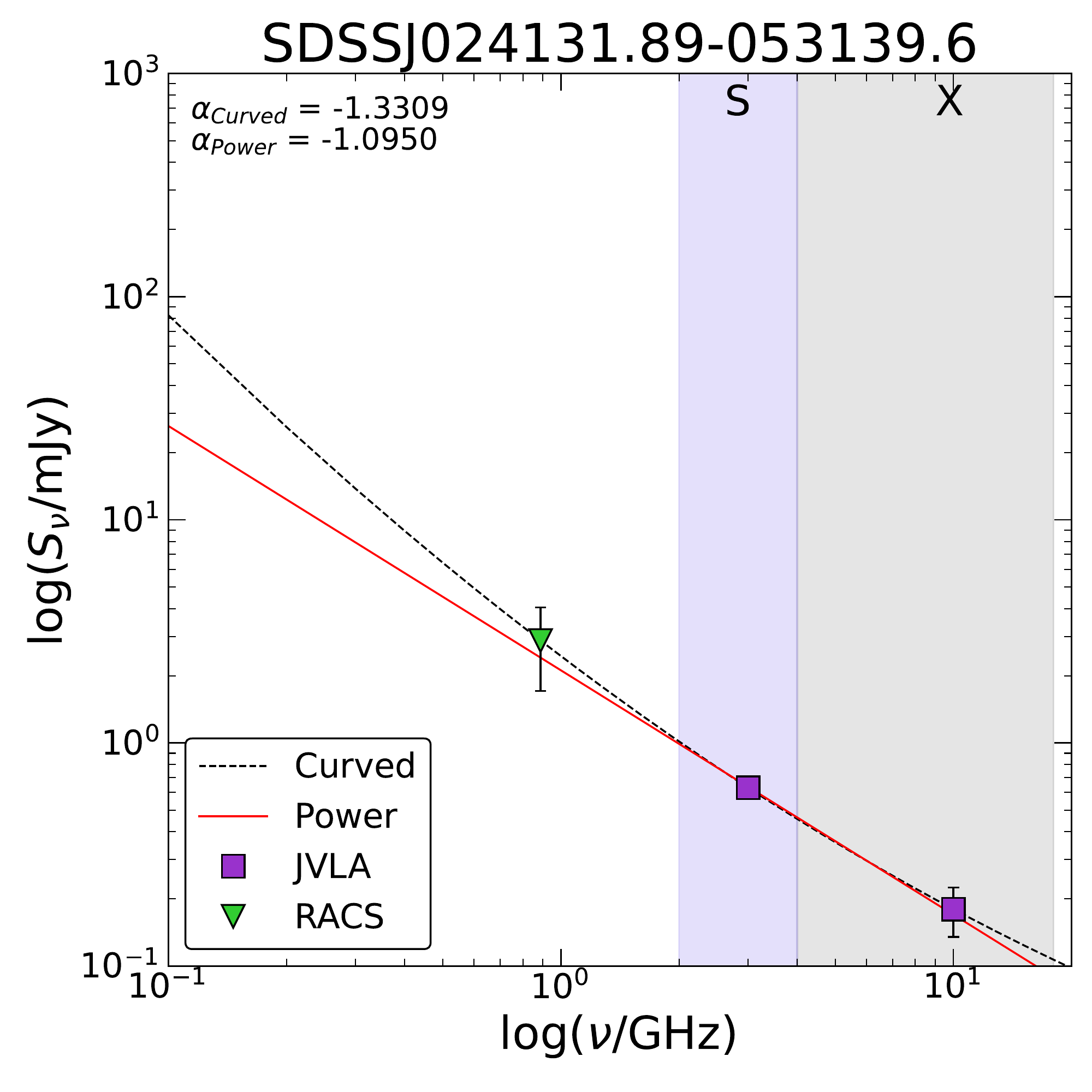}
\end{minipage} &
\begin{minipage}[t]{0.45\linewidth}
  \centering
  \includegraphics[width=0.95\linewidth]{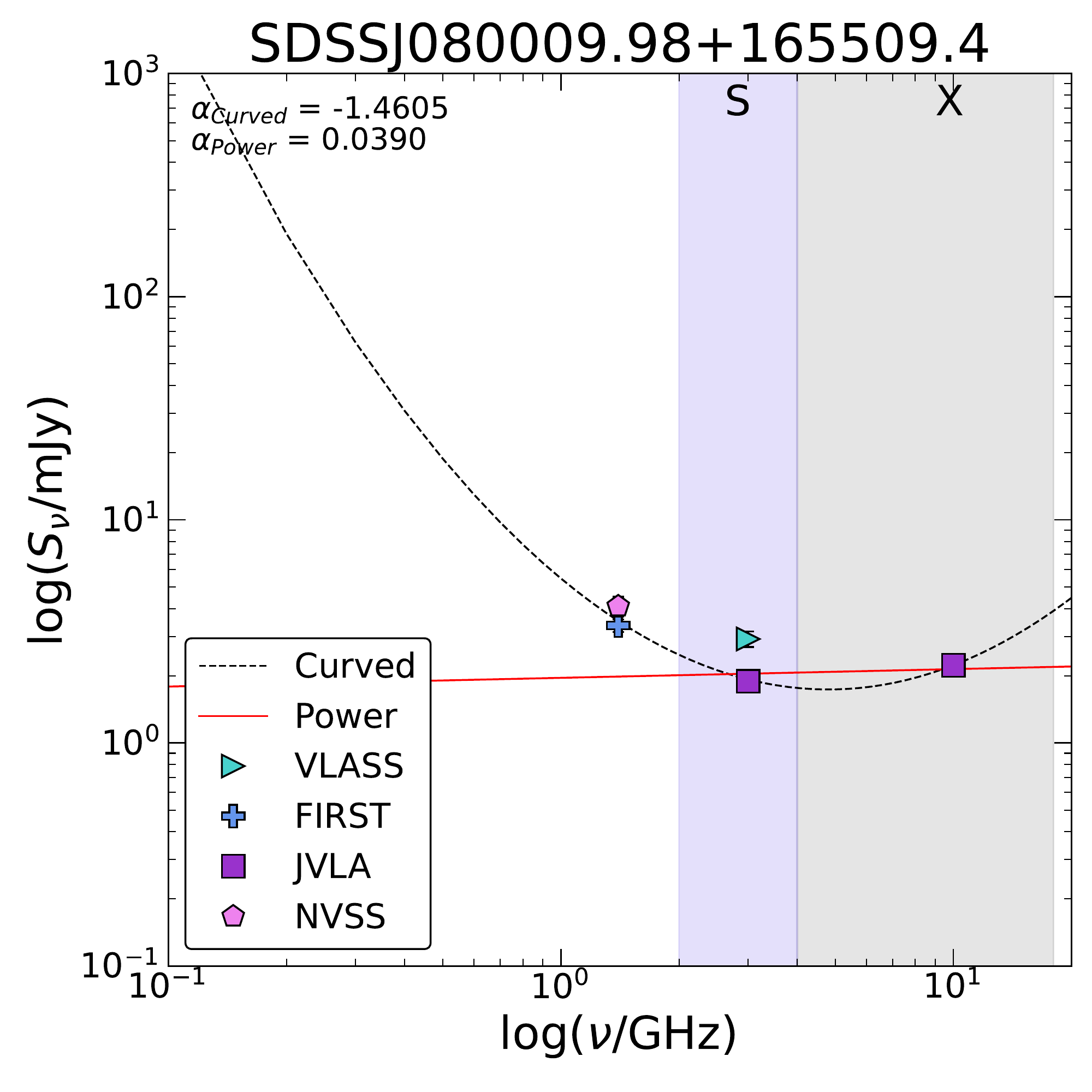}
\end{minipage} \\
\vspace{-0.2cm}
\begin{minipage}[t]{0.45\linewidth}
  \centering
  \includegraphics[width=0.95\linewidth]{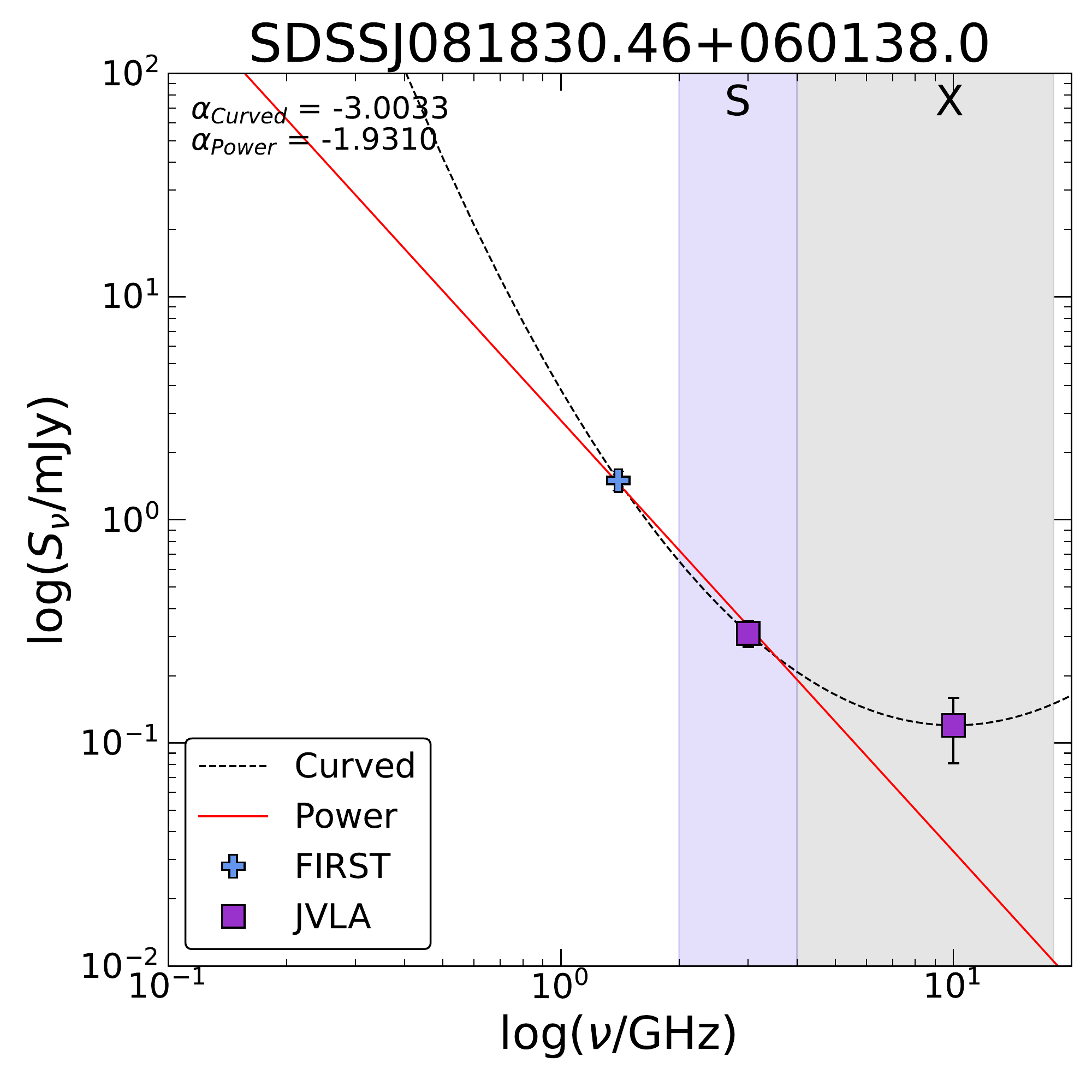}
\end{minipage} &
\begin{minipage}[t]{0.45\linewidth}
  \centering
  \includegraphics[width=0.95\linewidth]{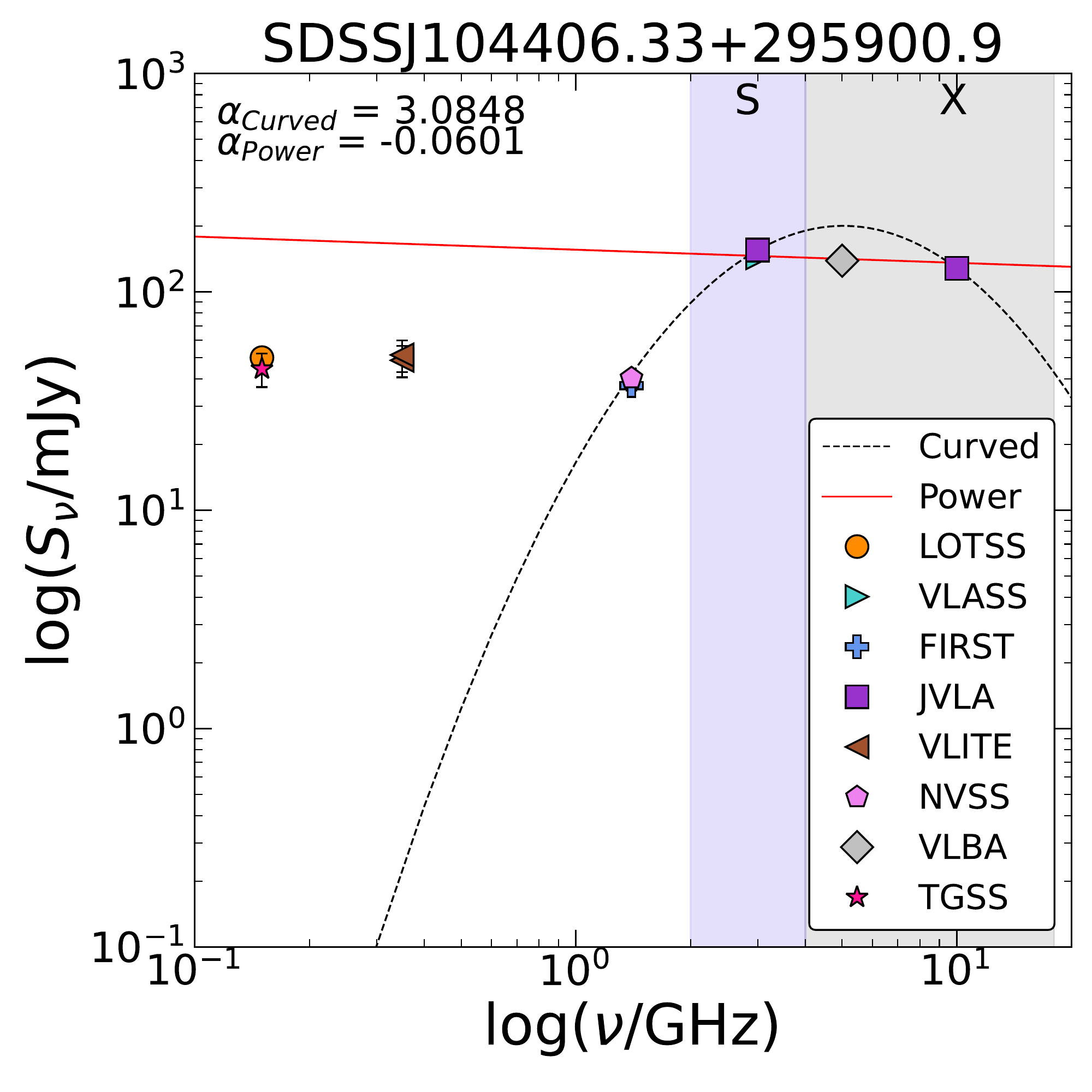}
\end{minipage} \\
\vspace{-0.2cm}
\begin{minipage}[t]{0.45\linewidth}
  \centering
  \includegraphics[width=0.95\linewidth]{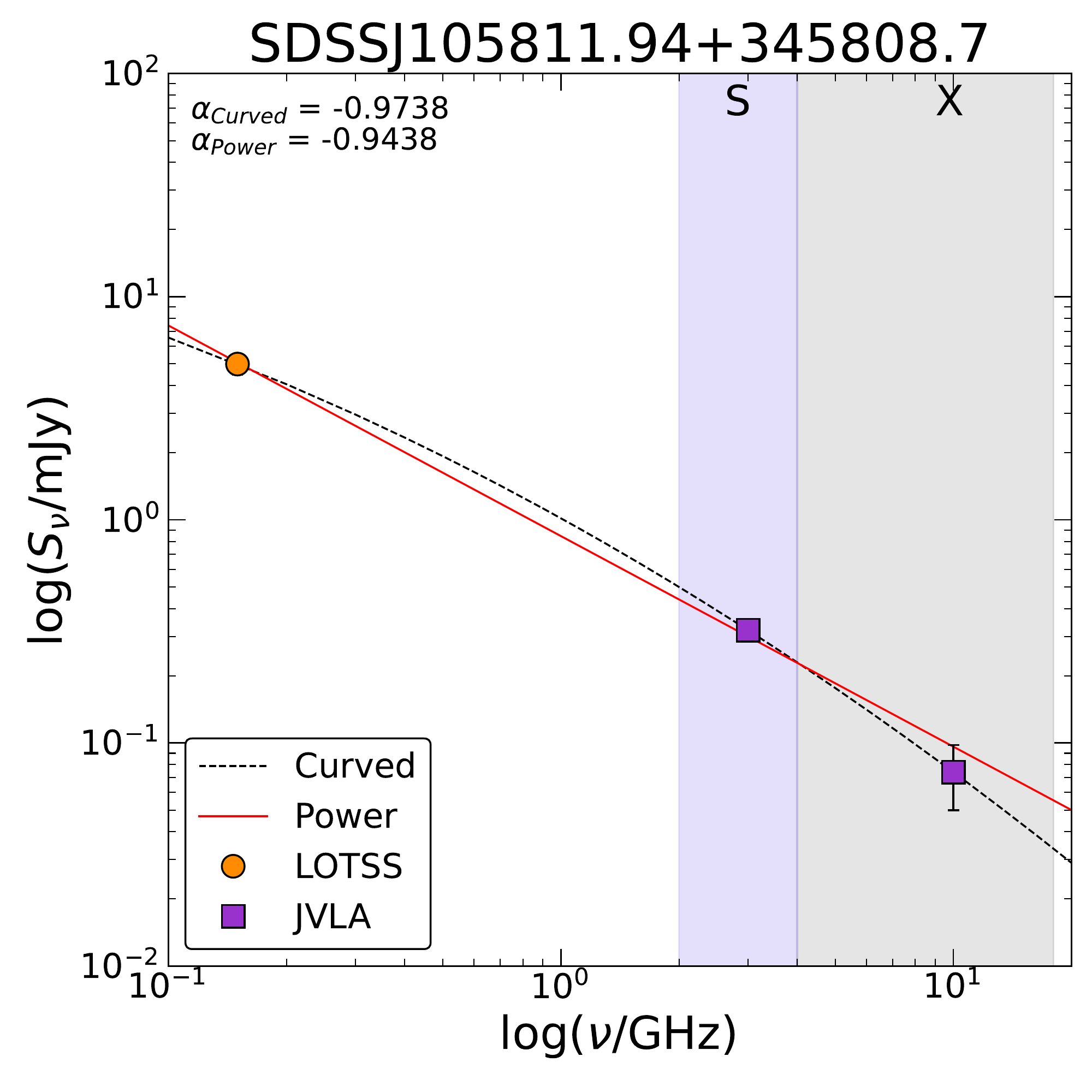}
\end{minipage} &
\begin{minipage}[t]{0.45\linewidth}
  \centering
  \includegraphics[width=0.95\linewidth]{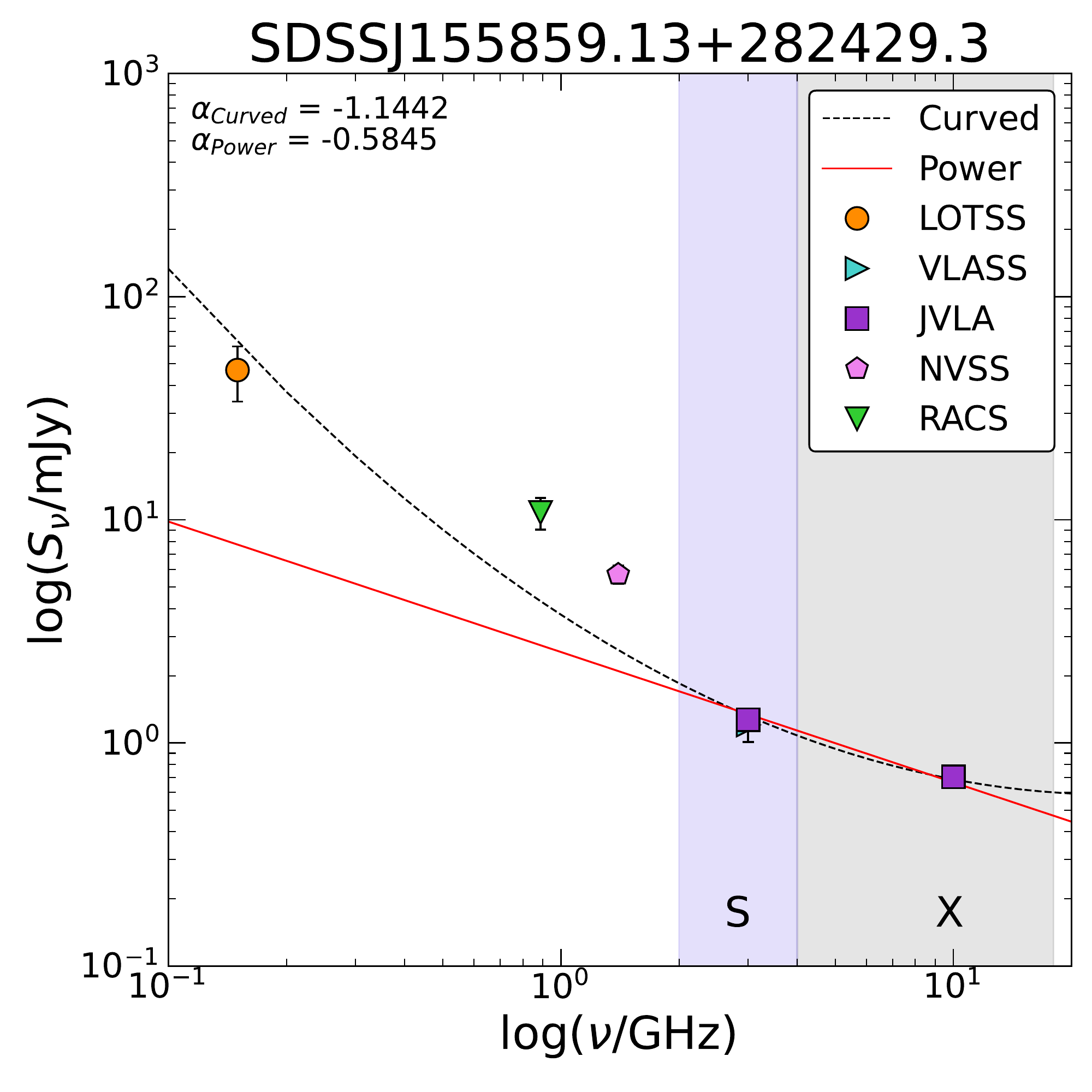}
\end{minipage} \\
\vspace{-0.2cm}
\begin{minipage}[t]{0.45\linewidth}
  \centering
  \includegraphics[width=0.95\linewidth]{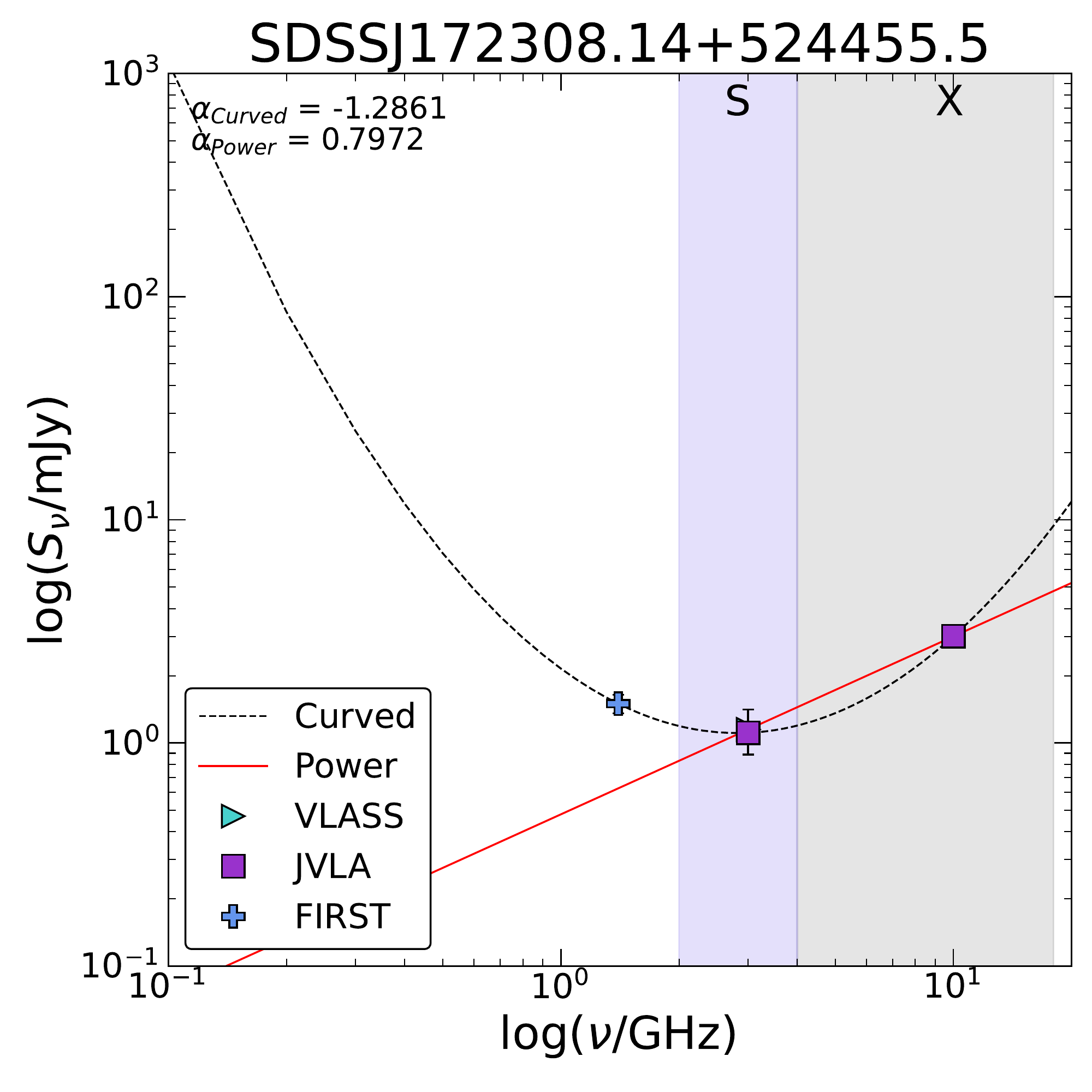}
\end{minipage} &
\begin{minipage}[t]{0.45\linewidth}
  \centering
  \includegraphics[width=0.95\linewidth]{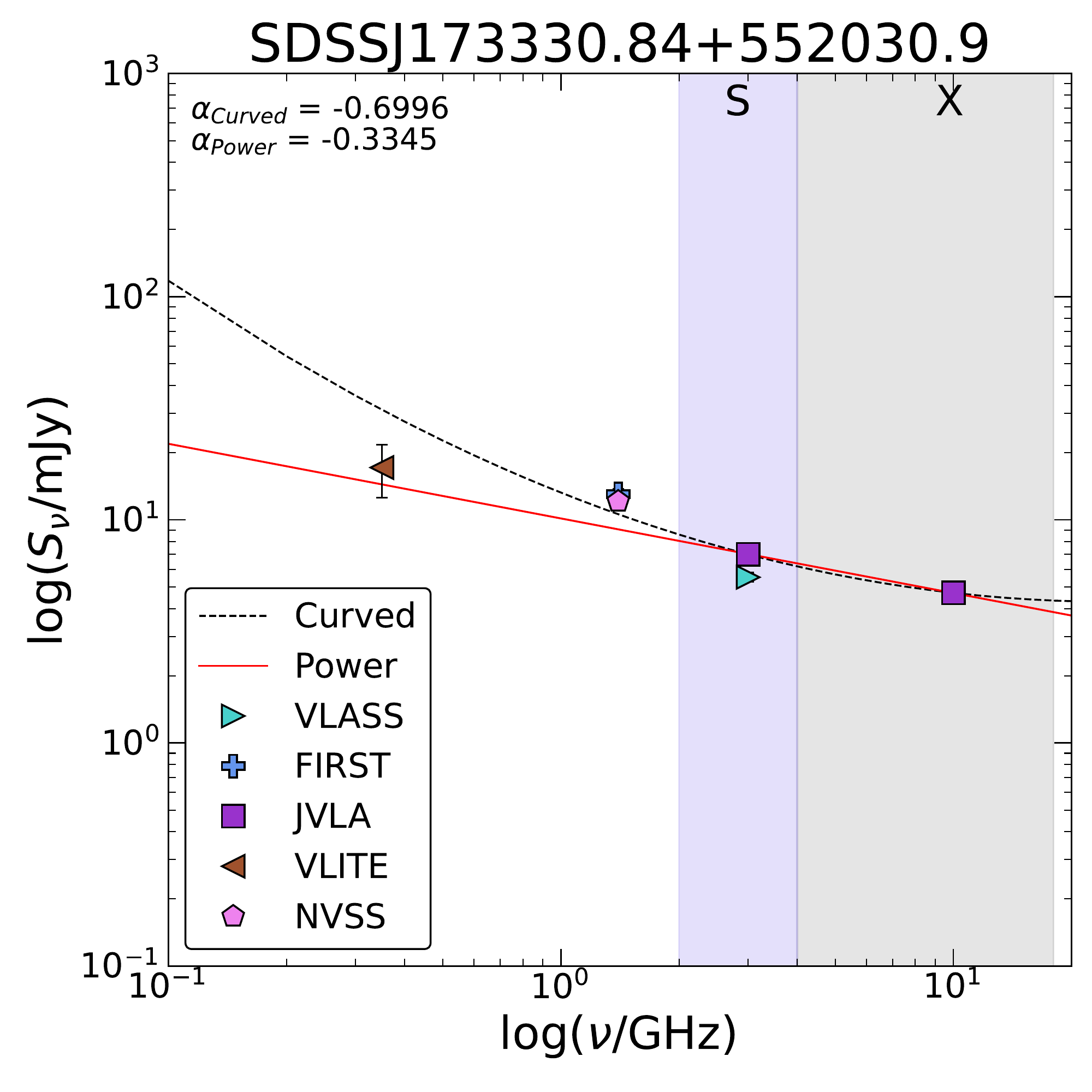}
\end{minipage} \\
\begin{minipage}[t]{0.45\linewidth}
  \centering
  \includegraphics[width=0.95\linewidth]{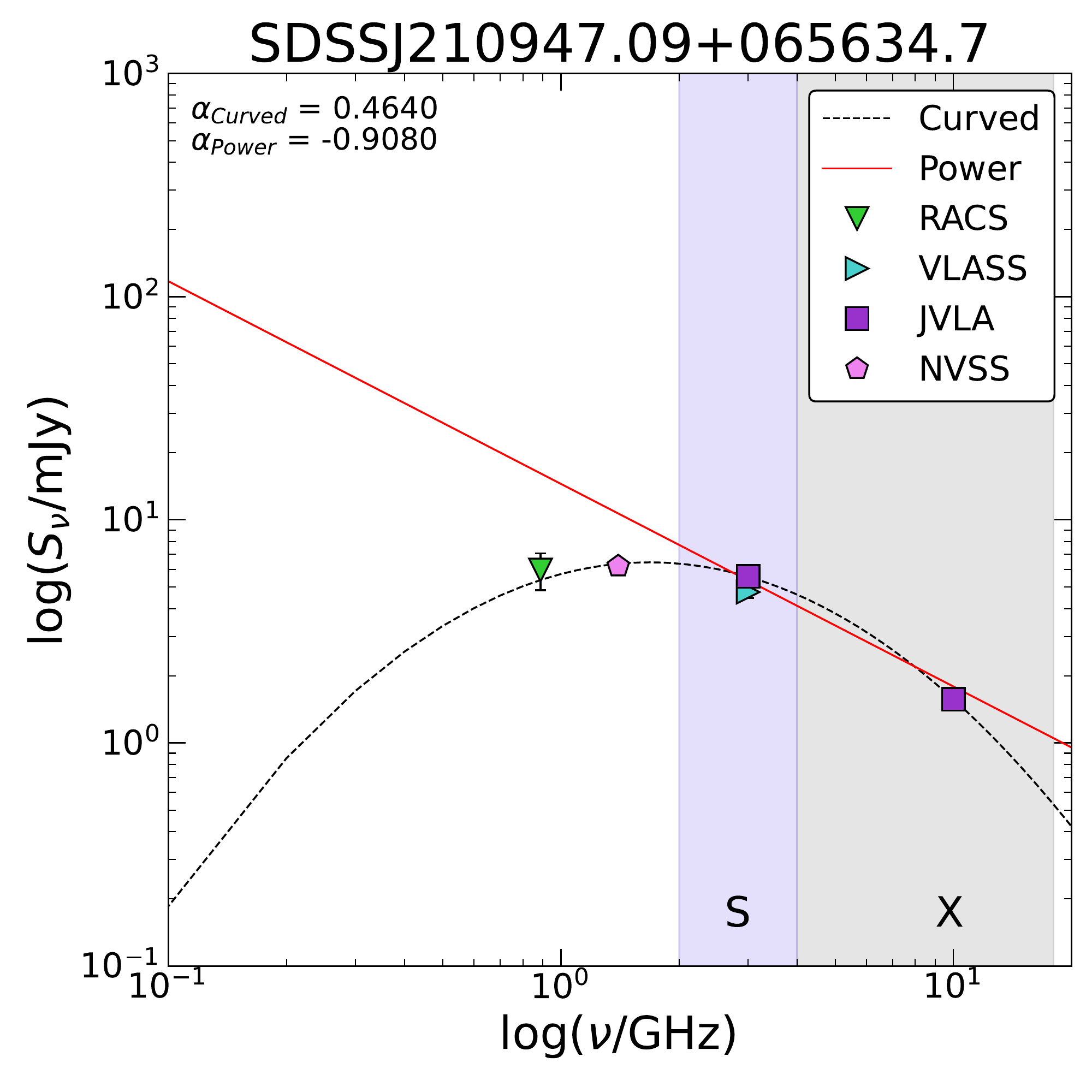}
\end{minipage} \\
  \end{longtable}%
  \captionof{figure}{Broadband radio spectra of unresolved targets showing the new, quasi-simultaneous VLA observations in purple squares, and the following archival measurements: VLASS (cyan, right-facing triangle), LOTSS (orange circle), FIRST (blue cross), VLITE (brown, left-facing triangle), NVSS (pink pentagon), VLBA (grey diamond), TGSS (hot pink star), and RACS (green, downward-facing triangle). For each source, the spectrum has been modeled with both a standard power law (red line) and a curved power law (dashed black line). The resultant spectral indices are listed.}%
  \addtocounter{table}{-1}%
 \label{fig:radiospecunresolved} 
\end{center}

\clearpage
\setcounter{figure}{13}

\begin{center}
  \begin{longtable}{cc}
      \begin{minipage}[t]{0.45\linewidth}
  \centering
  \includegraphics[width=0.95\linewidth]{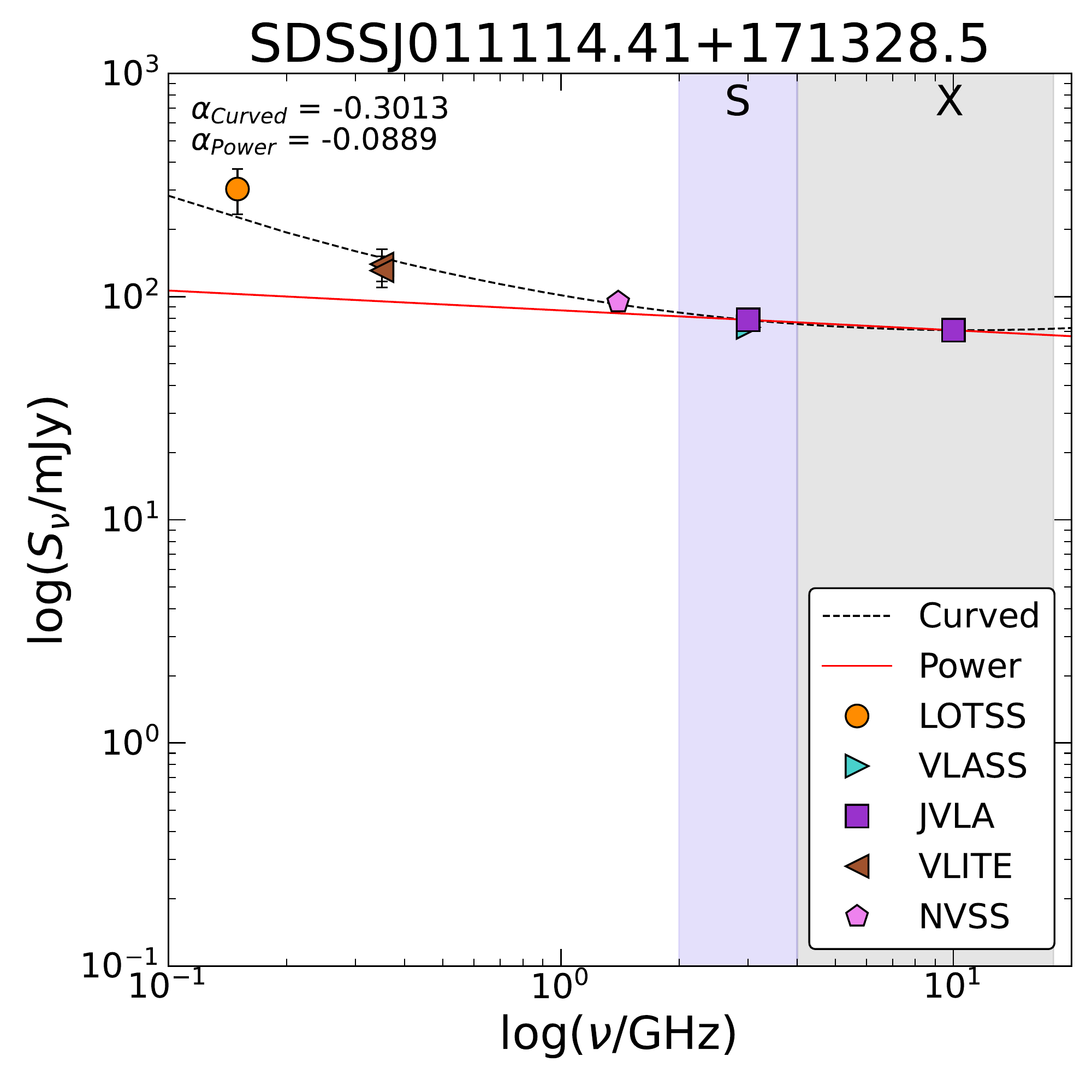}
\end{minipage} &
\begin{minipage}[t]{0.45\linewidth}
  \centering
  \includegraphics[width=0.95\linewidth]{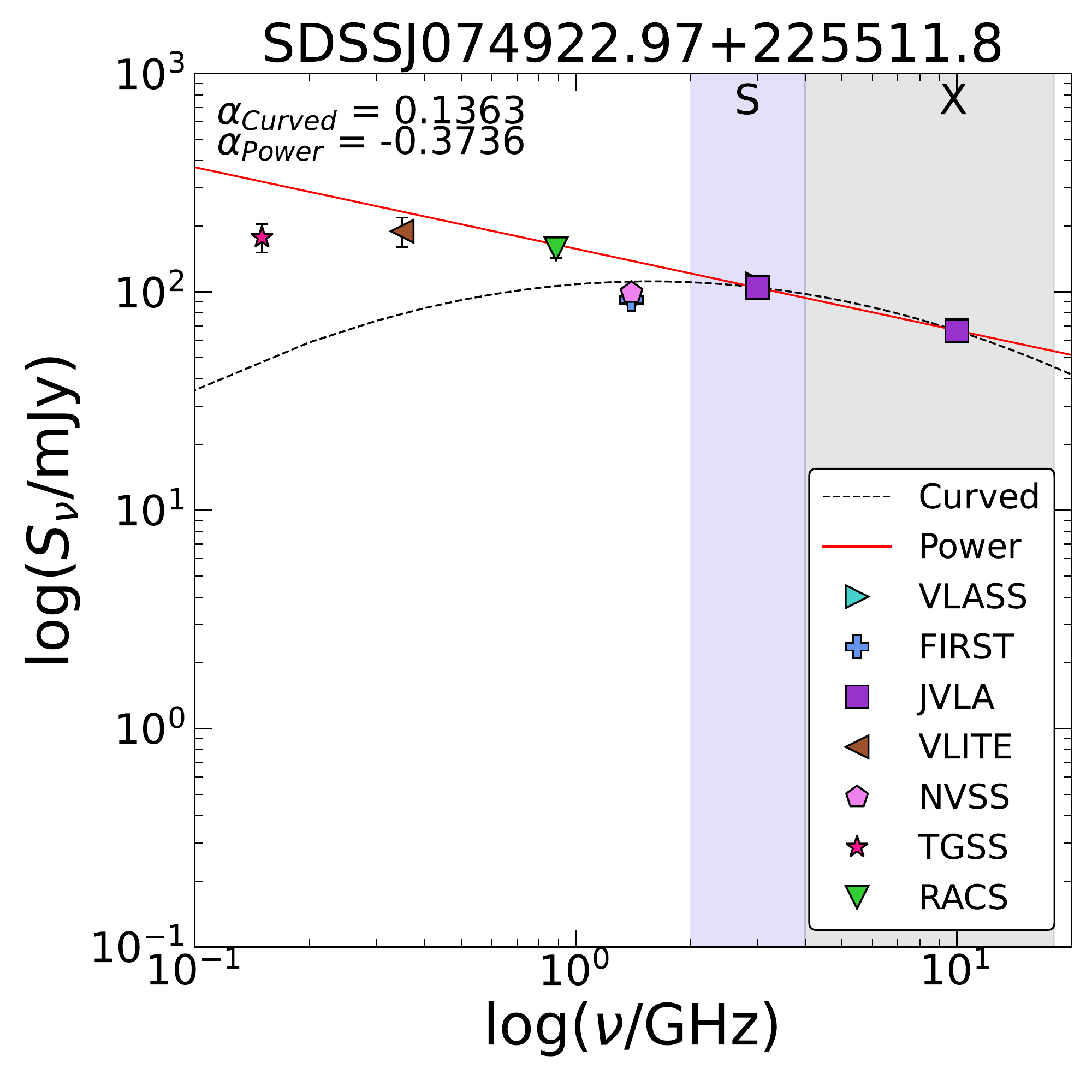}
\end{minipage} \\
\vspace{-0.2cm}
\begin{minipage}[t]{0.45\linewidth}
  \centering
  \includegraphics[width=0.95\linewidth]{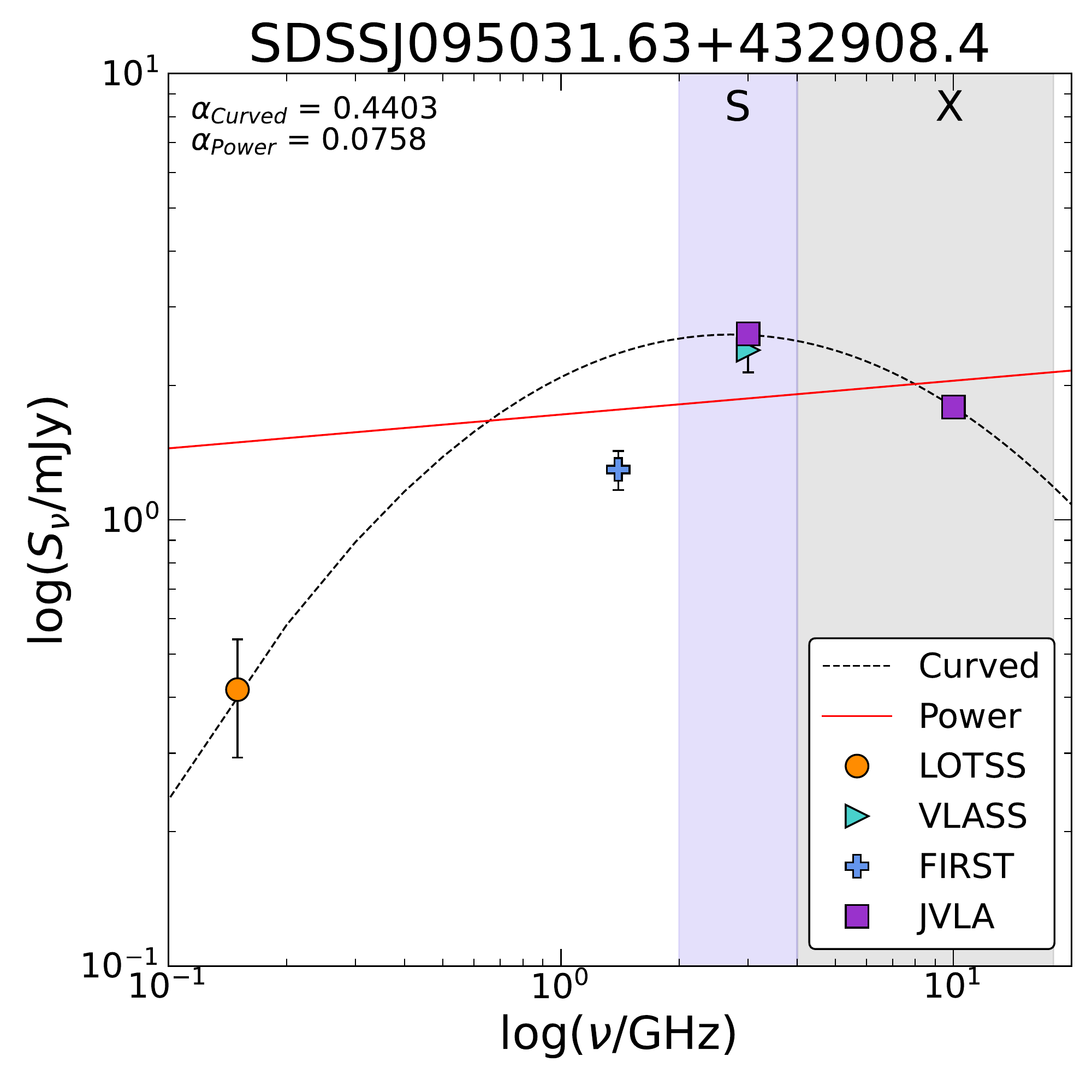}
\end{minipage} &
\begin{minipage}[t]{0.45\linewidth}
  \centering
  \includegraphics[width=0.95\linewidth]{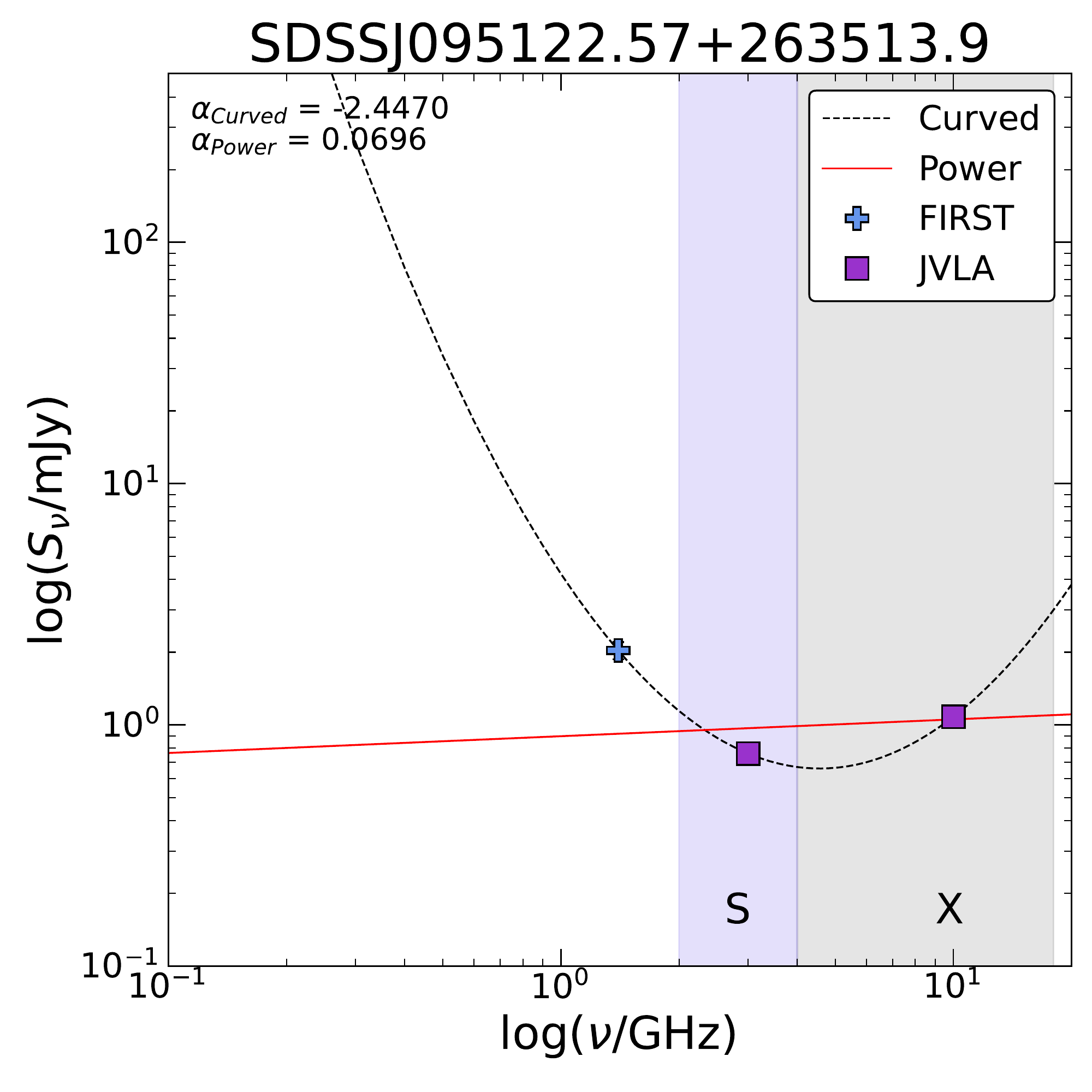}
\end{minipage} \\
\vspace{-0.2cm}
\begin{minipage}[t]{0.45\linewidth}
  \centering
  \includegraphics[width=0.95\linewidth]{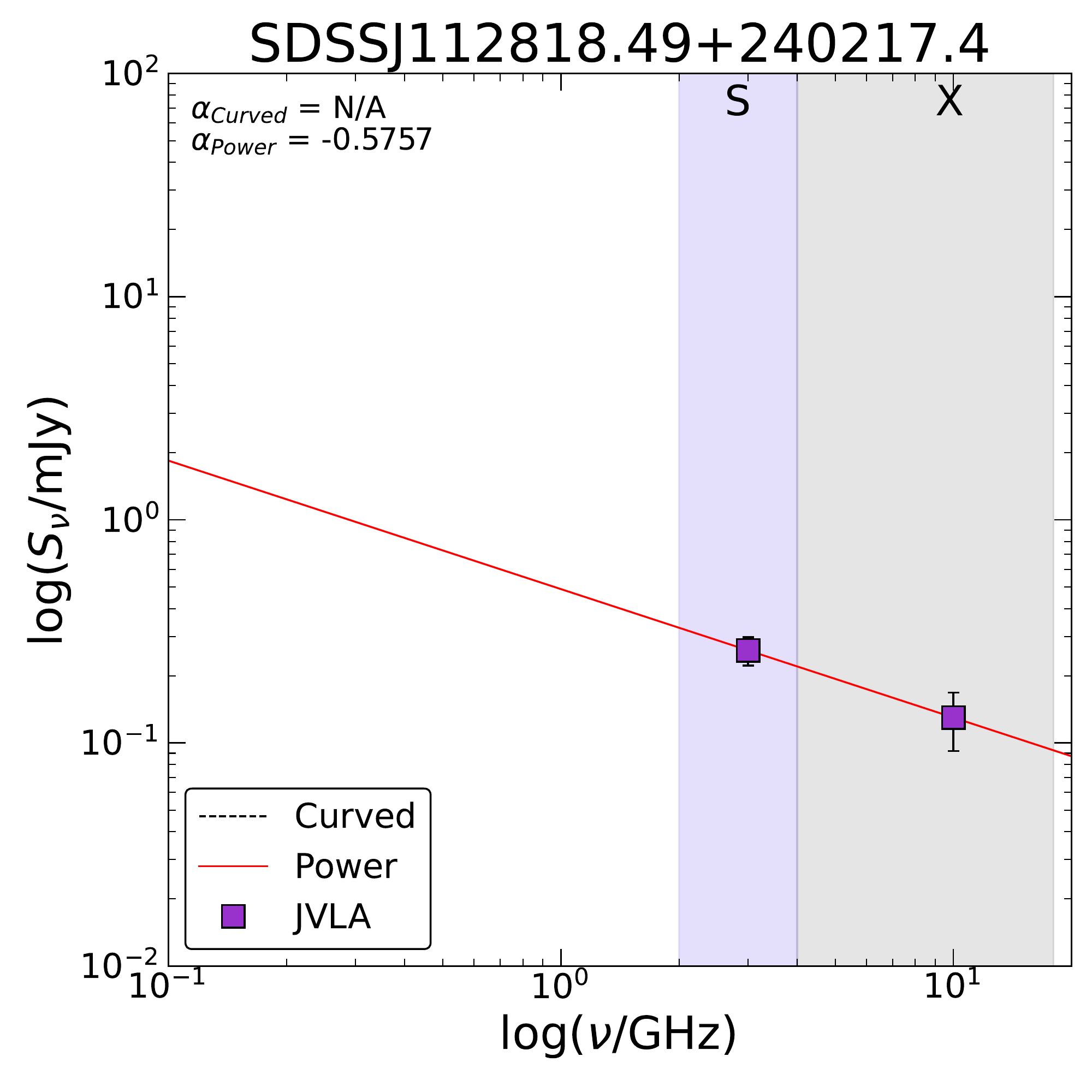}
\end{minipage} &
\begin{minipage}[t]{0.45\linewidth}
  \centering
  \includegraphics[width=0.95\linewidth]{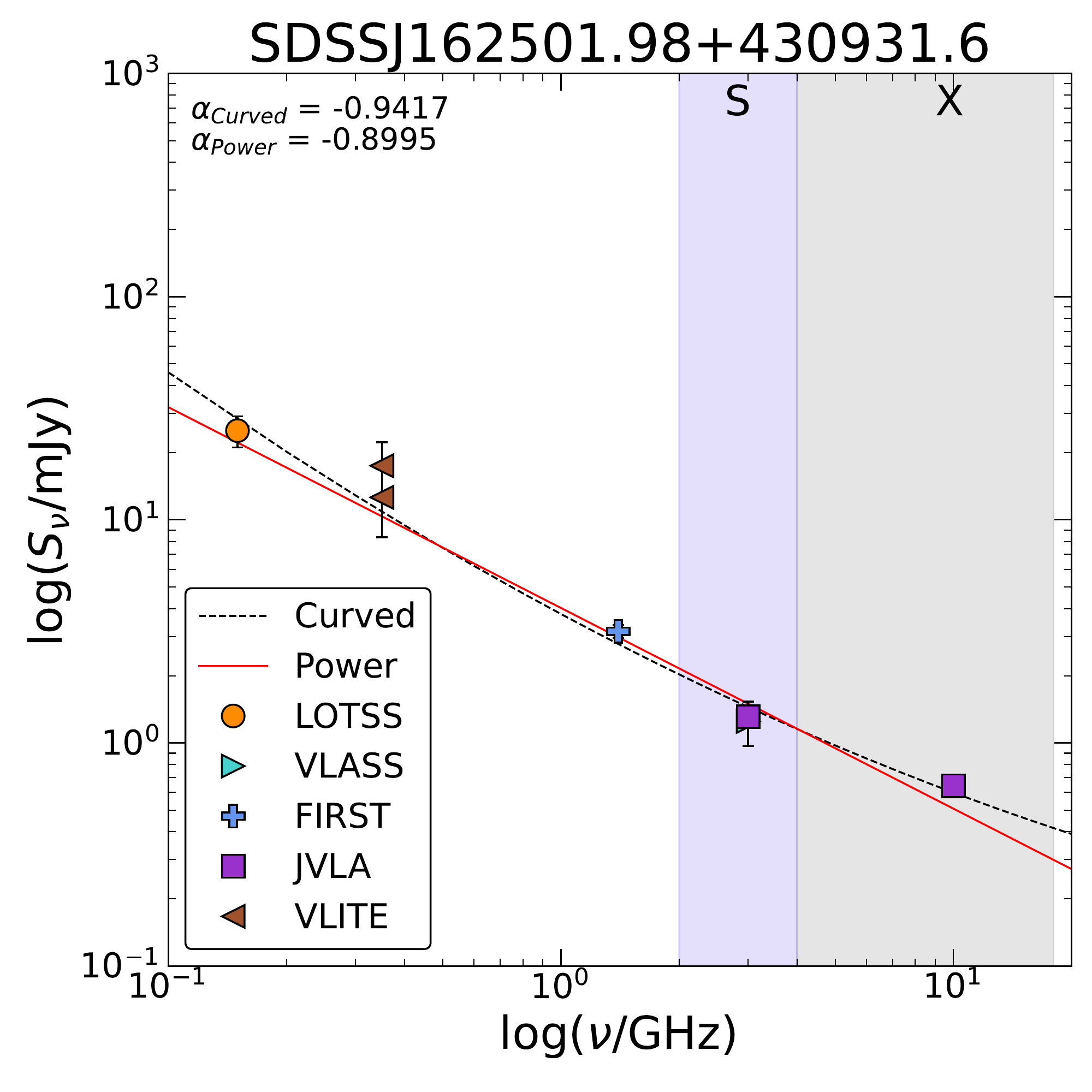}
\end{minipage} \\
  \end{longtable}%
  \captionof{figure}{Broadband radio spectra of multiple component targets showing the new, quasi-simultaneous VLA observations in purple squares, and the following archival measurements: VLASS (cyan, right-facing triangle), LOTSS (orange circle), FIRST (blue cross), VLITE (brown, left-facing triangle), NVSS (pink pentagon), VLBA (grey diamond), TGSS (hot pink star), and RACS (green, downward-facing triangle). For each source, the spectrum has been modeled with both a standard power law (red line) and a curved power law (dashed black line). The resultant spectral indices are listed.}%
  \addtocounter{table}{-1}%
 \label{fig:radiospecmulticomp} 
\end{center}

\clearpage
\setcounter{figure}{14}

\begin{center}
  \begin{longtable}{cc}
\begin{minipage}[t]{0.45\linewidth}
  \centering
  \includegraphics[width=0.95\linewidth]{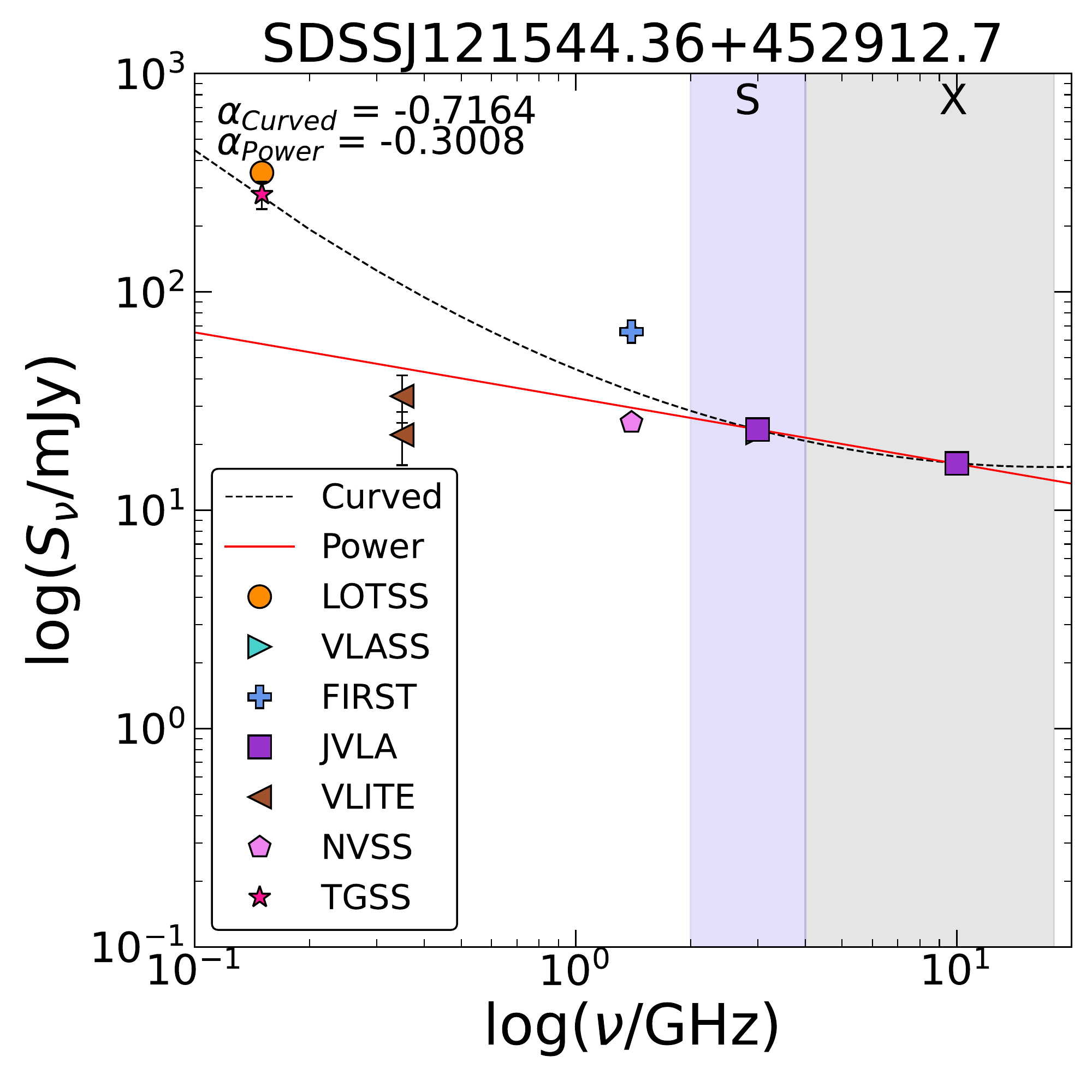}
\end{minipage} &
\begin{minipage}[t]{0.45\linewidth}
  \centering
  \includegraphics[width=0.95\linewidth]{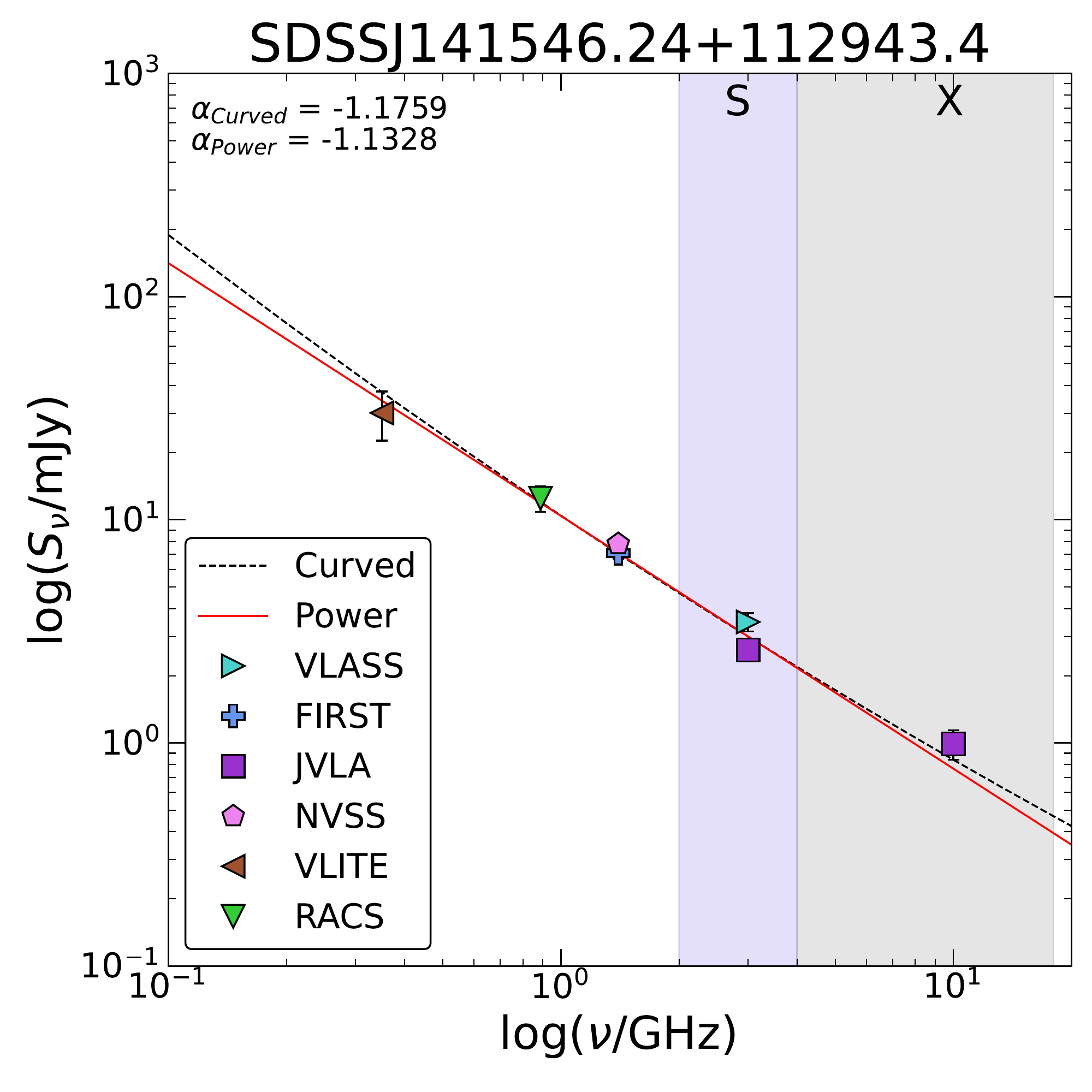}
\end{minipage} \\
\vspace{-0.2cm}
\begin{minipage}[t]{0.45\linewidth}
  \centering
  \includegraphics[width=0.95\linewidth]{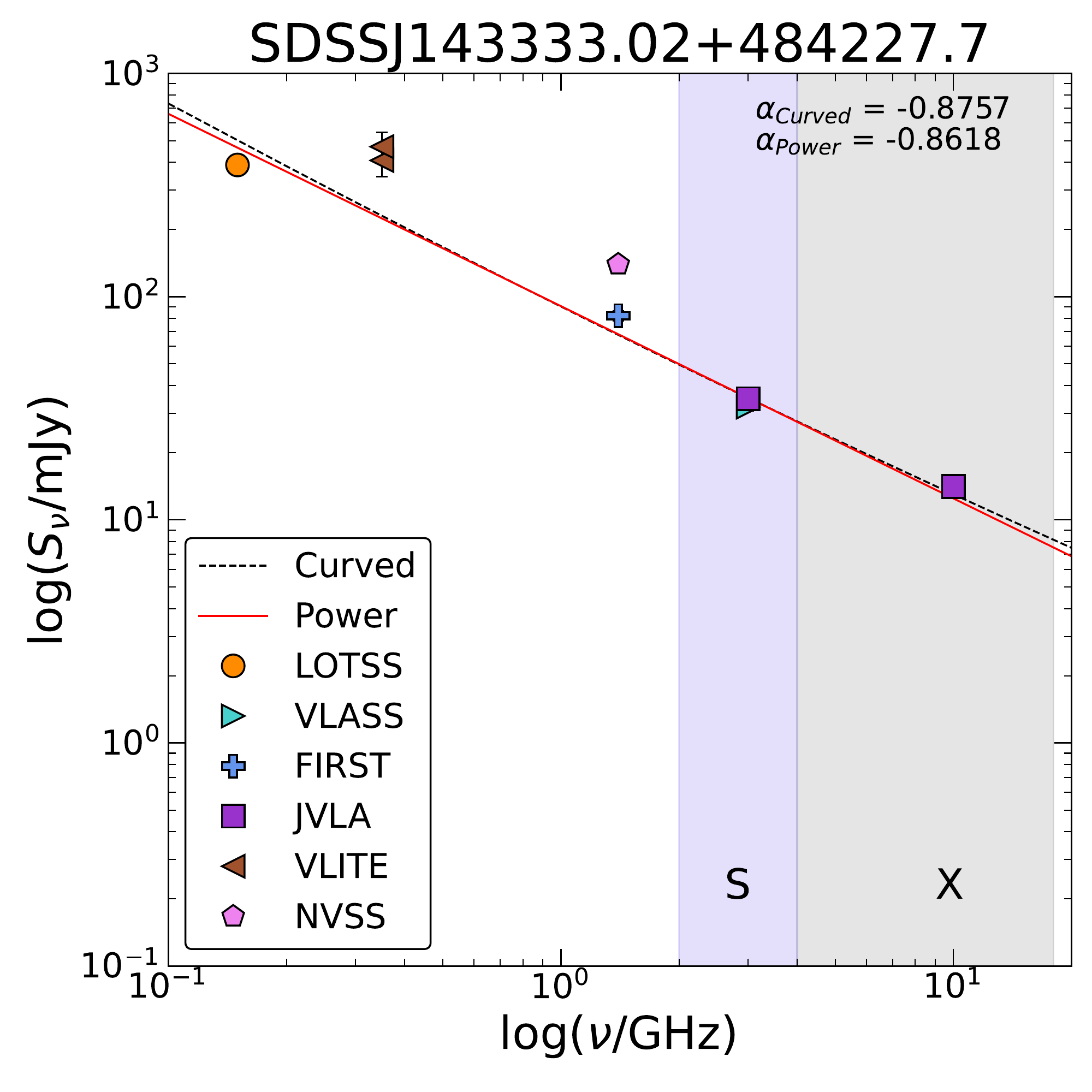}
\end{minipage} \\
  \end{longtable}%
  \captionof{figure}{Broadband radio spectra of extended targets showing the new, quasi-simultaneous VLA observations in purple squares, and the following archival measurements: VLASS (cyan, right-facing triangle), LOTSS (orange circle), FIRST (blue cross), VLITE (brown, left-facing triangle), NVSS (pink pentagon), VLBA (grey diamond), TGSS (hot pink star), and RACS (green, downward-facing triangle). For each source, the spectrum has been modeled with both a standard power law (red line) and a curved power law (dashed black line). The resultant spectral indices are listed.}%
  \addtocounter{table}{-1}%
 \label{fig:radiospecextended} 
\end{center}

\end{document}